\documentclass[a4paper,preprint,11pt]{elsarticle}
\usepackage[utf8]{inputenc}

\journal{arXiv}

\usepackage[margin=2cm]{geometry}
\usepackage{booktabs}
\usepackage{amsmath,amssymb,amsfonts}
\usepackage{hyperref}
\usepackage{bm}
\hypersetup{
    colorlinks=true,
    linkcolor=blue,
    filecolor=magenta,      
    urlcolor=cyan!80!blue,
    citecolor = green!80!black, 
    }

\usepackage{siunitx}
\usepackage{xcolor}
\definecolor{pyblue}{HTML}{1F77B4}
\definecolor{pyorange}{HTML}{FF7F0E}
\definecolor{pygreen}{HTML}{2CA02C}
\definecolor{pyred}{HTML}{D62728}
\definecolor{pycyan}{HTML}{17BECF}

\usepackage{algorithm}
\usepackage{algpseudocode}

\usepackage{caption}
\usepackage{subcaption}

\usepackage{xfrac} 
\usepackage{chemformula}
\usepackage{subcaption}

\begin{document}

\begin{frontmatter}
\title{Neural Physics: Using AI Libraries to Develop Physics-Based Solvers for Incompressible Computational Fluid Dynamics}
\author[IC]{Boyang Chen} 
\ead{boyang.chen16@imperial.ac.uk}
\author[IC,IX]{Claire E. Heaney\corref{cor1}}
\ead{c.heaney@imperial.ac.uk}
\author[IC,IX,DA]{Christopher C. Pain} 
\ead{c.pain@imperial.ac.uk}

\cortext[cor1]{Corresponding author}

\address[IC]{Applied Modelling and Computation Group, Department of Earth Science and Engineering, Imperial College London, SW7 2AZ, UK}
\address[IX]{Centre for AI-Physics Modelling, Imperial-X, Imperial College London, W12 7SL, UK}
\address[DA]{Data Assimilation Laboratory, Data Science Institute, Imperial College London, SW7 2AZ, UK}

\begin{abstract}
Numerical discretisations of partial differential equations (PDEs) can be written as discrete convolutions, which, themselves, are a key tool in AI libraries and used in convolutional neural networks (CNNs). We therefore propose to implement numerical discretisations as convolutional layers of a neural network, where the weights or filters are determined analytically rather than by training. Furthermore, we demonstrate that these systems can be solved entirely by functions in AI libraries, either by using Jacobi iteration or multigrid methods, the latter realised through a U-Net architecture. Some advantages of the Neural Physics approach are that (1)~the methods are platform agnostic; (2)~the resulting solvers are fully differentiable, ideal for optimisation tasks; and (3)~writing CFD solvers as (untrained) neural networks means that they can be seamlessly integrated with trained neural networks to form hybrid models. We demonstrate the proposed approach on a number of test cases of increasing complexity from advection-diffusion problems, the non-linear Burgers equation to the Navier-Stokes equations. We validate the approach by comparing our results with solutions obtained from traditionally written code and common benchmarks from the literature. We show that the proposed methodology can solve all these problems using repurposed AI libraries in an efficient way, without training, and presents a new avenue to explore in the development of methods to solve PDEs with implicit methods. 
\end{abstract}


\end{frontmatter}

\section{Introduction}
In recent years, Artificial intelligence (AI) has provided the ability to address challenges that previously could not be solved by classical means, such as the identification of images, spam control, predictive writing and developing self-driving cars~\citep{Dhall2019}. This huge success has encouraged not only the development of better and more sophisticated methods within the field of AI, but also the development of open source libraries devoted to simplify the entrance to and the use of AI. Some of the most popular libraries, such as PyTorch~\citep{pytorch}, TensorFlow~\citep{tensorflow}, scikit-learn~\citep{scikit}, XGBoost~\citep{xgboost} and JAX~\citep{Jax}, are maintained by large companies yet are also contributed to by their associated communities. These libraries are well documented, highly optimised, easily deployable on different hardware architectures and, consequently, are widely used. To use AI libraries on different computer architectures, the user needs only to make minimal changes to the code, which is not the case for codes written in C\texttt{++} and Fortran, where programmers have to grapple with rewriting code or writing additional code in other languages in order to exploit the benefits of GPUs. As well as driving the continued development of accessible platform-independent code, AI has also been the inspiration behind the design of a number of specific computer architectures that are highly optimised for certain tasks commonly performed in AI, such as dense matrix-vector multiplication. Researchers have begun to exploit such developments in AI hardware and software for scientific computation, and have found matrix-vector multiplications can be performed much faster, demonstrated for reconstruction of magnetic resonance images~\citep{Lu2020,Lu2021}, the dynamics of many-body quantum systems~\citep{Morningstar2022}, quantum chemistry~\citep{Pederson2022}. The latest computers specially designed for AI workflows include Google's TPUs~\citep{Lewis2022}, Cerebras's CS-2~\citep{CS2} and CS-3~\citep{CS3}, and Graphcore's Intelligence Processing Units (IPU)~\citep{GraphCore_IPU}, referred to as AI chips, AI accelerators or AI processors. When designing these AI processors, companies are looking to reduce the large energy overheads seen in large clusters of CPUs and GPUs. For example, with nearly 1~million cores on a single chip, Cerebras's CS-2 has been reported to be up to 10~times as energy efficient as some other GPUs~\citep{reuters_cerebras}. Energy efficiency is becoming particularly important as we may not be able to afford to run large petascale or exascale simulations on current CPU clusters due to the need to reduce our energy usage and the associated impact on our planet~\citep{MIT_cerebras}. The potential for these new computers to speed up scientific calculations is significant. Cerebras's CS-3 has been applied to a molecular dynamics problems and achieved 1.2 million simulation steps per second, which is 700 times faster than the world's fastest supercomputer~\citep{Forbes}. This corresponds to performing 2 years' worth of GPU-based simulations in a single day. The approach proposed here, of numerically solving Partial Differential Equations (PDEs) using methods in machine-learning libraries, therefore brings the use of these new AI processors closer to scientific applications and makes the solution of the large systems of equations formed by the discretisation of PDEs, potentially more environmentally friendly.  

Solving numerical discretisations of PDEs, exactly, using machine-learning libraries, relies on the observation that numerical discretisations and convolutional neural networks (CNNs) have in common the discrete convolution. \citet{Trefethen} observed that numerical approximations of differential operators are equivalent to  convolutions whose filters are chosen in such a way that the particular numerical approximation is replicated. This idea of representing numerical discretisations as convolutional layers has been used by \citet{Zhao2020} and \citet{Wang2022} to enable CFD problems to benefit from the  computational efficiency of GPUs. \citet{Zhao2020} implement an explicit high-order finite-difference discretisation of the Navier-Stokes equations using TensorFlow. A constrained interpolation profile scheme is used to solve the advection term and preconditioned conjugate gradient method is used to solve the Poisson equation. The authors obtain good qualitative agreement when validating their results against the literature for 2D lid-driven cavity flow and flow past a cylinder. For numbers of grid points up to just over~\num{600000}, their proposed method is over 10 times faster (running on GPUs) than their original Fortran-based code (running on CPUs). \citet{Wang2022} present a similar idea, also using TensorFlow, to implement a finite-difference discretisation of the variable-density Navier-Stokes equations. They deploy their models on TPUs and achieve good weak and strong scaling. 2D and 3D Taylor-Green vortex benchmarks are used for validation, and the expected convergence rates are obtained as the spatial resolution is increased using up to 10$^{9}$~grid points. The capability to model turbulent flows is demonstrated by modelling a planar jet, the results of which show good statistical agreement with reference solutions. We will extend this idea and develop multigrid solvers within an AI framework. 

We propose a novel way of developing CFD software based on methods and techniques available in AI libraries, which can simplify both development and deployment whilst performing the same operations that CFD software would normally do. In this way, for example, a convolutional neural network (CNN) can be repurposed to become a structured multigrid solver, where downsampling and upsampling convolutional layers are equivalent to the restriction and prolongation operators, and the smoothing steps are performed using Jacobi iterations also realised through a convolutional layer. By repurposing readily available machine-learning software using Neural Physics, we expect that the development of CFD can benefit in at least four areas: (i)~deployment to different computer architectures (such as new AI processors with distributed memory) will be more efficient, and the resulting code will be specifically optimised for that architecture; (ii)~optimisation tasks will be more viable due to the differentiability of the solvers; (iii)~the development of hybrid physics-based and data-driven methods~\citep{Um2020,Beck2021} becomes very convenient within our framework, as both types of method can be represented by neural networks (untrained and trained) removing any communication issues between programs in different languages;  (iv)~reduction of the implementation time and lowering of the entry barrier by making the code more ``high-level'' than current CFD software implementations. 

The approach we outline is different to that of physics-informed neural networks (PINNs)~\citep{Zheng2020} and equation discovery~\citep{Long2019}. We specify the weights of convolutional layers in order to realise, exactly, a particular numerical discretisation. Both PINNs and PDENet~2.0~\citep{Long2019} determine weights of neural networks by training, which introduces an additional approximation to that introduced by discretisation alone. In physics-informed approaches~\citep{Zheng2020}, researchers have used operators that are related (within a multiplicative constant) to finite difference discretisations. Nevertheless, a neural network is trained with a loss that attempts to constrain the network to obey particular physical laws, so the discretisation will not be satisfied exactly. In PDENet~2.0, \citet{Long2019} attempt to discover new physical relationships from data, by encoding certain derivatives exactly through analytically specifying some of the weights of the filters of convolutional operations but training to find the other weights. The approach we outline is also different to other recent differentiable CFD codes. \citet{Bezgin2023} use JAX's extended NumPy module (a fully-differentiable version of the popular NumPy package) to build a fully differentiable CFD solver with a more classical approach.  \citet{Farsi2025} exploit the differentiability of symbolic operators in FireDrake. Although both these approaches will enable linkage with neural networks, by embedding discretisations and solution algorithms directly into neural network architectures, Neural Physics allows a closer integration of physics-based NN layers (weights determined analytically) and machine-learning based NN layers (weights found through learning), where the resulting coupled model can even be solved implicitly. 

Our primary contribution in this paper is the development of a CFD solver, written as a neural network, that solves the governing equations and is capable of predicting complex fluid dynamics. By formulating discretised PDEs as CNNs, the system can be solved with an accuracy matching that of conventional Fortran or \verb|C++| solvers. We are able to mimic a multigrid solver with a U-Net in order to solve incompressible flows. No training is required, as weights are determined by the chosen discretisation scheme. We refer to this approach as Neural Physics. 

This paper is the first in a set of related papers~\citep{Phillips2023,Chen2023,Phillips2022,Chen2024,Nadimy2025,Naderi2024,Li2024DGFEM,Li2024ERT,Huang2025} on the Neural Physics approach, that is, solving discretised systems with (untrained) neural networks. After benchmarking and validating the approach for single-phase flows in this paper, we have extended the idea to model more complex physics, including neutron diffusion~\citep{Phillips2023} and multiphase flows~\citep{Chen2023}. To use higher-order finite element schemes, we have introduced a new finite element method which has the same stencil for every node and is therefore amenable to representation by convolutional layers~\citep{Phillips2022}. We have also applied the method to discretisations of the 2D shallow water equations to model flooding~\citep{Chen2024,Nadimy2025} and the Discrete Element Method to model particle systems~\cite{Naderi2024}. All these examples use structured grids, however we have extended the work to unstructured meshes using space-filling curves and graph neural networks~\cite{Li2024DGFEM}. Benefitting from the automatic differentiability of neural-network discretisations, we have also used this approach to solve inverse problems, including characterising the subsurface with electrical resistivity tomography~\citep{Li2024ERT} and estimating parameters for land-surface models~\citep{Huang2025}. 

The structure of the remaining paper is as follows. The methods involved in expressed  discretised systems of PDEs as neural networks, and solving these systems also with neural networks is explained in Section~\ref{sec:methods}. Results are presented in Section~\ref{sec:results}, a discussion follows in Section~\ref{sec:discussion} and concluding remarks are given in Section~\ref{sec:conclusion}.

\section{Methods}\label{sec:methods}

The methods in this section are derived largely in 2D, but can be easily modified for 1D and 3D. 

\subsection{Advection-diffusion equation}
\label{AD} 
The advection-diffusion equation solved here is given by:
\begin{equation}
\frac{\partial T}{\partial t} + u\frac{\partial T}{\partial x} +  v\frac{\partial T}{\partial y} + 
\sigma T - \nu \nabla^2 T =s \,,
\label{adv-diff-eqn}
\end{equation}
in which $T$ is a scalar concentration field, $(u,v)$ is the advection velocity, $\sigma$~is an absorption term, $\nu$~is the constant diffusivity and $s$~is the source. In the advection-diffusion simulations presented here, $\sigma=s=0$ although non-zero values of $\sigma$ will be used when this equation is solved as the momentum equations in the Navier-Stokes Equations~\eqref{NS-eqn}.

For conciseness we derive the discretisation of the advection-diffusion equation with a simple central difference scheme for advection, a second-order central difference scheme for diffusion and a second order scheme in time; although we do present results for a first order upwind advection scheme and more complex central difference discretisations of advection and diffusion for flow past a bluff body. For a regular grid ($\Delta x = \Delta y$) and an advection velocity of $(u,v)= (1,1)$, the spatial derivatives can be written as
\begin{eqnarray}
\left(\frac{\partial T}{\partial x}\right)_{\ell,\,m} &\approx& \frac{T_{\ell+1,\,m}-T_{\ell-1,\,m}}{2\Delta x} \,, \\[2mm]
\left(\frac{\partial T}{\partial y}\right)_{\ell,\,m} &\approx& \frac{T_{\ell,\,m+1}-T_{\ell,\,m-1}}{2\Delta x} \,, \\[2mm]
- \left(\frac{\partial^2 T}{\partial x^2}\right)_{\ell,\,m} &\approx& \frac{-T_{\ell+1,\,m}+2T_{\ell,\,m}-T_{\ell-1,\,m}}{\Delta x^2} \,, \\[2mm]
- \left(\frac{\partial^2 T}{\partial y^2}\right)_{\ell,\,m} &\approx&\frac{-T_{\ell,\,m+1}+2T_{\ell,\,m}-T_{\ell,\,m-1}}{\Delta x^2} \,,
\end{eqnarray}
in which the solution field at the $\ell$th node in the increasing $x$ direction and the $m$th node in the $y$-direction is written as $T_{\ell,\,m}$. To implement a predictor-corrector scheme, we define the following
\begin{subequations}\label{eq:predictor_equations}
\begin{eqnarray}
\left(\frac{\partial T}{\partial x}\right)_{\ell,\,m}^{n+\sfrac{1}{2}} &\approx& \frac{(\widetilde{T}_{\ell+1,\,m}^{n+1}+T_{\ell+1,\,m}^n)-(\widetilde{T}_{\ell-1,\,m}^{n+1}+T_{\ell-1,\,m}^n)}{4\Delta x} \,, \\[2mm]
\left(\frac{\partial T}{\partial y}\right)_{\ell,\,m}^{n+\sfrac{1}{2}} &\approx& \frac{(\widetilde{T}_{\ell,\,m+1}^{n+1}+T_{\ell,\,m+1}^n)-(\widetilde{T}_{\ell,\,m-1}^{n+1}+T_{\ell,\,m-1}^n)}{4\Delta x} \,,\\[2mm]
- \left(\frac{\partial^2 T}{\partial x^2}\right)_{\ell,\,m}^{n+\sfrac{1}{2}} &\approx& \frac{-(\widetilde{T}_{\ell+1,\,m}^{n+1}+T_{\ell+1,\,m}^n)+2(\widetilde{T}_{\ell,\,m}^{n+1}+T_{\ell,\,m}^n)-(\widetilde{T}_{\ell-1,\,m}^{n+1}+T_{\ell-1,\,m}^n)}{2\Delta x^2} \,,\\[2mm]
- \left(\frac{\partial^2 T}{\partial y^2}\right)_{\ell,\,m}^{n+\sfrac{1}{2}} &\approx&\frac{-(\widetilde{T}_{\ell,\,m+1}^{n+1}+T_{\ell,\,m+1}^n)+2(\widetilde{T}_{\ell,\,m}^{n+1}+T_{\ell,\,m}^n)-(\widetilde{T}_{\ell,\,m-1}^{n+1}+T_{\ell,\,m-1}^n)}{2\Delta x^2} \,,
\end{eqnarray}
\end{subequations}
in which $\widetilde{T}_{\ell,\,m}^{n+1}$ represents the best guess for the solution at node $\ell,\,m$ at time level~$n+1$. Initially, the best guess for $\widetilde{T}_{\ell,\,m}^{n+1}$ is $T_{\ell,\,m}^{n}$. A prediction for $\widetilde{T}_{\ell,\,m}^{n+1}$ is calculated by using this best guess, Equations~\eqref{eq:predictor_equations} and the following
\begin{equation}\label{eq:predictor}
    T_{\ell,\,m}^{n+1} = T_{\ell,\,m}^n - \Delta t\left( \left(\frac{\partial T}{\partial x}\right)_{\ell,\,m}^{n+\sfrac{1}{2}} + \left(\frac{\partial T}{\partial y}\right)_{\ell,\,m}^{n+\sfrac{1}{2}} - \nu\left(\frac{\partial^2 T}{\partial x^2}\right)_{\ell,\,m}^{n+\sfrac{1}{2}} - \nu\left(\frac{\partial^2 T}{\partial y^2}\right)_{\ell,\,m}^{n+\sfrac{1}{2}} \right) \,  ,
\end{equation}
in which $\Delta t$ is the time step size. 
Using Equation~\eqref{eq:predictor} to update the best guess, this is substituted back into the equations in system~\eqref{eq:predictor_equations} and the result substituted into Equation~\eqref{eq:predictor} to give the corrected approximation to the solution at node $\ell,\,m$ at time level $n+1$. Using square brackets, we define a 3~by 3 matrix of values centred around cell $\ell,\,m$ as
\begin{equation}\label{eq:grid_values}
[T]_{\ell,\,m}^{n} := \left( \begin{array}{lll} T_{\ell-1,\,m-1}^n & T_{\ell-1,\,m}^n & T_{\ell-1,\,m+1}^n\\ T_{\ell,\,m-1}^n & T_{\ell,\,m}^n & T_{\ell,\,m+1}^n\\
T_{\ell+1,\,m-1}^n & T_{\ell+1,\,m}^n & T_{\ell+1,\,m+1}^n
\end{array}\right)\,.
\end{equation}
With this, we can rewrite the predictor-corrector scheme from Equations~\eqref{eq:predictor_equations} and~\eqref{eq:predictor} as 
\begin{equation}\label{eq:pred-corr-adv-diff}
    T_{\ell,\,m}^{n+1} = T_{\ell,\,m}^{n} - \left(\frac{\Delta t}{2\Delta x}\left( \begin{array}{rrr} 0&-1&0\\-1&0&1\\0&1&0\\\end{array}\right) 
    + \frac{\nu\Delta t}{\Delta x^2}\left( \begin{array}{rrr} 0&-1&0\\-1&4&-1\\0&-1&0\\\end{array}\right)\right)\ast \frac{1}{2}\left( [\widetilde{T}]_{\ell,\,m}^{n+1} + [T]_{\ell,\,m}^{n}\right)\,,
\end{equation} 
where the symbol $\ast$ represents a discrete convolution, which, for cell $\ell,\,m$, is defined as follows:
\begin{equation}\label{define_convolution}
\left.A\ast T\right|_{\ell,m} :=  \sum_{i=-N}^N\sum_{j=-N}^N  A_{i,j} \,T_{\ell+i,m+j}\,.
\end{equation}
The matrix $A$ is of dimension $2N+1$ by $2N+1$ and represents the coefficients associated with the numerical discretisation. For a 5-point stencil represented by a 3 by 3 matrix, $N=1$. The definition in~\eqref{eq:pred-corr-adv-diff} could be considered to be that of the cross-correlation rather than a convolution, as, for some, the definition of the discrete convolution reverses the sign of the indices $i$ and $j$ in the sum for one term, e.g. $A_{i,j} T_{\ell-i,m-j}$. In fact, the PyTorch manual does note that their convolution functions, \texttt{Conv1D}, \texttt{Conv2D} and \texttt{Conv3D}, are actually cross-correlations~\citep{pytorch-crosscorrelation}. We adopt the definition widely used in machine-learning communities, that Equation~\eqref{define_convolution} is taken as the discrete convolution~\citep{Long2019}. 

Equation~\eqref{eq:pred-corr-adv-diff} represents both predictor and corrector stages of the time iteration. For the predictor stage, Equation~\eqref{eq:pred-corr-adv-diff} is solved with $\widetilde{T}^{n+1}_{\ell,m}=T^n_{\ell,m}$. To calculate the corrector stage,  Equation~\eqref{eq:pred-corr-adv-diff} is solved using the solution of the predictor stage for $\widetilde{T}^{n+1}_{\ell,m}$. Writing the predictor-corrector scheme in this way reveals the equivalence between this discretisation and that of a convolutional layer which has linear activation functions, and a 3~by 3 kernel or filter with weights of
\begin{equation}
\label{A}
A:=
  \frac{\Delta t}{2\Delta x}\left(\begin{array}{rrr} 0&-1&0\\-1&0&1\\0&1&0\\\end{array}\right) +   \frac{\nu\Delta t}{\Delta x^2}\left(\begin{array}{rrr} 0&-1&0\\-1&4&-1\\0&-1&0\\\end{array}\right)\,,
\end{equation}
operating on a grid of values $ [\widetilde{T}]_{\ell,\,m}^{n+1} + [T]_{\ell,\,m}^{n}$ for all values of $\ell,\,m$. With this definition of $A$, Equation~\eqref{eq:pred-corr-adv-diff} can be written as: 
\begin{equation}\label{disc-as-filter}
    {T}_{\ell,\,m}^{n+1} = T_{\ell,\,m}^{n} 
    - 
    A \ast \frac{1}{2}\left( [\widetilde{T}]_{\ell,\,m}^{n+1} + [T]_{\ell,\,m}^{n}\right)\,,
\end{equation}
which can then be implemented by CNNs. For upwind differencing of the advection terms, Equation~\eqref{A} becomes the 3~by 3 filter with weights of
\begin{equation}
\label{A-up}
A:=
  \frac{\Delta t}{\Delta x}\left(\begin{array}{rrr} 0&0&0\\0&-2&1\\0&1&0\\\end{array}\right) + \frac{\nu\Delta t}{\Delta x^2}\left(\begin{array}{rrr} 0&-1&0\\-1&4&-1\\0&-1&0\\\end{array}\right)\,. 
\end{equation}

The boundary conditions are enforced in 2D by adding two additional rows of nodes, one at the top of the grid and the other at the bottom of the grid, and two additional columns of nodes, one at the left of the grid and the other at the right of the grid. These additional ghost cells or halo nodes are known as padding by the machine-learning community. For our advection-diffusion test cases, we use zero padding, which enforces a zero solution on the boundaries. We also apply zero Dirichlet boundary conditions when solving the Burgers equation and Navier-Stokes equations. Bearing this in mind, we can rewrite Equation~\eqref{disc-as-filter} for the entire solution field as
\begin{equation}\label{disc-as-filter}
    \bm{\widetilde{T}}^{n+1} = \bm{T}^{n} 
    - 
    A \ast\left( \bm{\widetilde{T}}_{\text{pad}} ^{n+1} + \bm{T}_{\text{pad}}^{n}\right)\,,
\end{equation}
where $\bm{T}$ is the solution field on an $L$ by $M$ grid and $\bm{T}_{\text{pad}}$ is the solution field on an $L+2$ by $M+2$ grid (i.e., with the additional rows and columns of halo nodes or padding):
\begin{equation}
\bm{T}^n := \left( \begin{array}{lll} T_{1,1}^n & \cdots & T_{1,\,M}^n\\ \vdots & \ddots & \vdots\\
T_{L,\,1}^n & \cdots & T_{L,\,M}^n \end{array}\right)
\quad
\bm{T}_{\text{pad}} := \left( \begin{array}{lll} T_{0,\,0}^n & \cdots & T_{0,\,M+1}^n\\ \vdots & \ddots & \vdots\\
T_{L+1,\,0}^n & \cdots & T_{L+1,\,M+1}^n \end{array}\right). 
\end{equation}
In general, a solution field will have padding if a convolution is to be applied to it.


\subsection{Multigrid methods based on CNNs} 
\label{MGMG}

For incompressible flow, the most computationally expensive part of the simulation is solving the linear system of equations. For large systems of equations, multigrid methods are amongst the most efficient solvers. They scale linearly with the number of degrees of freedom, maintaining a constant number of iterations as the problem size increases~\citep{Osterlee,Gerya2009} (unlike for Jacobi or Gauss-Seidel methods). The idea behind multigrid methods is that it is beneficial to transfer information between solutions at different resolutions in order to increase the convergence rate~\cite{Osterlee,Thomas2001,Wesseling2001}.  This concept has similarities with a family of encoder-decoder neural networks, which includes the U-Net~\citep{Ronneberger2015}. In the encoder of a U-Net, an image is repeatedly downsampled enabling the network to learn progressively coarser-scale behaviour. The decoder acts in reverse and increases the resolution of the feature maps. In addition, there are skip connections which link layers of similar resolution in the encoder and decoder. As the decoder is reconstructing the images, the skip connections give it direct access to information at the same resolution from the encoder, which is thought to improve convergence of the network. Although there are examples of using ideas from multigrid to benefit neural networks~\citep{Thuerey2020}, we believe our paper is the first implementation of a multigrid method using a U-Net architecture. 
 
A geometric multigrid method works by creating a hierarchy of nested meshes and solving the equations at all levels, using the coarser meshes to correct the solution of the finer meshes. Multigrid methods consist of three basic operations: (i)~Smoothing: typically a couple of iterations of a linear solver such as Jacobi or Gauss-Seidel; (ii)~Restriction: interpolation of the residual from a given mesh to a coarser mesh; (iii)~Prolongation: extrapolation of the error correction from a coarse mesh to a finer mesh. A multigrid iteration or cycle consists of performing the following steps:
\begin{enumerate}[(1)]
    \item Smooth $\eta$ times;
    \item Compute residual and restrict to a coarser mesh;
    \item Smooth $\eta$ times;
    \item Repeat steps (2) and (3) until reaching the coarsest mesh desired;
    \item Solve the system (exactly or approximately);
    \item Prolongate from coarse mesh to a finer mesh and correct the solution from the finer mesh;
    \item Smooth $\phi$ times;
    \item Repeat steps (6) and (7) until the finest mesh is reached.
\end{enumerate}

In this work, we use the simplest multigrid method, often referred to as a sawtooth cycle~\citep{Tsuruga2018}, in which we restrict from the finest to the coarsest mesh, and then prolongate and smooth at each level from the coarsest to the finest mesh during each multigrid iteration, see Figure~\ref{fig:sawtooth}. 

\begin{figure}[htbp]
\centering
\begin{tikzpicture}
 \draw[color=gray!35!,ultra thick,dashed,dash pattern=on 6pt off 3pt] (-1.5,0) node[left, text=black] {finest grid} -- (10,0);
 \draw[color=gray!35!,ultra thick,dashed,dash pattern=on 6pt off 3pt] (-1.5,-1.5) node[left, text=black] {intermediate grid} -- (10,-1.5);
 \draw[color=gray!35!,ultra thick,dashed,dash pattern=on 6pt off 3pt] (-1.5,-3) node[left, text=black] {coarsest grid} -- (10,-3) ; 
    \node (A) at (0,0) [circle,draw=pyblue,fill=pyblue,scale=0.7] {};
    \node (B) at (1,-1.5) [circle,draw=pyblue,fill=pyblue,scale=0.7] {};
    \draw[->, ultra thick, >=latex, draw=pyblue] (A) -- node[midway, left] {\small restriction} (B);
    
    \node (C) at (2,-3) [circle,draw=pyblue,fill=pyblue,scale=0.7] {};
    \draw[->, ultra thick, >=latex, draw=pyblue] (B) -- node[midway, left] {\small restriction} (C);
    
    \node (D) at (4,-3) [circle,draw=pyblue,fill=pyblue,scale=0.7] {};
    \draw[->, ultra thick, >=latex, draw=pyblue] (C) -- node[midway, below] {\small smooth} (D);
    
    \node (E) at (5,-1.5) [circle,draw=pyblue,fill=pyblue,scale=0.7] {};
    \draw[->, ultra thick, >=latex, draw=pyblue] (D) -- node[midway, right] {\small prolongation} (E);
    
    \node (F) at (7,-1.5) [circle,draw=pyblue,fill=pyblue,scale=0.7] {};
    \draw[->, ultra thick, >=latex, draw=pyblue] (E) -- node[midway, above] {\small smooth} (F);
    
    \node (G) at (8,0) [circle,draw=pyblue,fill=pyblue,scale=0.6] {};
    \draw[->, ultra thick, >=latex, draw=pyblue] (F) -- node[midway, right] {\small prolongation} (G);
    
    \node (H) at (10,0) [circle,draw=pyblue,fill=pyblue,scale=0.6] {};
    \draw[->, ultra thick, >=latex, draw=pyblue] (G) -- node[midway, above] {\small smooth} (H);
\end{tikzpicture}
\caption{\label{fig:sawtooth}A schematic diagram showing how the smoothing, restriction and prolongation operations are arranged in one multigrid iteration or cycle.}
\end{figure}

\begin{figure}[htbp] 
\centering
\includegraphics[width=16cm,angle=0, trim=0mm 0mm 0mm 0.0mm, clip]{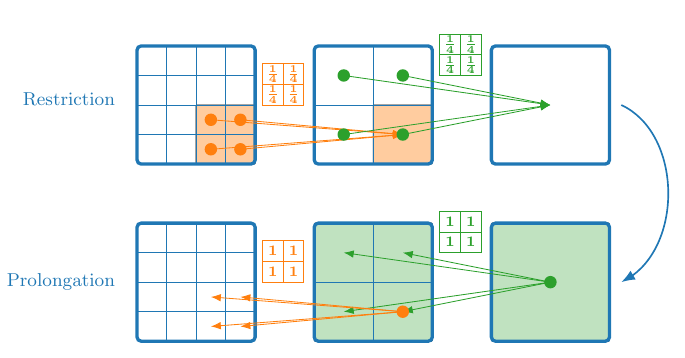}
\caption{Coarsening and restriction operations shown in 2D. Three grid levels are shown with the weights of the convolutional filters required for restriction and prolongation. The values at orange nodes are mapped from grid level~1 to~2 (and vice verse) whilst values at green nodes are mapped from level~2 to~3 (and vice versa).}
\label{c-and-r} 
\end{figure}

Figure~\ref{c-and-r} shows the restriction and prolongation operations for three mesh levels in a 2D example. During restriction, the values from 4 cells are averaged and sent to one cell in the grid at the next (coarser) level: see the orange nodes and the orange shaded area, for example. The restriction operator is represented by a 2~by 2 filter (of a convolutional layer) of values one quarter. During prolongation, the value in one cell is copied to four cells in the next (finer) grid: see the green nodes and green shaded area, for example. The prolongation operator is represented by a 2~by 2 filter of values one and stride~2. In Figure~\ref{c-and-r}, the smoothing and skip connections are not indicated for clarity.  

\begin{figure}[htbp] 
\centering
\includegraphics[width=17cm,angle=0, trim=0mm 15mm 0mm 1mm, clip]{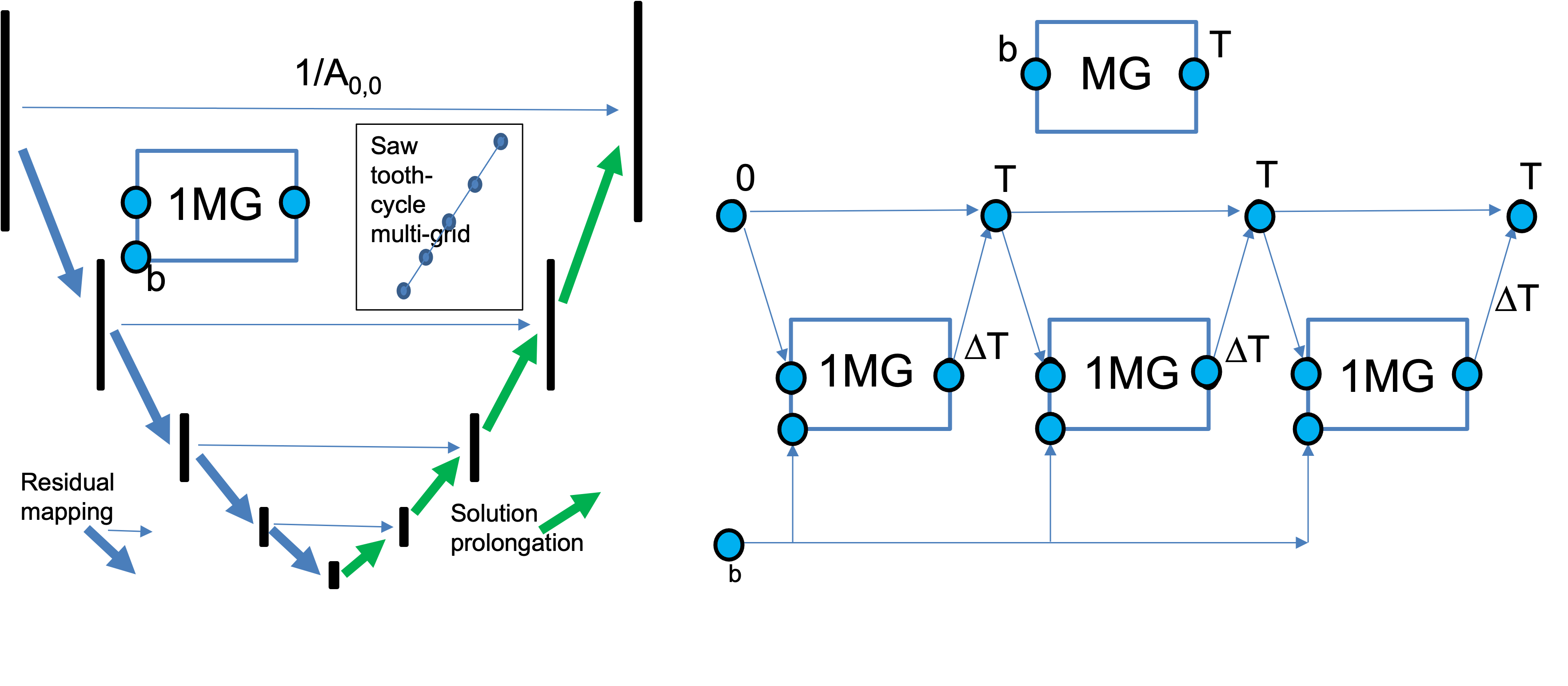} 
\caption{Left: a schematic diagram showing how one multigrid cycle (1MG) uses skip connections to pass the residual between the layers using a U-Net (blue horizontal arrows). Restrictions are indicated by blue arrows and prolongations by green arrows. Right bottom: a schematic diagram showing how three multigrid cycles are brought together to form an overall solution method. The input of this multigrid method is the source~$\bm{b}$. The output is the solution $\bm{T}$ after three cycles. (The initial estimate of the solution is taken as zero.) Right top: a box labelled ``MG'' is used to represent three MG cycles, and its input and output, which are shown in detail beneath it (bottom right).} 
\label{pages1-4-btm} 
\end{figure}

In \ref{appendi_MGMG}, Figures~\ref{pages1-4-top} and \ref{pages1-4-top_new} show a detailed example of how a multigrid method can be represented by a CNN, indicating the smoothing and skip connections, which we use  to send the residuals from the restriction layers of the convolutional neural network to the prolongation layers. Such an architecture (a CNN with skip connections) is known as a U-Net~\citep{Ronneberger2015}, see Figure~\ref{pages1-4-btm} (left). This figure (left) shows the structure of the neural network for one sawtooth multigrid cycle or iteration (``1MG''), where four restrictions are followed by four prolongations and the skip connections are shown as horizontal arrows. Several multigrid cycles can be joined together to form an overall multigrid method and Figure~\ref{pages1-4-btm} (right) shows three single multigrid cycles linked together to form the multigrid method (``MG'') that we apply at a particular time step. The input to the multigrid is the source $\bm{s}$ and the first estimate of the solution $\bm{T}$ is zero.

To illustrate the smoothing, restriction and prolongation, we consider the advection-diffusion problem, Equation~\eqref{adv-diff-eqn}, with a constant advection velocity $(u,\,v)$, a constant grid spacing of $\Delta x = \Delta y$ and zero source and absorption terms $s=0=\sigma$. An implicit backward Euler time differencing scheme for this case can be written as 
\begin{equation}\label{eq:BE-advection-concise}
     A  \ast [T]_{\ell,\,m}^{n+1} = T_{\ell,\,m}^{n} \,,
\end{equation}
where 
\begin{equation}
    A = \left(\, \begin{array}{rrr} 0&0&0\\0&1&0\\0&0&0\\\end{array}\right) + \frac{ u \Delta t}{2\Delta x}\left( \begin{array}{rrr} 0&-1&0\\0&0&0\\0&1&0\\\end{array}\right) + \frac{v\Delta t}{2\Delta x}\left( \begin{array}{rrr} 0&0&0\\-1&0&1\\0&0&0\\\end{array}\right) 
    + \frac{\nu\Delta t}{\Delta x^2}\left( \begin{array}{rrr} 0&-1&0\\-1&4&-1\\0&-1&0\\\end{array}\right).
\end{equation} 
The central entry of matrix $A$ is written as $A_{0,0}$ and corresponds to the diagonal of the matrix that would result from writing this in matrix-vector form. Applying relaxation with a coefficient of $\alpha\in(0,2]$ to Jacobi iteration results in:  
\begin{equation}
T_{\ell,\,m}^{n+1}  = {(1-\alpha)} \widetilde{T}_{\ell,\,m}^{n+1}  +  \frac{\alpha}{A_{0,0}} \left(A \ast [\widetilde{T}]_{\ell,\,m}^{n+1}  \right), 
\label{Jacobi-1-relax}
\end{equation}
where $\widetilde{T}_{\ell,\,m}^{n+1}$ represents the best estimate of ${T}_{\ell,\,m}^{n+1}$. To generate our results, we use $\alpha=1$. We use one smoothing step in our implementation. If the advection velocity were to vary spatially, then the matrix $A$ would also vary and would require indices indicating this in Equations~\eqref{eq:BE-advection-concise} to~\eqref{Jacobi-1-relax}. 

The restriction, for coarsening level $k$, can be written as
\begin{eqnarray}
T_{\ell,m}^{(k)} & = & \frac{1}{4} \left(T_{2\ell,2m}^{(k-1)} + T_{2\ell+1,2m}^{(k-1)} + T_{2\ell,2m+1}^{(k-1)} + T_{2\ell+1,2m+1}^{(k-1)} \right) \\[2mm]
&=& \frac{1}{4} \left( \begin{array}{rr} 1 & 1 \\ 1 & 1 \\ \end{array}\right) \ast \left( \begin{array}{cc}  T_{2\ell,2m+1}^{(k-1)}& T_{2\ell+1,2m+1}^{(k-1)}\\[2mm] T_{2\ell,2m}^{(k-1)}& T_{2\ell+1,2m}^{(k-1)}\\ \end{array}\right) 
\end{eqnarray}
and the prolongation operator is defined by: 
\begin{eqnarray}
 \left( \begin{array}{cc}  T_{2\ell,2m+1}^{(k-1)}& T_{2\ell+1,2m+1}^{(k-1)}\\[2mm] T_{2\ell,2m}^{(k-1)}& T_{2\ell+1,2m}^{(k-1)}\\ \end{array}\right) =  T_{\ell,m}^{(k)}\left( \begin{array}{cc} 1& 1\\ 1 & 1\\ \end{array}\right), 
\end{eqnarray}
where larger values of $k$ correspond to a coarser grid (see Figure~\ref{c-and-r}).

\subsection{Burgers equation}
\label{method-burgers}
The 2D Burgers equation can be written as
\begin{eqnarray}
\frac{\partial \bm{q}}{\partial t} + u\frac{\partial \bm{q}}{\partial x} +  v\frac{\partial \bm{q}}{\partial y} + 
\sigma \bm{q} - \nu \nabla \cdot \nabla \bm{q} =\bm{s} \,, 
\label{Burgers-eqn}
\end{eqnarray}
in which $\bm{q}=(u\;\,v)^T$ is the velocity, $\bm{s} = (s_x\ s_y)^T$ is a source (taken as zero), $\sigma$ is an absorption term (taken as zero) and $\nu$ is the viscosity coefficient. This equation is discretised in a similar way to the advection-diffusion equation described in the previous section. We employ the same discretisation in space (either central or upwind differencing) and in time (two step second-order accurate scheme) using $3\times 3$ filters on a 2D regular grid. We discretise each of the velocity components, $(u \; v)^T$ and thus, to maintain consistency with the previous section, we can denote a velocity component (either $u$ or $v$) as $T$ and solve for each one separately (first $T\equiv u$, then $T\equiv v$), see Figures~\ref{pages5-6-tikzversion-left} and~\ref{pages5-6-tikzversion-right}.

Applying the second-order time discretisation to the Burgers equation but treating the absorption term implicitly gives
\begin{equation}\label{eq:predictor-Burgers}
    (1+\sigma_{\ell,m} \Delta t) T_{\ell,\,m}^{n+1} = T_{\ell,\,m}^n - \Delta t\left( u^{n+\sfrac{1}{2}}_{\ell,m}\left(\frac{\partial T}{\partial x}\right)_{\ell,\,m}^{n+\sfrac{1}{2}} + v^{n+\sfrac{1}{2}}_{\ell,m}\left(\frac{\partial T}{\partial y}\right)_{\ell,\,m}^{n+\sfrac{1}{2}} - \nu\left(\frac{\partial^2 T}{\partial x^2}\right)_{\ell,\,m}^{n+\sfrac{1}{2}} - \nu\left(\frac{\partial^2 T}{\partial y^2}\right)_{\ell,\,m}^{n+\sfrac{1}{2}} \right) \,  ,
\end{equation}
where $T$ represents either $u$ or $v$. Applying central differencing in space results in 
\begin{eqnarray}
(1+\sigma_{\ell,m} \Delta t)\, T_{\ell,\,m}^{n+1} &=& T_{\ell,\,m}^{n} - u^{n+\sfrac{1}{2}}_{\ell,m} \odot A^u \ast \frac{1}{2}\left( [\widetilde{T}^{n+1}]_{\ell,m}+[T^{n}]_{\ell,m} \right) \nonumber \\[2mm]
&& \phantom{T_{\ell,\,m}^{n}} - v^{n+\sfrac{1}{2}}_{\ell,m} \odot A^v \ast \frac{1}{2}\left( [\widetilde{T}^{n+1}]_{\ell,m}+[T^{n}]_{\ell,m} \right) \nonumber\\[2mm]
&& \phantom{T_{\ell,\,m}^{n}\ v^{n+\sfrac{1}{2}}_{\ell,m} \odot} - A^\nu \ast \frac{1}{2}\left( [\widetilde{T}^{n+1}]_{\ell,m}+[T^{n}]_{\ell,m} \right) \,, \label{eq:burgers_point}
\end{eqnarray}
where the stencils of filters are as follows:
\begin{equation}
A^u:=\frac{\Delta t}{2\Delta x} \left( \begin{array}{rrr} 0&-1&0\\ 0&0&0\\ 0&1&0\end{array}\right),\   A^v:=\frac{\Delta t}{2\Delta x} \left( \begin{array}{rrr} 0&0&0\\ -1&0&1\\ 0&0&0\end{array}\right),\   A^\nu:=\frac{\nu\Delta t}{\Delta x^2} \left( \begin{array}{rrr} 0&-1&0\\ -1&4&-1\\ 0&-1&0\end{array}\right).\label{AAA} 
\end{equation}
Other FEM based filters as well as 
3D filters are shown in \ref{A1}. 

As in the previous section, we can write this equation for the entire solution field, bearing in mind that any solution field to which we apply a convolution will include padding:
\begin{equation}\label{eq:burgers_all_points_compact}
(\bm{b}+\bm{\sigma} \Delta t) \odot \bm{T}^{n+1} = \bm{T}^{n} - \left(\bm{u}^{n+\sfrac{1}{2}} \odot \left(A^u \ast 
\bm{T}_{\text{pad}}^{n+\sfrac{1}{2}}\right) + \bm{v}^{n+\sfrac{1}{2}} \odot \left( A^v \ast 
\bm{T}_{\text{pad}}^{n+\sfrac{1}{2}}\right) + A^\nu \ast 
\bm{T}_{\text{pad}}^{n+\sfrac{1}{2}}\right) \, ,
\end{equation}
in which $\bm{b}$ is a tensor of the same dimension as $\bm{T}^{n+1}$ with every entry equal to~1, and $\bm{T}_{\text{pad}}^{n+\sfrac{1}{2}}=\frac{1}{2}\left( \bm{\widetilde{T}}_{\text{pad}}^{n+1} + \bm{T}_{\text{pad}}^{n}\right)$, similarly for $\bm{u}^{n+\sfrac{1}{2}}$ and $\bm{v}^{n+\sfrac{1}{2}}$. 
Rewriting this equation by using entrywise division, indicated by~$\oslash$, gives
\begin{eqnarray}
\bm{T}^{n+1} = \textcolor{pygreen}{\bm{T}^{n} \oslash (\bm{b}+\bm{\sigma} \Delta t)} &-& \textcolor{pyorange}{\bm{u}^{n+\sfrac{1}{2}} \odot \left(A^u \ast 
\bm{T}_{\text{pad}}^{n+\sfrac{1}{2}}\right) \oslash (\bm{b}+\bm{\sigma} \Delta t)} 
\nonumber \\ 
&-&  \textcolor{pycyan}{\bm{v}^{n+\sfrac{1}{2}} \odot \left( A^v \ast \bm{T}_{\text{pad}}^{n+\sfrac{1}{2}}\right) \oslash (\bm{b}+\bm{\sigma} \Delta t)} 
\nonumber \\ &-& A^\nu \ast 
\bm{T}_{\text{pad}}^{n+\sfrac{1}{2}}\oslash (\bm{b}+\bm{\sigma} \Delta t) \, ,
\label{Tn+1} 
\end{eqnarray}
The terms coloured green, orange and cyan correspond to the vertical bars in Figures~\ref{pages5-6-tikzversion-left} which illustrates how the filters are applied to the variables. For simplicity, in the diagram diffusivity is not shown. We need to use this CNN twice to achieve second order accuracy in time: once for the predictor step and once for the corrector, see Figures~\ref{pages5-6-tikzversion-left} and~\ref{pages5-6-tikzversion-right}. 

\subsection{Navier-Stokes equations}
\label{NS}

\begin{figure}[htbp]
    \centering

    \begin{subfigure}[t]{0.48\textwidth}
        \centering
        \includegraphics[width=\linewidth]{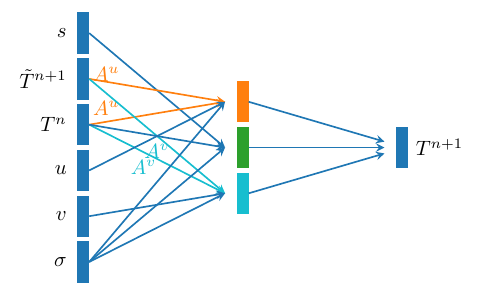}
        \caption{\label{pages5-6-tikzversion-left}Top: Applying CNN filters associated with advection ($A^u$, in orange, and $A^v$, in cyan) to a scalar field $T$. For the Burgers equation or the momentum equations of the Navier-Stokes equations, $T$ could represent one velocity component.}
    \end{subfigure}
    \hfill
    \begin{subfigure}[t]{0.48\textwidth}
        \centering
        \includegraphics[width=\linewidth]{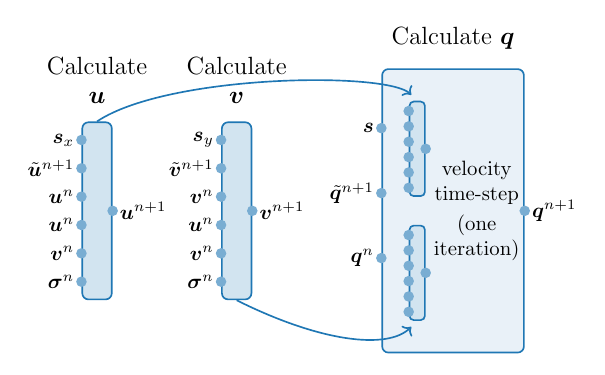}
        \caption{\label{pages5-6-tikzversion-right}How the velocities are calculated during an iteration (either the predictor or corrector) where the tilde sign represents the best current approximation.}
    \end{subfigure}

    \vspace{1em}

    \begin{subfigure}[t]{0.95\textwidth}
        \centering
        \includegraphics[width=\linewidth]{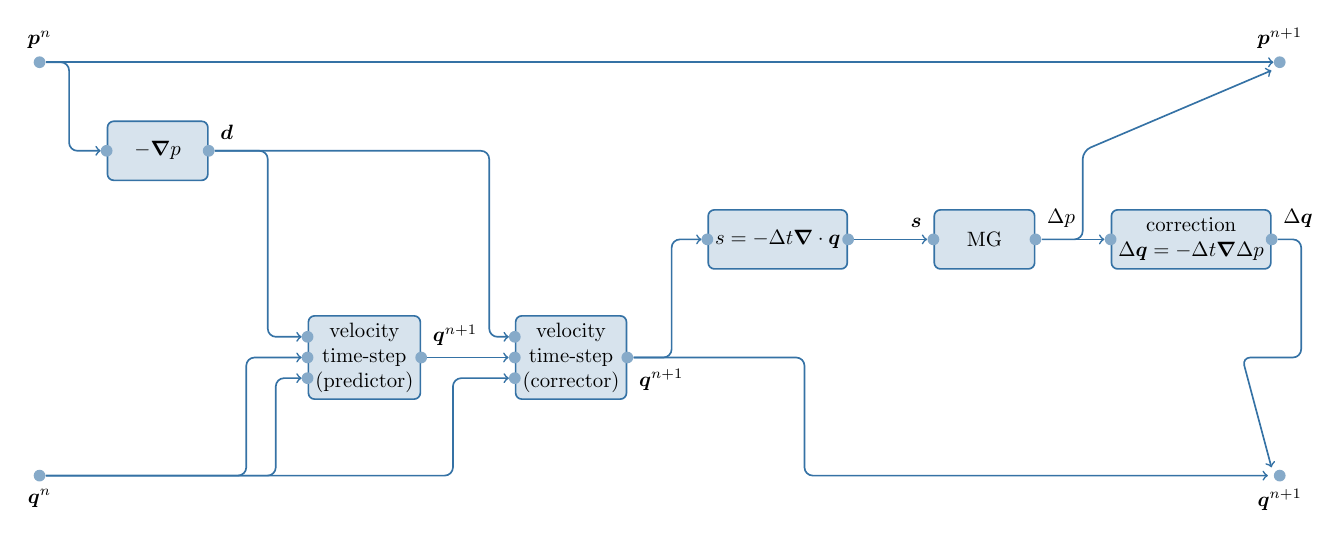}
        \caption{\label{pages5-6-tikzversion-btm}A single time step in the solution of the Navier-Stokes equations with a velocity predictor and corrector, a multigrid solution for pressure and a correction to the velocity field with the updated pressure solution.}
    \end{subfigure}
    \caption{Schematic diagrams showing how a neural network can be used for time stepping. The plot at the top left shows this for the advection equation. On the top right we indicate how this can be repurposed to solve equations involving the advection of momentum. The bottom plot shows a single time step in the solution of the Navier-Stokes equations. In these diagrams, $\bm{s}$ the source term for two different equations.}
    \label{pages5-6-tikzversion} 
\end{figure}

The Navier-Stokes equations can be written as 
\begin{eqnarray}
\frac{\partial \bm{q}}{\partial t} + u\frac{\partial \bm{q}}{\partial x} +  v\frac{\partial \bm{q}}{\partial y} + 
\sigma \bm{q} - \nu \nabla^2 \bm{q} &=& -\nabla p \,,
\label{NS-eqn}\\
\nabla  \cdot \bm{q} &=& \bm{0} \,,
\label{NS-eqn-cty}
\end{eqnarray}
in which $\bm{q}=(u \;v)^T$ in 2D and $\bm{q}=(u\;v\;w)^T$ in 3D, $p$ is the pressure, $\sigma$ is an absorption term (taken as zero) and $\nu$ is the viscosity coefficient. The momentum equations are discretised similarly to the Burgers equation (see Figure~\ref{pages5-6-tikzversion-left}), but with a source of $\bm{s}=-\nabla p$. We solve for both the velocity components and combine the resulting convolutional operations within one architecture as shown in Figure~\ref{pages5-6-tikzversion-right}. To solve for pressure, we use a projection-based method formed by manipulating the discretised equations according to Algorithm~\ref{alg:projection}, see also Figure~\ref{pages5-6-tikzversion-btm}. Boundary conditions are enforced as before, through padding. We populate the padding with the specified value if we have Dirichlet boundary conditions. If we have a zero normal-derivative boundary condition, we populate the corresponding field with the values that are adjacent to the padding region. 

We use a finite element discretisation based on bilinear rectangular (in 2D) and quadrilateral (in 3D) finite elements for the advection terms. We use the 27 point stencil for the diffusion and advection operators in 3D implemented with $3\times 3\times 3$ filters. The advection and diffusion operators are described in the appendix. 

\begin{algorithm}[H]
\caption{\label{alg:projection} Projection-based solution method for velocity and pressure.}
\begin{algorithmic}[1]
\State Solve for $\bm{q}^{n+1}$ using the two-step approach outlined for the Burgers and advection-diffusion equation but treating the term involving $\sigma$ fully implicitly: 
\begin{eqnarray}
\frac{\bm{q}^{n+1} -\bm{q}^{n} }{\Delta t} + u^n\frac{\partial \bm{q}^{n+\frac{1}{2}} }{\partial x} +  v^n\frac{\partial \bm{q}^{n+\frac{1}{2}} }{\partial y} + 
\sigma \bm{q}^{n+1} - \nu \nabla^2 \bm{q}^{n+\frac{1}{2}} = -\nabla p^n \,.
\label{NS-eqn-disc}
\end{eqnarray}
\State Solve for pressure correction $\Delta p$: 
\begin{eqnarray}
\nabla^2 \Delta p = -\frac{1}{\Delta t} \nabla \cdot \bm{q}^{n+1} . 
\label{PC-DP}
\end{eqnarray}
\State Solve for the velocity correction $\Delta \bm{q}$ using the multigrid solver (Figure~\ref{pages1-4-btm}):
\begin{equation}
\Delta \bm{q} = -\Delta t \nabla \Delta p. 
\end{equation}
\State Update pressure solution: $p^{n+1} = p^n + \Delta p$
\State Update velocity solution: $\bm{q}^{n+1} \leftarrow \bm{q}^{n+1} + \Delta \bm{q}$.
\end{algorithmic}
\end{algorithm}

\section{Results}\label{sec:results}
In this section we present results for an explicit time discretisation of the advection-diffusion equation; a multigrid solver applied to the advection-diffusion equation; a solution of the nonlinear Burgers equation and finally we combine the methods used in solving the three previous problems to generate results for incompressible flow past a bluff body. Throughout this section, lengths are in metres, times are in seconds and velocities are in metres per second. The code is available from the  \href{https://github.com/ImperialCollegeLondon/AI4PDEs}{AI4PDEs/NN4PDEs repository}. 

\subsection{Advection-diffusion equation}
\label{sec:Advection}
\begin{figure}[htbp]
\centering

\begin{minipage}[]{0.40\textwidth}
\includegraphics[width=\textwidth]{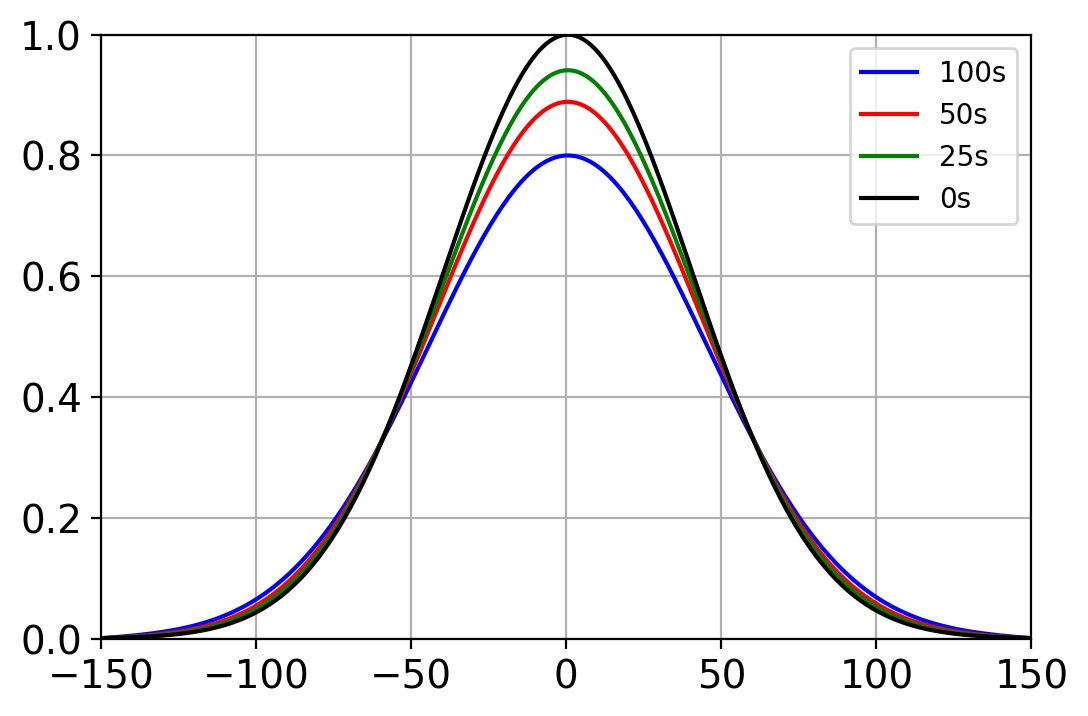}
\put(-100,20){{\textcolor{blue}{\small 0.7998}}}
\put(-100,40){{\textcolor{red}{\small 0.8887}}}
\put(-100,60){{\textcolor{green}{\small 0.9409}}}
\end{minipage}
\begin{minipage}[]{0.40\textwidth}
\includegraphics[width=\textwidth]{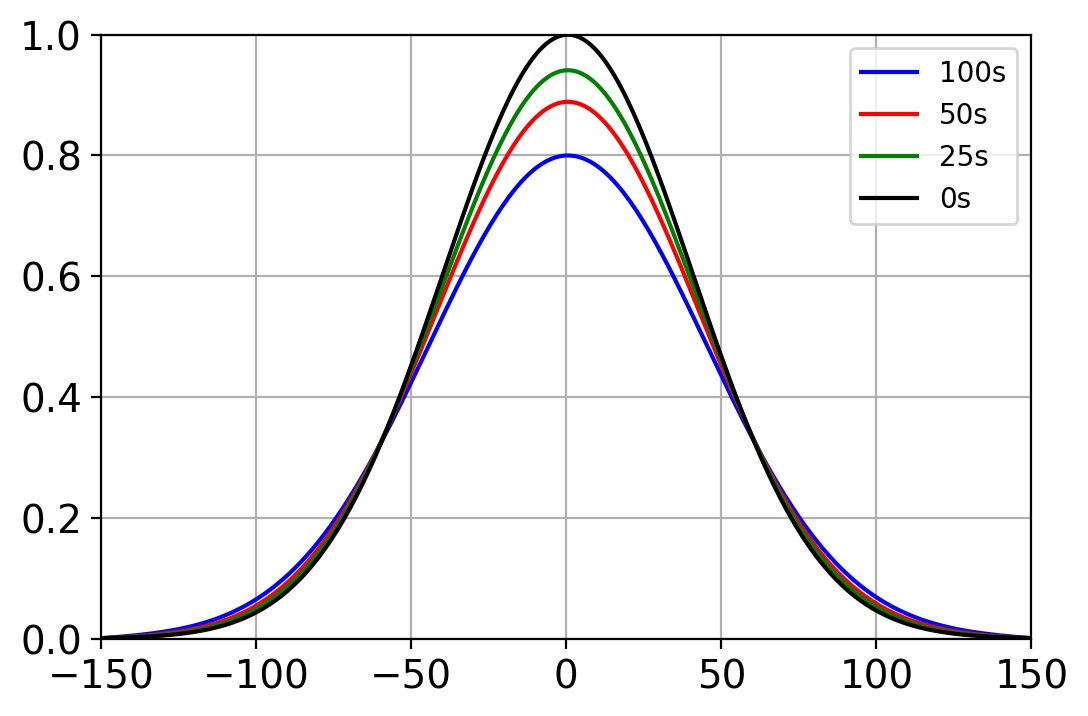}
\put(-100,20){{\textcolor{blue}{\small 0.7998}}}
\put(-100,40){{\textcolor{red}{\small 0.8887}}}
\put(-100,60){{\textcolor{green}{\small 0.9409}}}
\end{minipage}
\caption{Numerical solution of the 2D diffusion equation at times 0, 25, 50 and 100 seconds. The initial condition of a Gaussian distribution is diffused by a traditional computational fluid dynamics code (left) and the Neural Physics approach based on CNNs (right). The three values shown in each graph correspond to the maximum values of the solution which occur at $x$=0.}
\label{fig:advection_diffusion_with_no_advection}
\end{figure}

We solve the 2D time-dependent diffusion equation in a domain measuring 300~by 300 with the origin located at the centre of the domain. A Gaussian distribution centred at $(x_{0},y_{0})=(0,0)$ and with a width of $\gamma=10$ is taken as the initial condition (see Equation~\eqref{Gaussian_distribution_T}). The diffusion coefficient has value $\SI{2}{m^{2}s^{-1}}$. We set the weights of a CNN based on a second-order time discretisation scheme with central differencing in space for the diffusion operator. The nodal spacing in the $x$ and $y$ directions is constant at $\Delta x=\Delta y=1$. The proposed CNN model is compared with a traditionally coded PDE solver using the same numerical methods and tested on different grids, showing that the results are identical to within round-off error (see the line plots in Figure~\ref{fig:advection_diffusion_with_no_advection}). 

\begin{figure}[htbp]
\centering
\includegraphics[width=0.5\textwidth]{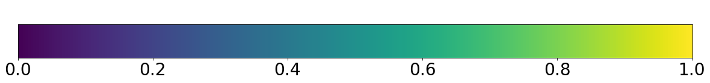}\\
\begin{subfigure}[t]{0.2\textwidth}
\centering
\includegraphics[width=\textwidth]{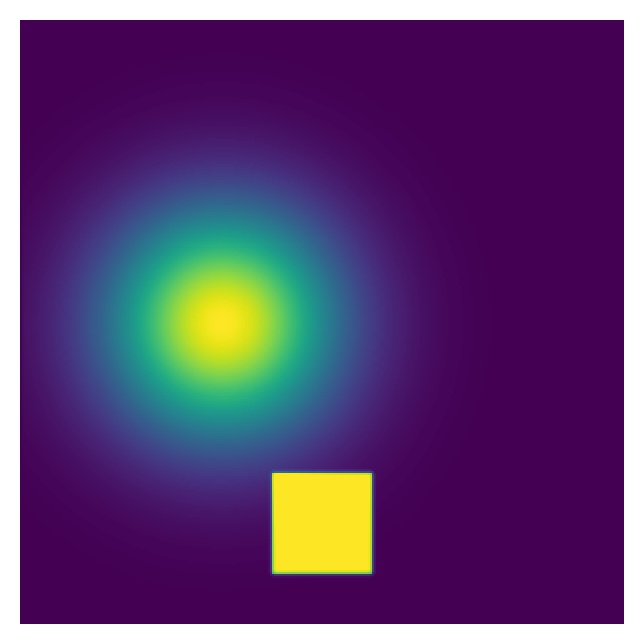}
\put(-110,30){{\Large (a)}}
\end{subfigure}
\begin{subfigure}[t]{0.2\textwidth}
\centering
\includegraphics[width=\textwidth]{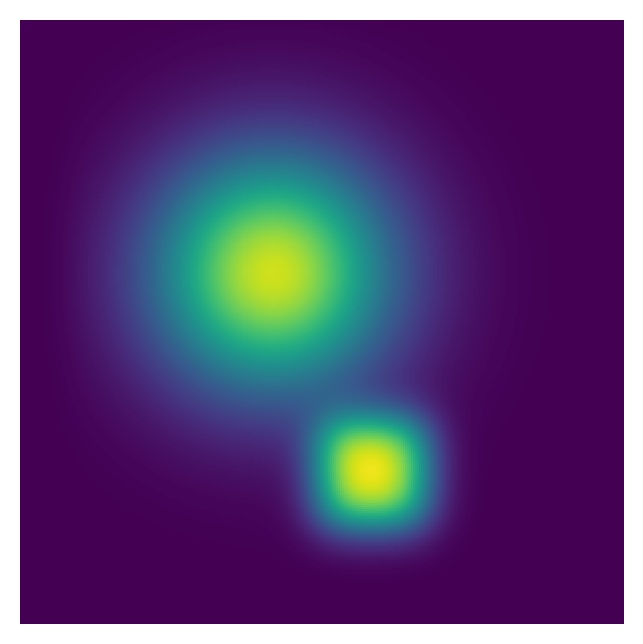}
\end{subfigure}
\begin{subfigure}[t]{0.2\textwidth}
\centering
\includegraphics[width=\textwidth]{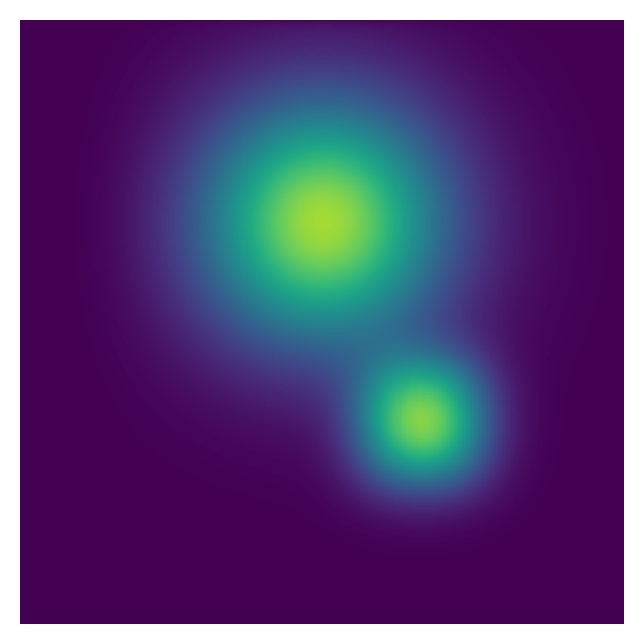}
\end{subfigure}
\begin{subfigure}[t]{0.2\textwidth}
\centering
\includegraphics[width=\textwidth]{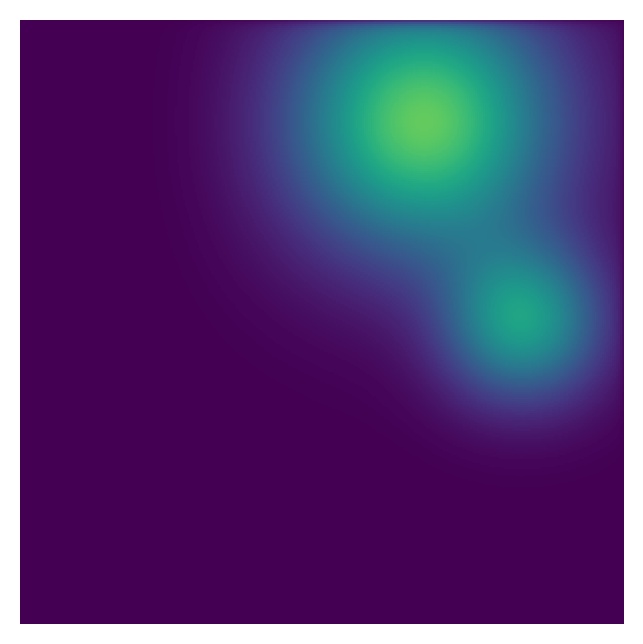}
\end{subfigure}

\begin{subfigure}[t]{0.2\textwidth}
\centering
\includegraphics[width=\textwidth]{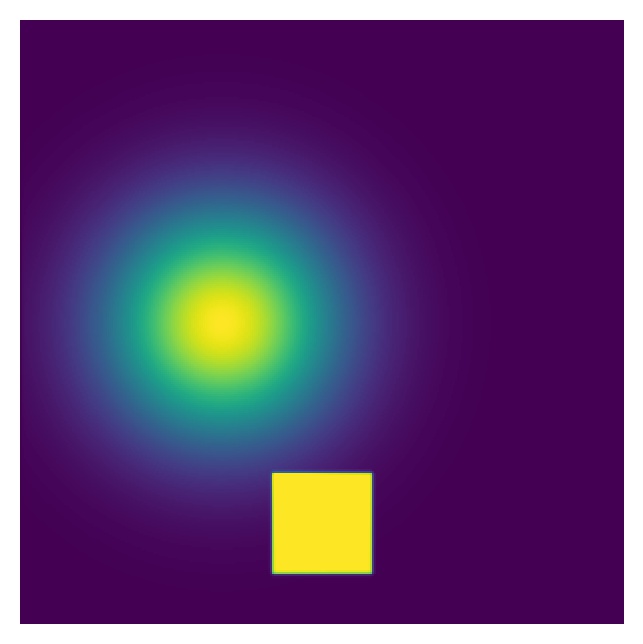}
\put(-110,30){{\Large (b)}}
\caption*{\Large \SI{0}{s}}
\end{subfigure}
\begin{subfigure}[t]{0.2\textwidth}
\centering
\includegraphics[width=\textwidth]{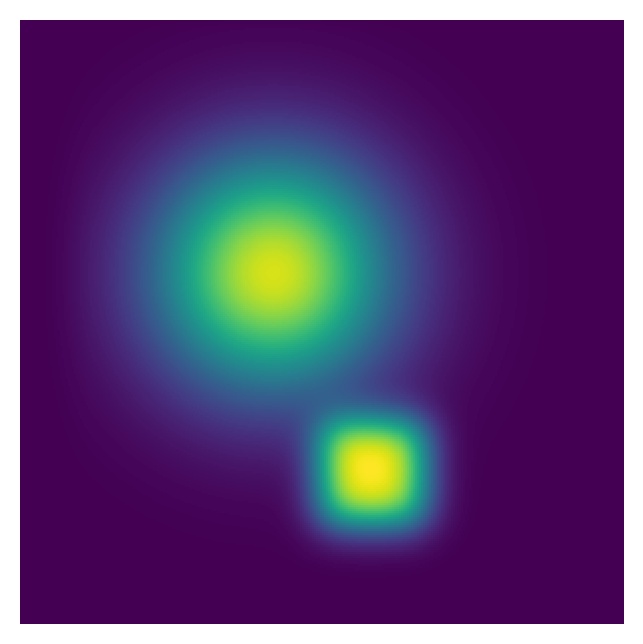}
\caption*{\Large \SI{25}{s}}
\end{subfigure}
\begin{subfigure}[t]{0.2\textwidth}
\centering
\includegraphics[width=\textwidth]{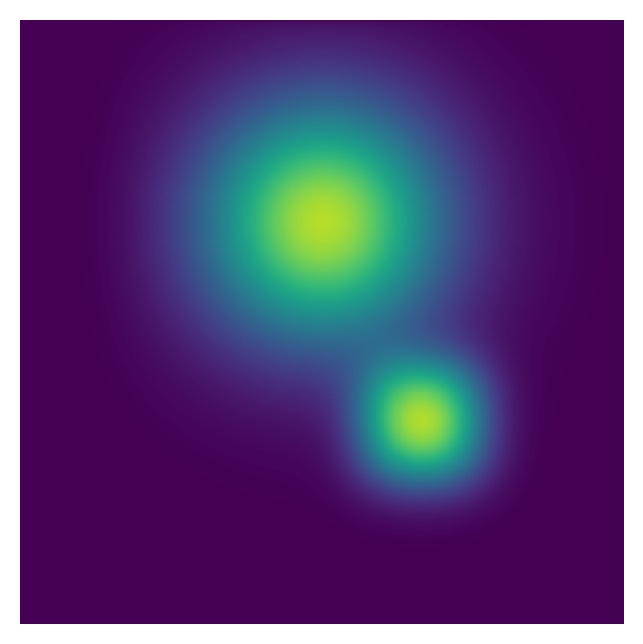}
\caption*{\Large \SI{50}{s}}
\end{subfigure}
\begin{subfigure}[t]{0.2\textwidth}
\centering
\includegraphics[width=\textwidth]{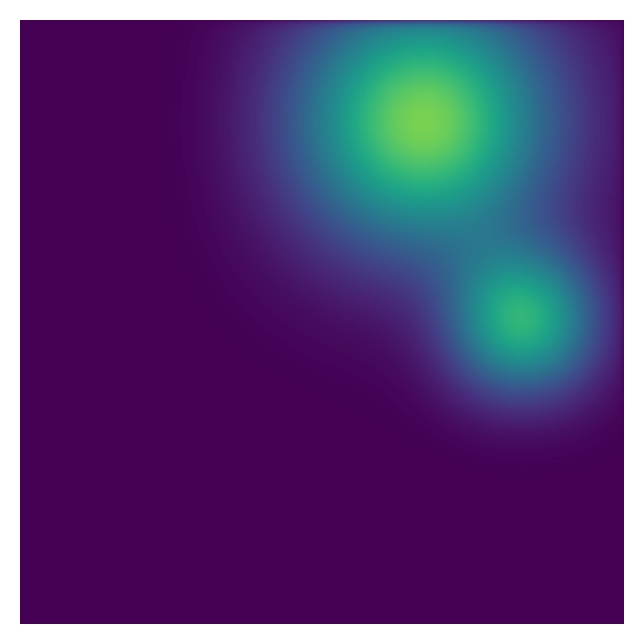}
\caption*{\Large \SI{100}{s}}
\end{subfigure}
\caption{Numerical solution of the advection-diffusion equation by a CNN with analytically determined weights at times 0, 25, 50 and 100 seconds. The initial condition is a combination of a square and a Gaussian centred at $(x_{0},y_{0})=(0,-100)$ and $(x_{0},y_0)=(-50,0)$ respectively. Both sets of results have a second-order time discretisation scheme with central differencing in space for the diffusion operator. The plots on row~(a) have upwind differencing for the advection operator; and on row~(b) have central differencing for the advection operator.}
\label{fig:advection_diffusion_explicit}
\end{figure}

For the same domain, we advect an initial condition with a uniform velocity field $(u,v)=(1,1)$ and a uniform diffusion coefficient $\nu=\SI{2}{m^{2}s^{-1}}$. The initial condition is a combination of a Gaussian distribution and a square, which can be written as
\begin{align}
T^0(x,y) &=  \exp \left(- \left(\frac{(x-x_{0})^{2}}{2 \gamma^2} + \frac{(y-y_{0})^{2}}{2 \gamma^2} \right) \right) \  + \ \begin{cases}
    1,      & \quad      x\in\text{[-25,25]}, \; y\in\text{[-75,-125]} \\
    0,      & \quad      \text{otherwise}
         \end{cases}  
\label{Gaussian_distribution_T}
\end{align}

in which $T^0$ represents the initial condition of the scalar field $T$, $x,y\in [-150,\,150]$, the centre of the Gaussian distribution is at $(x_{0},y_0)=(-50,0)$, and the coefficient representing the width of the distribution is $\gamma=40$. The square distribution is centred at $(x_{0},y_{0})=(0,-100)$. The nodal spacing in the $x$ and $y$ directions are constant at $\Delta x=\Delta y=1$. The Courant number is $C=0.1$ and the time-step size is $\Delta t=0.1$. We investigate two discretisation methods, both with a predictor-corrector time discretisation that is second order accurate: (i)~a first order upwind differencing for advection and central differencing for diffusion; (ii)~a second order central differencing for both advection and diffusion. Both discretisation methods are implemented by specifying the weights of the filter of a convolutional layer, see Section~\ref{AD} (and not by performing any training). From Figure~\ref{fig:advection_diffusion_explicit}, we see that both methods produce similar results, and, for longer times, the effect of diffusion becomes more pronounced as expected. One can, however, see that the diffusion seems to be slightly more pronounced in the upwind solution because of the increasing dissipation provided by upwinding of the advection terms. 

\subsection{Multigrid solver for the advection-diffusion equation} 
Here we show results for the advection-diffusion equation obtained by a sawtooth multigrid method implemented using a CNN. There is a uniform advection velocity of $(u,\; v)=(1,\; 1)$, the source strength $s$ is zero and $\nu$ is the constant diffusivity of $\SI{2}{m^{2}s^{-1}}$. For a domain measuring 128~by 128, we advect an initial condition of a Gaussian distribution centred at $(x_{0},y_{0})=(0,0)$ and the coefficient representing the width of the distribution is $\gamma=10$ (see Equation~\eqref{Gaussian_distribution_T}). The nodal spacing in the $x$ and $y$ directions is constant, $\Delta x=\Delta y=1$. The Courant number associated with the grid is $C=50$ and the time step size is $\Delta t=50$. In one time-step, the information will therefore advect over 50 grid points, and a multigrid solver is well-suited to this challenge, whereas other methods would struggle. We investigate the solution after just one time step, as the solver would behave in a similar manner for successive time steps. The initial guess for the solution at time level $n=1$ is $T^{n=1,0}=T^{0}=0$, i.e., the solution at the previous time level which is the initial condition. We use upwind differencing in space for advection (see Equation~\eqref{A-up} in Section~\ref{AD}) and central differencing for diffusion. The Jacobi iterative method is embedded in the proposed neural network and is used as the multigrid smoothing method. Skip connections are used to transfer the residuals between the convolutional layers through a U-Net architecture, see Section~\ref{MGMG}. The numerical methods used to generate these results are implemented by initialising the weights of the filter of a convolutional layer. No training of the network is performed. 

\begin{figure}[htbp]
\centering
\begin{subfigure}[t]{0.22\textwidth}
\centering
\includegraphics[width=\textwidth]{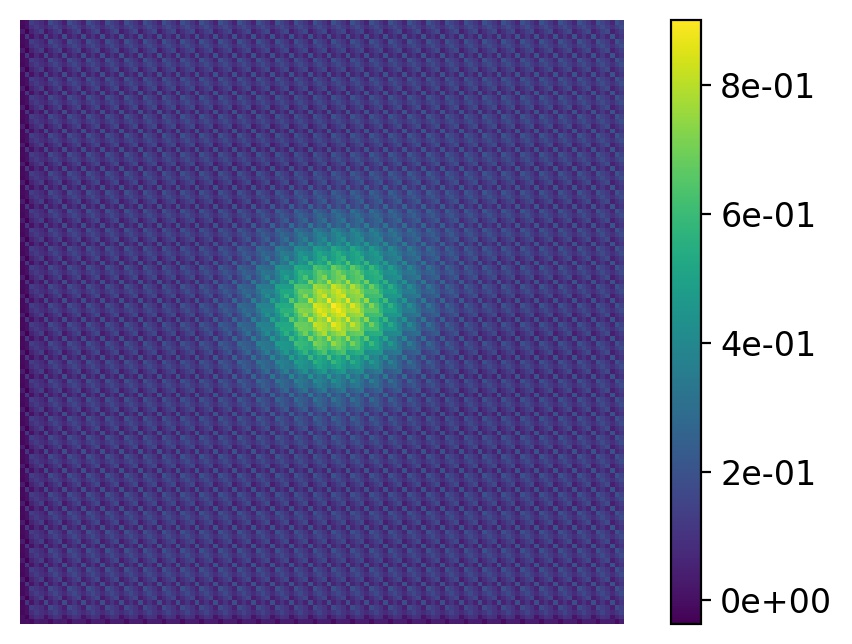}
\caption*{Initial condition, $T^0$}
\end{subfigure}
\begin{subfigure}[t]{0.22\textwidth}
\centering
\includegraphics[width=\textwidth]{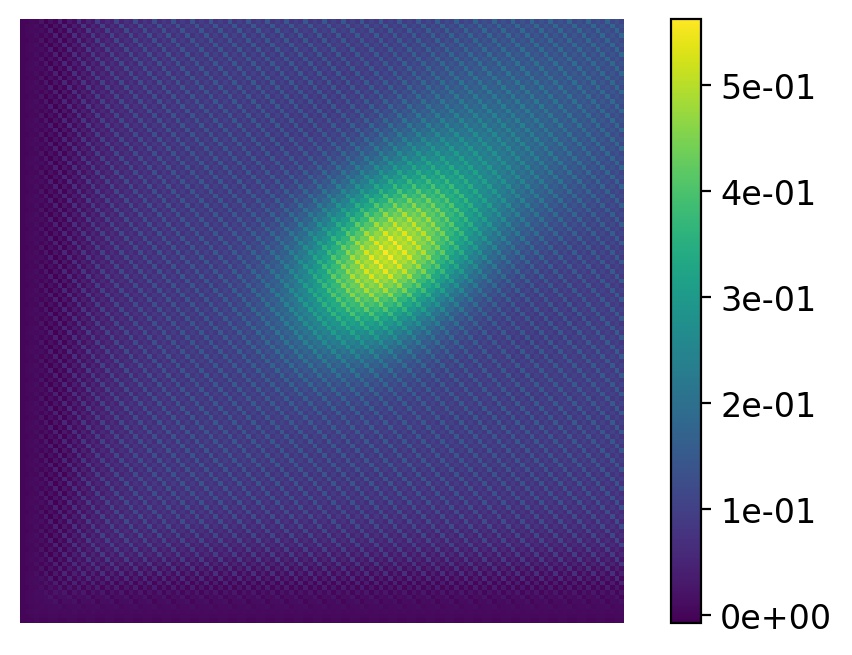}
\caption*{$T^{1,5}$}
\end{subfigure}
\begin{subfigure}[t]{0.22\textwidth}
\centering
\includegraphics[width=\textwidth]{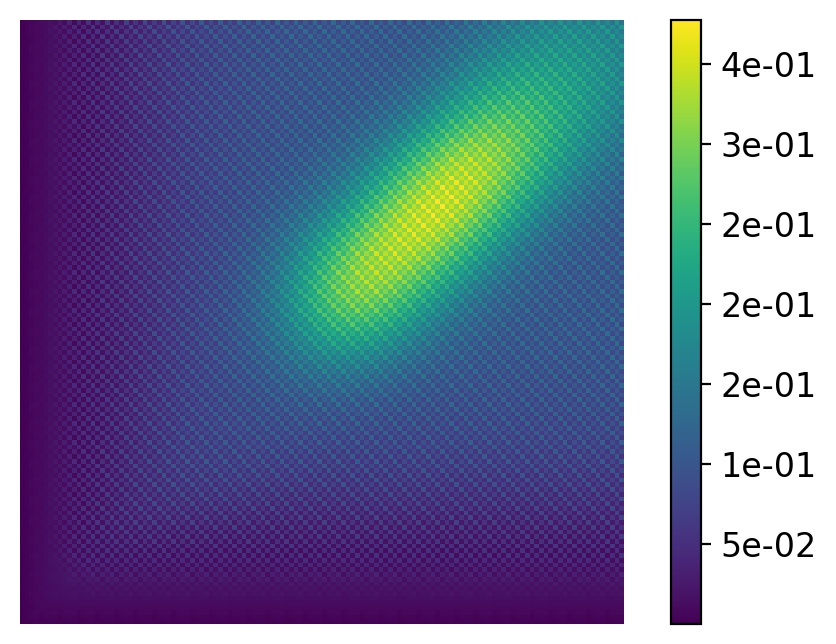}
\caption*{$T^{1,10}$}
\end{subfigure}
\begin{subfigure}[t]{0.22\textwidth}
\centering
\includegraphics[width=\textwidth]{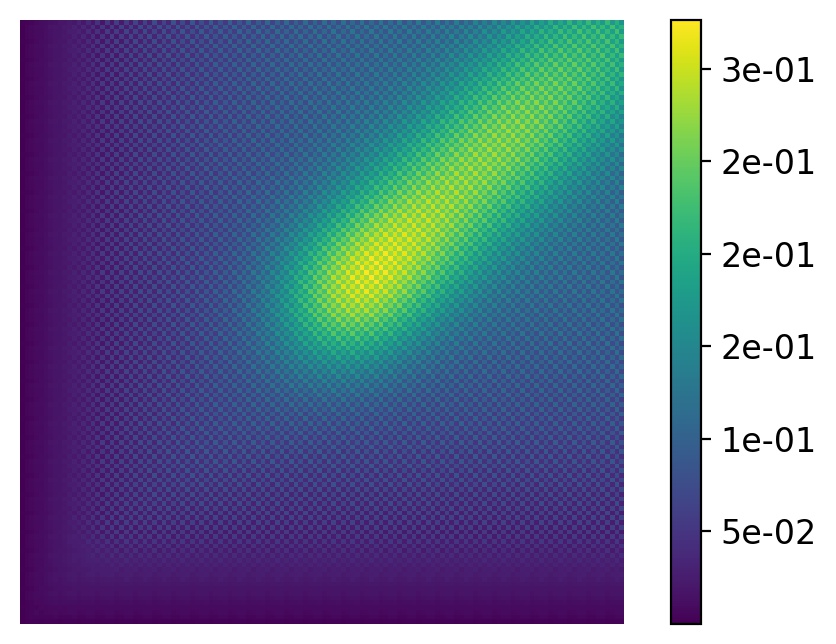}
\caption*{$T^{1,15}$}
\end{subfigure}
\begin{subfigure}[t]{0.22\textwidth}
\centering
\includegraphics[width=\textwidth]{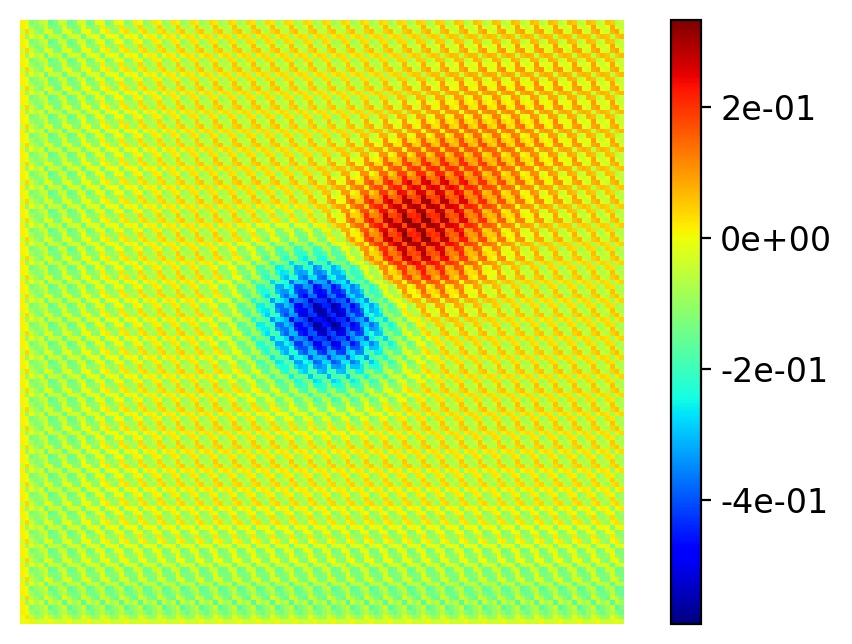}
\caption*{$T^{1,5}-T^{1,0}$}
\end{subfigure}
\begin{subfigure}[t]{0.22\textwidth}
\centering
\includegraphics[width=\textwidth]{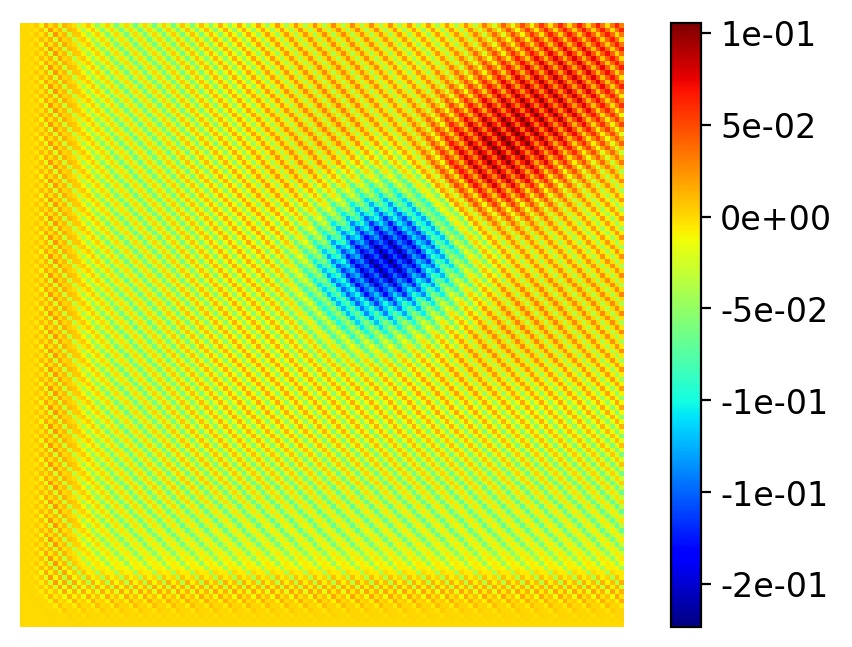}
\caption*{$T^{1,10}-T^{1,5}$}
\end{subfigure}
\begin{subfigure}[t]{0.22\textwidth}
\centering
\includegraphics[width=\textwidth]{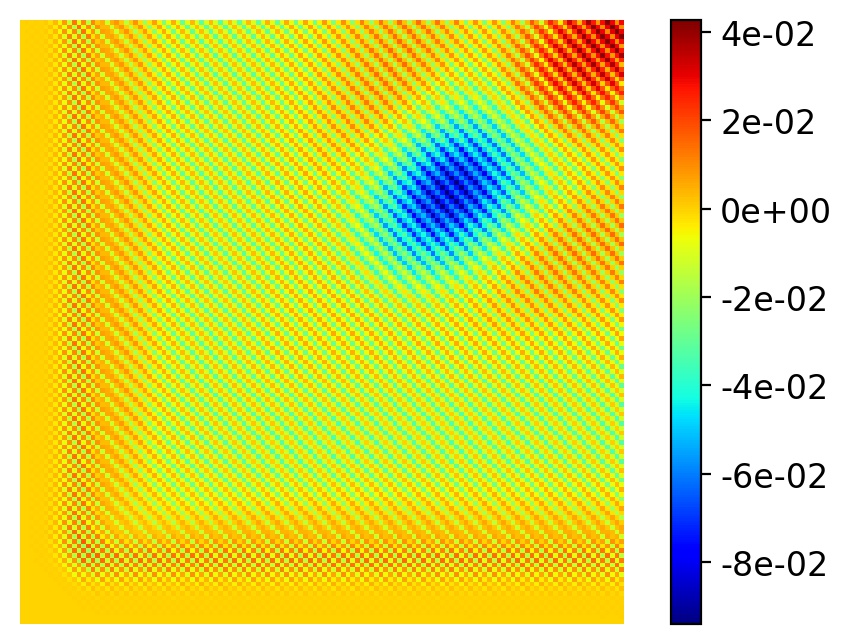}
\caption*{$T^{1,15}-T^{1,10}$}
\end{subfigure}
\begin{subfigure}[t]{0.22\textwidth}
\centering
\includegraphics[width=\textwidth]{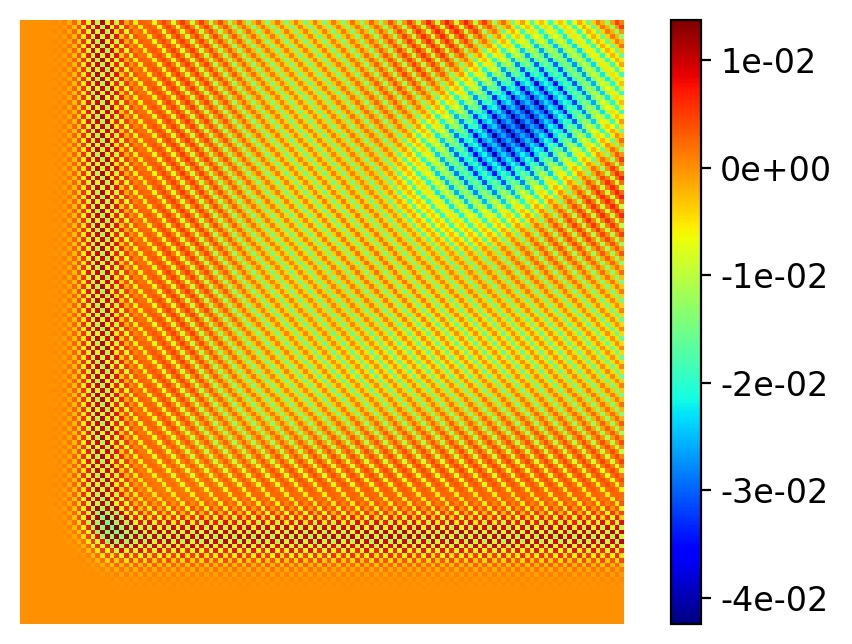}
\caption*{$T^{1,20}-T^{1,15}$}
\end{subfigure}
\caption{Multigrid solution of advection-diffusion equation by a CNN. The first row, at the top, shows the solution with an initial condition of a Gaussian distribution centred at $(x_{0},y_{0})=(0,0)$ after 5, 10 and 15 multigrid iterations ($T^{1,i}$ where $i$ represents the iteration number). The second row shows the difference between the certain solutions during the iterative process. \label{fig:multigrid_iterative_solutions}}
\end{figure}

\begin{figure}[htbp]
\centering
    \centering   \includegraphics[width=0.8\textwidth]{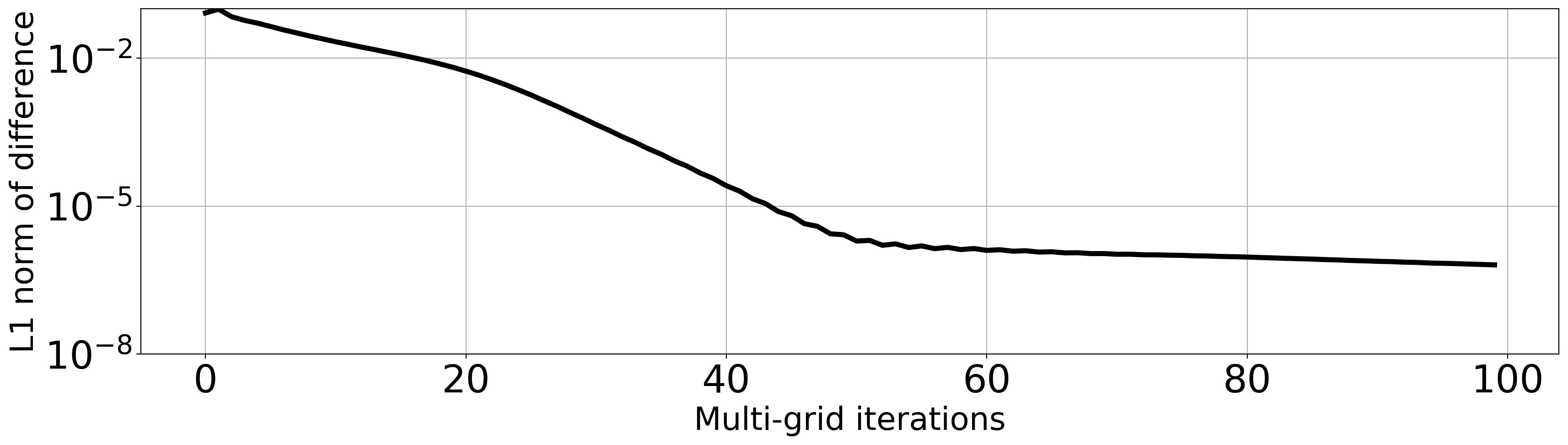}
    \caption{\label{fig:multigrid_convergence}A plot of the $L_1$ norm of difference between the solutions at the current and previous iterations for the multigrid solution of the advection-diffusion equation.}
\end{figure}

\begin{figure}[htbp]
\centering
\begin{subfigure}[t]{0.22\textwidth}
\centering
\includegraphics[width=\textwidth]{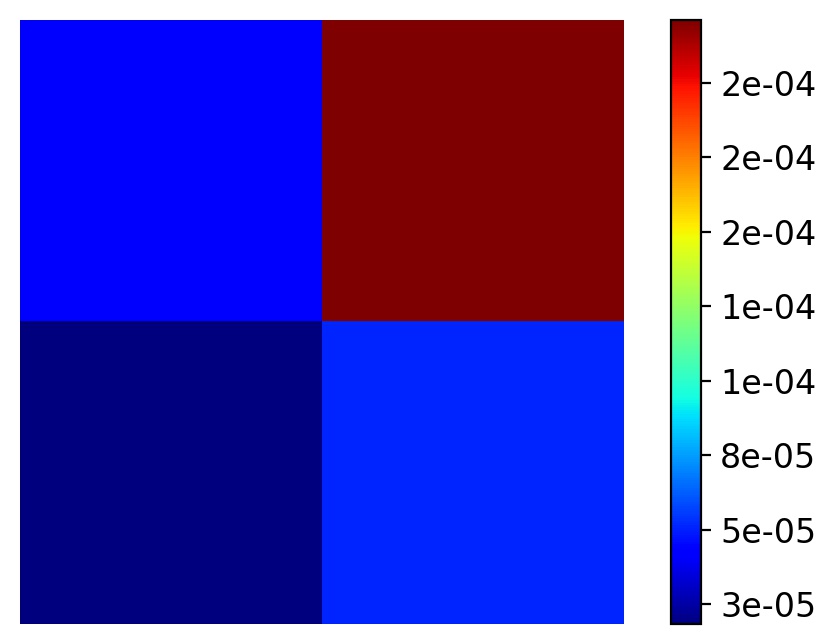}
\caption*{\text{\small Residual on $2 \times 2$}}
\end{subfigure}
\begin{subfigure}[t]{0.22\textwidth}
\centering
\includegraphics[width=\textwidth]{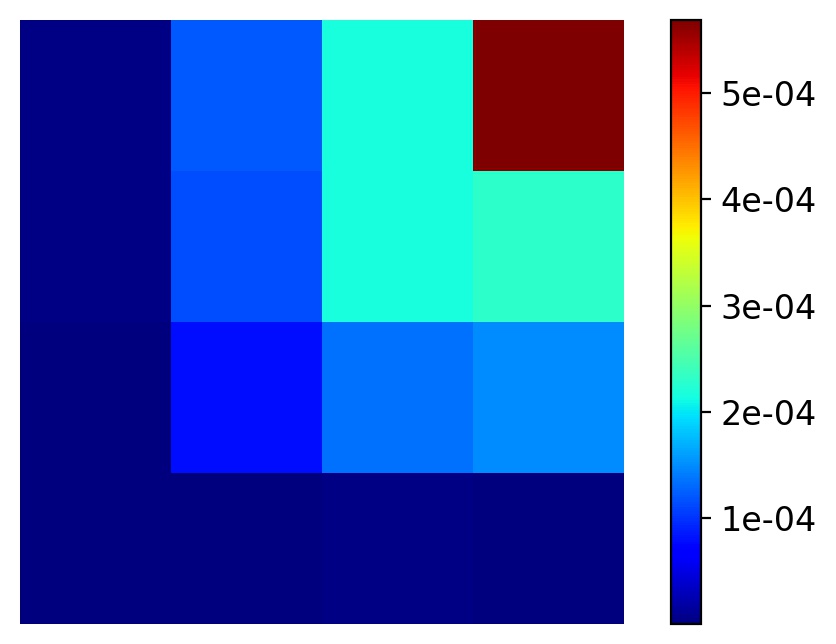}
\caption*{\text{\small Residual on $4 \times 4$}}
\end{subfigure}
\begin{subfigure}[t]{0.22\textwidth}
\centering
\includegraphics[width=\textwidth]{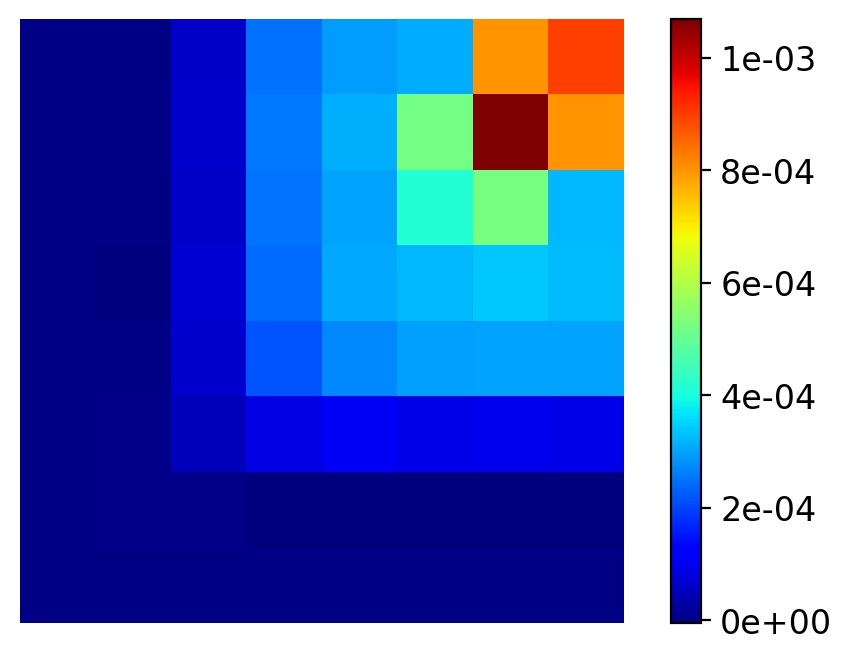}
\caption*{\text{\small Residual on $8 \times 8$}}
\end{subfigure}
\begin{subfigure}[t]{0.22\textwidth}
\centering
\includegraphics[width=\textwidth]{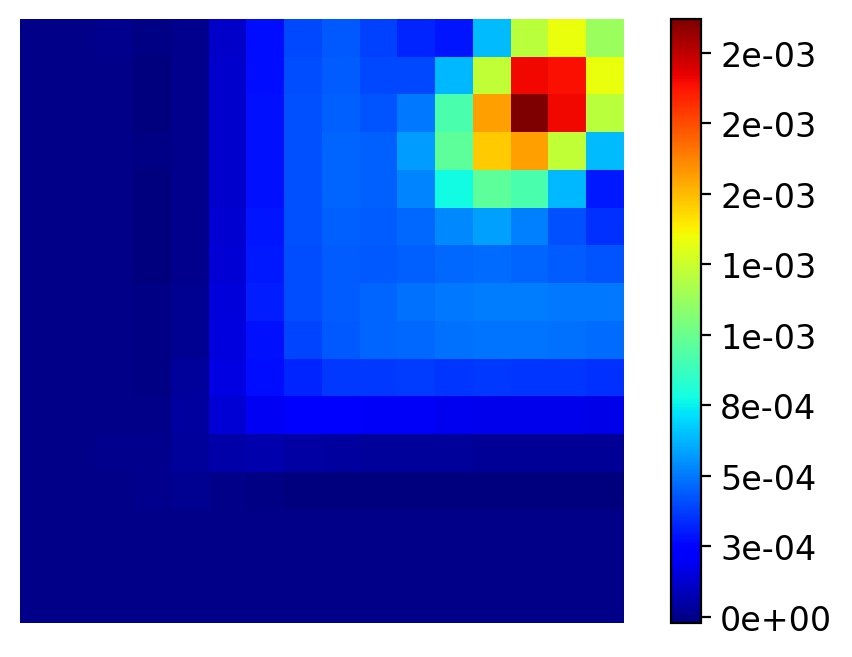}
\caption*{\text{\small Residual on $16 \times 16$}}
\end{subfigure}

\begin{subfigure}[t]{0.22\textwidth}
\centering
\includegraphics[width=\textwidth]{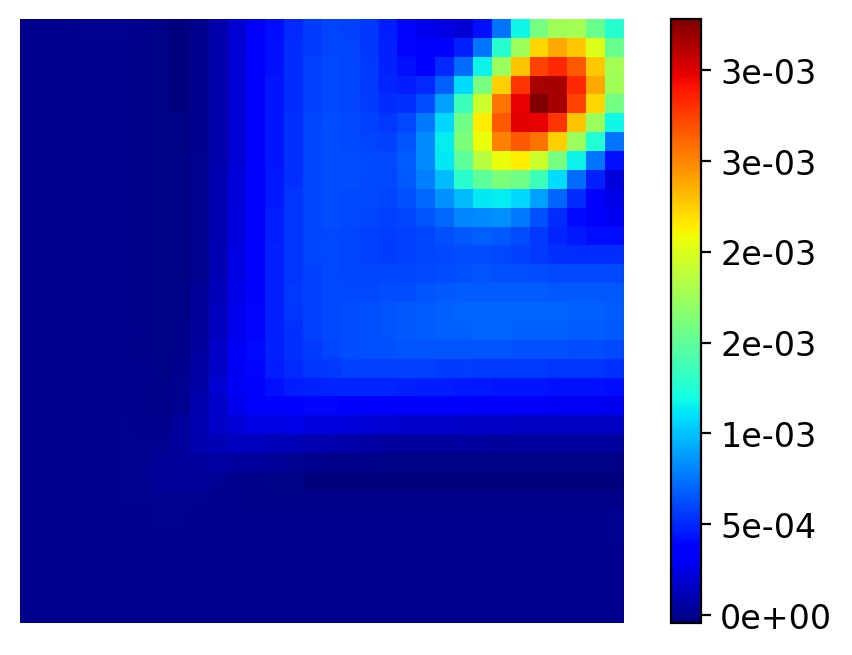}
\caption*{\text{\small Residual on $32 \times 32$}}
\end{subfigure}
\begin{subfigure}[t]{0.22\textwidth}
\centering
\includegraphics[width=\textwidth]{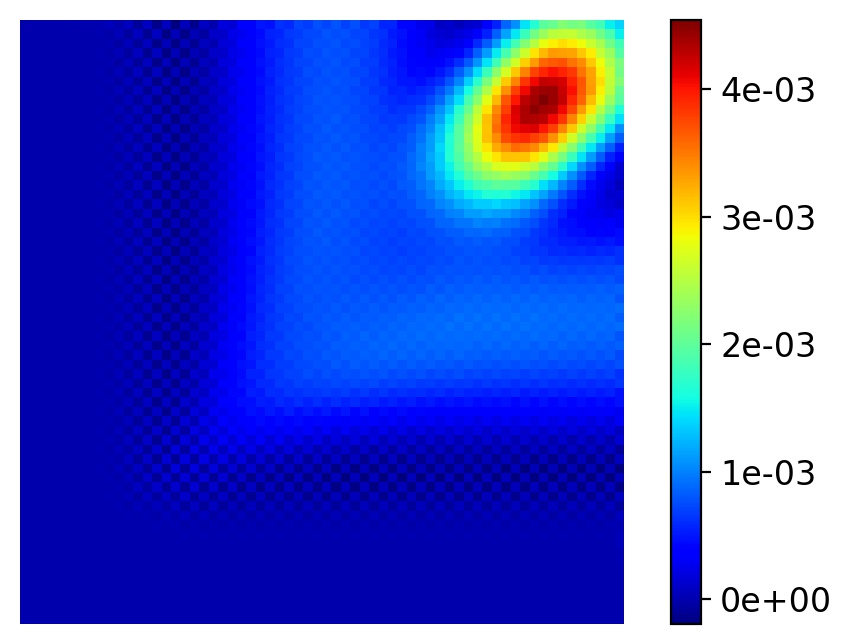}
\caption*{\text{\small Residual on $64 \times 64$}}
\end{subfigure}
\begin{subfigure}[t]{0.22\textwidth}
\centering
\includegraphics[width=\textwidth]{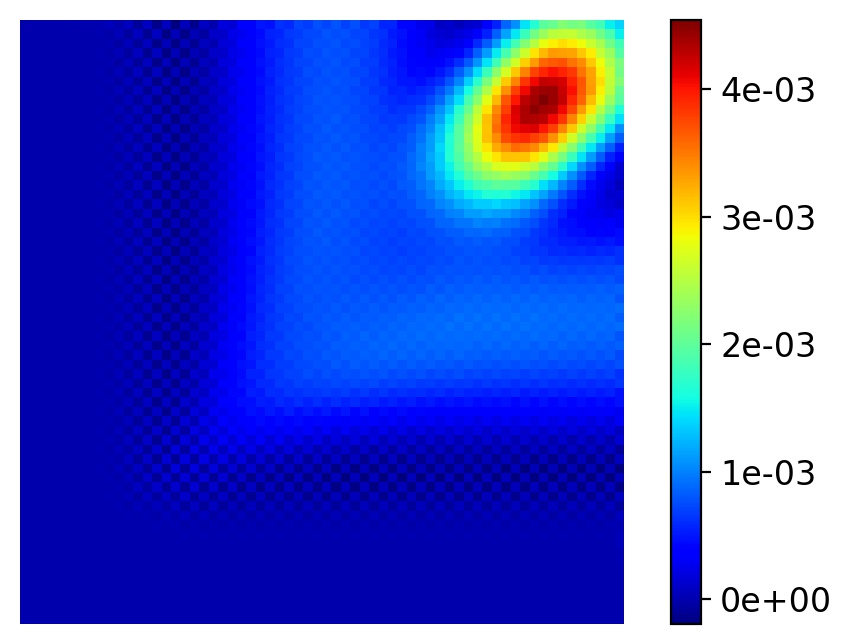}
\caption*{\text{\small Residual on $128 \times 128$}}
\end{subfigure}
\begin{subfigure}[t]{0.22\textwidth}
\centering
\includegraphics[width=\textwidth]{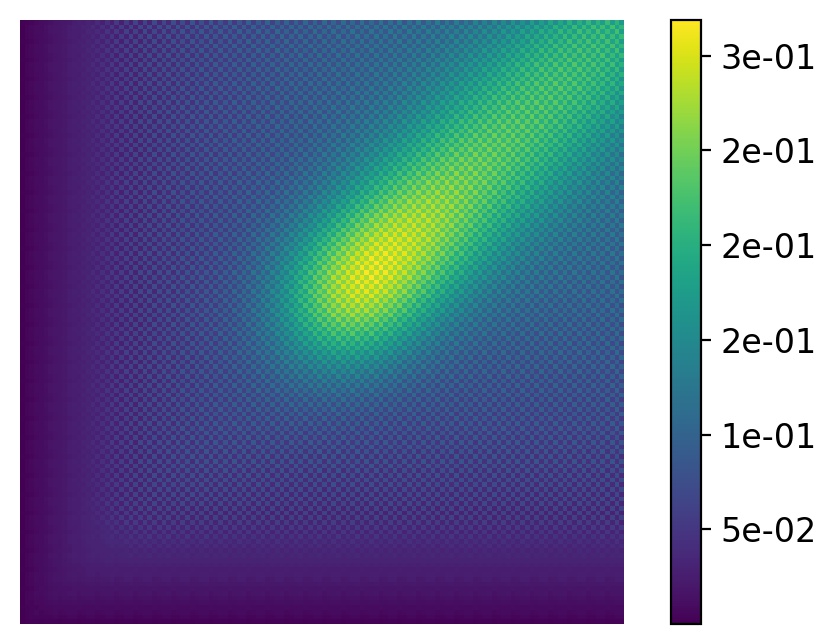}
\caption*{\text{\small Smoothed solution}}
\end{subfigure}
\caption{Multigrid solution of advection-diffusion equation by a CNN. The plots represent the spatial evolution of residuals on different grids within one multigrid iteration, with the final plot showing the converged solution after 100 multigrid iterations.}
\label{fig:multigrid_grids}
\end{figure}

The results of applying this network to the initial condition are shown in Figure~\ref{fig:multigrid_iterative_solutions}, plotting the solution at the 5th, 10th, and 15th multigrid iterations during the solution process. As the iterations increase, the advection across the domain is better represented. We also show the difference between the solutions at certain iterations to show how the multigrid method corrects the solution on its way to obtaining the final converged solution. One can notice that the method converges very quickly in the vicinity of where the initial solution is non-zero and, although it rapidly spreads information across the grid in the advection direction, it needs more iterations to make fine adjustments to information that it has advected a longer distance. In Figure~\ref{fig:multigrid_convergence} we show the $L_1$ norm of the difference of the solution over successive iterations. There is a reduction in the $L_1$ norm of over three orders of magnitude in the first 50 multigrid iterations, with only a small reduction thereafter. Here, we assume that the $L_1$ norm of the difference between two consecutive solutions is a gauge of convergence and set the tolerance for convergence to be  $10^{-5}$. Figure~\ref{fig:multigrid_grids} shows the residual of the 15th multigrid iteration on the different grid levels. This shows how the solution is corrected on each grid level, revealing that the coarse grid levels seem to do much of the work in representing the solution correction. 

\subsection{Burgers equation}
\begin{figure}[htbp]
\centering

\begin{minipage}[t]{1\textwidth}
\centering
\hspace{8em}
\includegraphics[width=0.5\textwidth]{figures/Figure_advection_diffusion/legend.png}
\put(-350,10){{\small Velocity magnitude (m/s)}}
\end{minipage}

\begin{subfigure}[t]{0.19\textwidth}
\centering
\includegraphics[width=\textwidth]{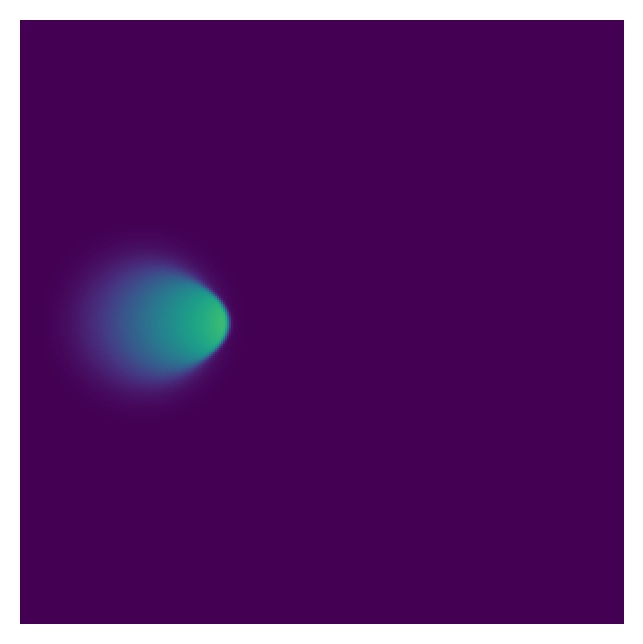}
\put(-110,30){\rotatebox{90}{\Large 10s}}
\end{subfigure}
\begin{subfigure}[t]{0.27\textwidth}
\centering
\includegraphics[width=\textwidth]{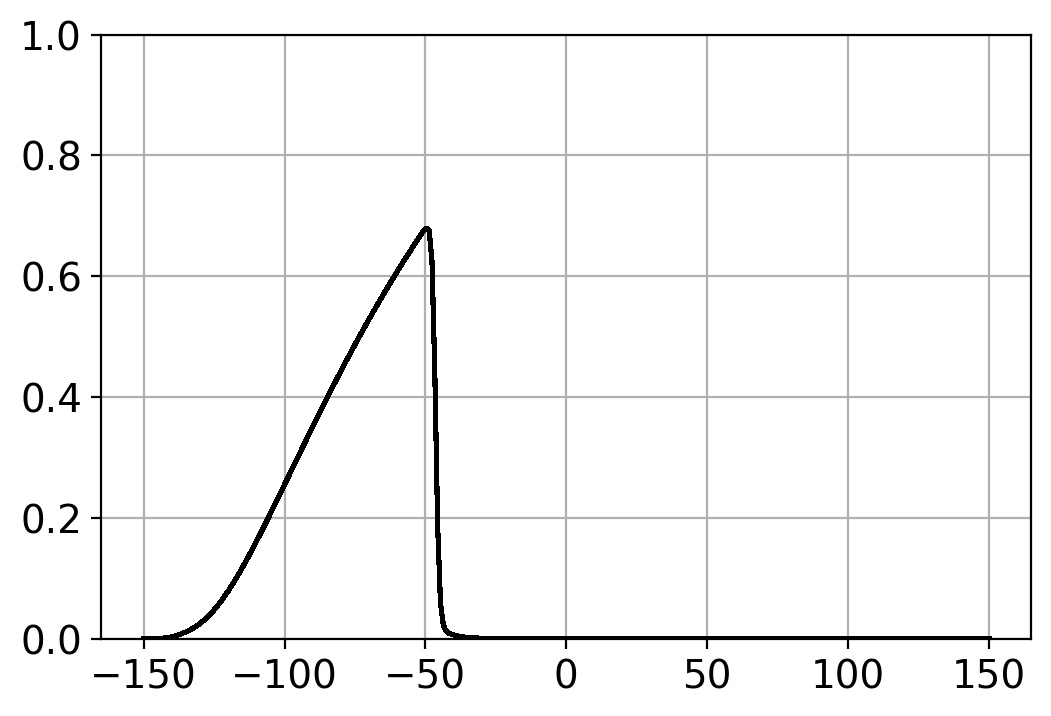}
\end{subfigure}
\begin{subfigure}[t]{0.19\textwidth}
\centering
\includegraphics[width=\textwidth]{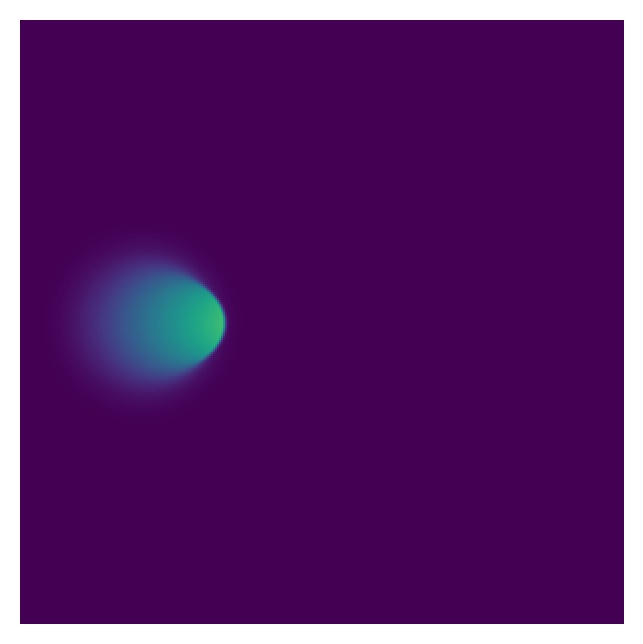}
\end{subfigure}
\begin{subfigure}[t]{0.27\textwidth}
\centering
\includegraphics[width=\textwidth]{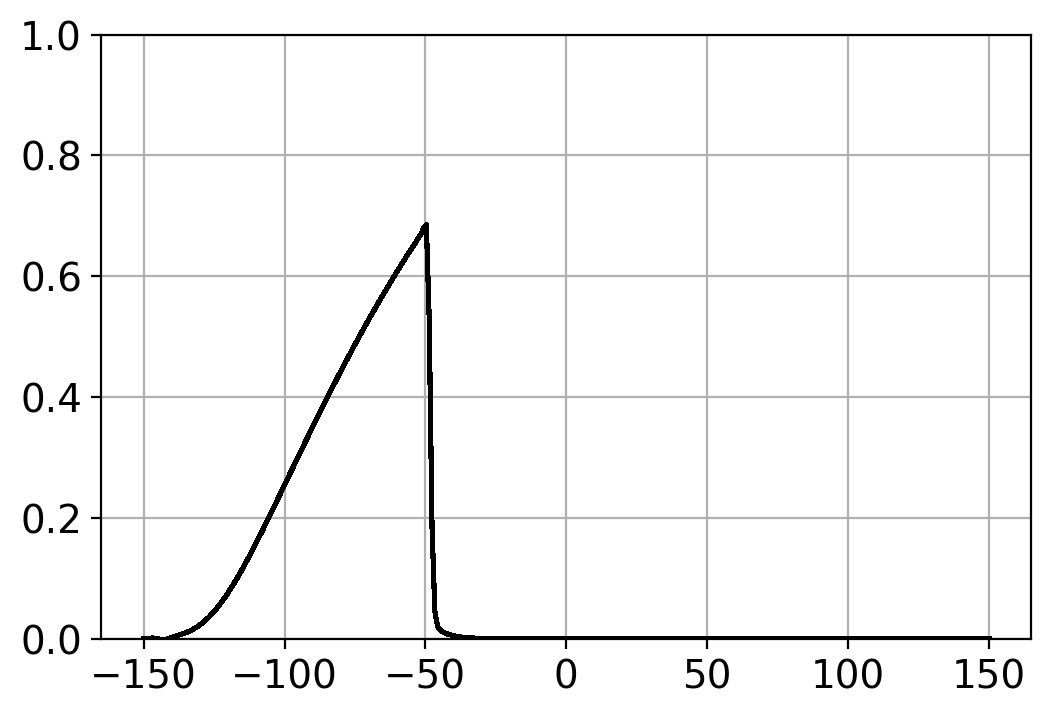}
\end{subfigure}

\begin{subfigure}[t]{0.19\textwidth}
\centering
\includegraphics[width=\textwidth]{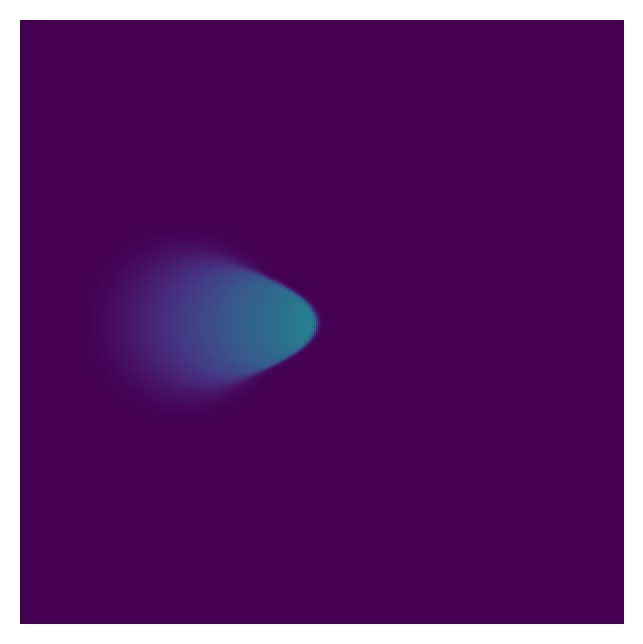}
\put(-110,30){\rotatebox{90}{\Large 30s}}
\end{subfigure}
\begin{subfigure}[t]{0.27\textwidth}
\centering
\includegraphics[width=\textwidth]{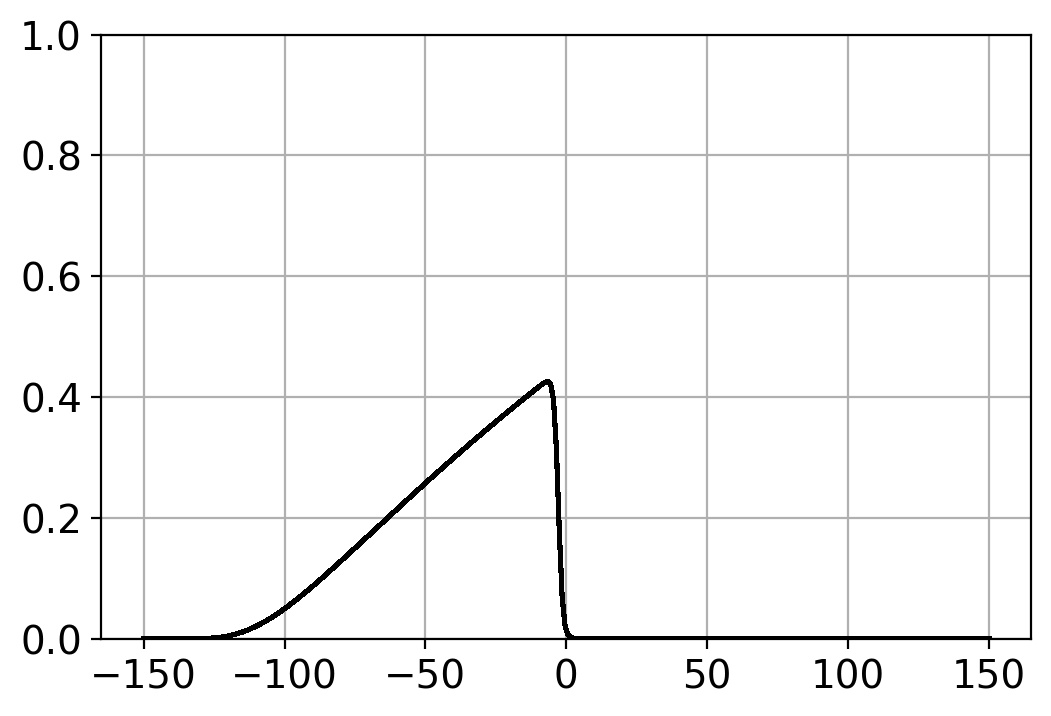}
\end{subfigure}
\begin{subfigure}[t]{0.19\textwidth}
\centering
\includegraphics[width=\textwidth]{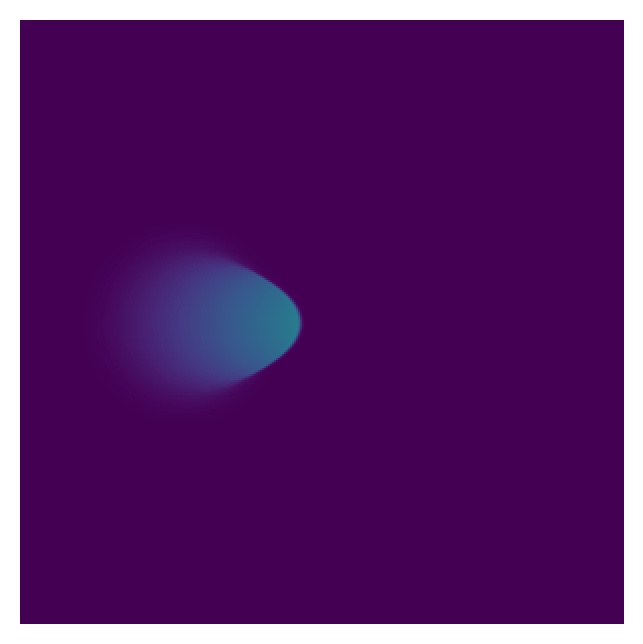}
\end{subfigure}
\begin{subfigure}[t]{0.27\textwidth}
\centering
\includegraphics[width=\textwidth]{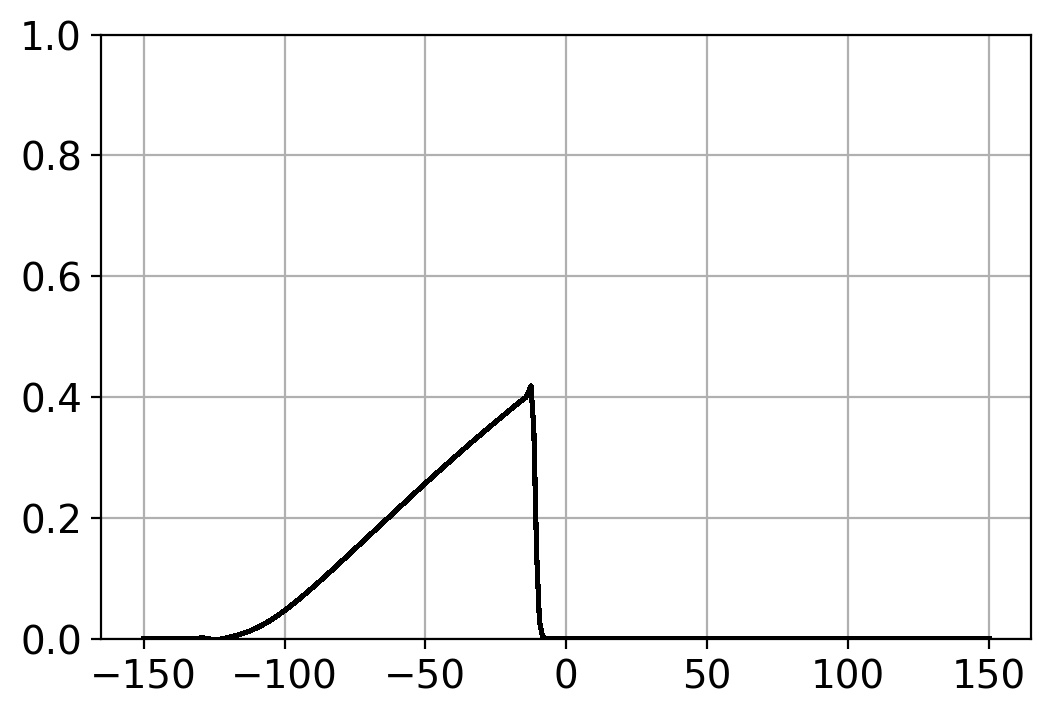}
\end{subfigure}

\begin{subfigure}[t]{0.19\textwidth}
\centering
\includegraphics[width=\textwidth]{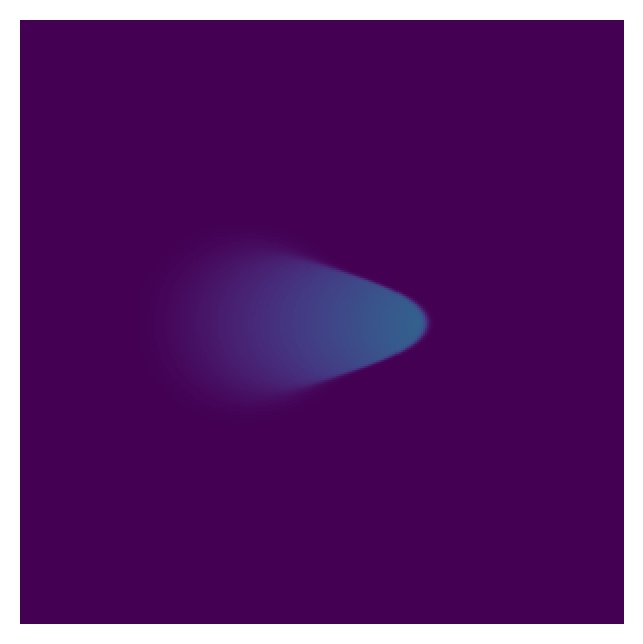}
\put(-110,30){\rotatebox{90}{\Large 60s}}
\end{subfigure}
\begin{subfigure}[t]{0.27\textwidth}
\centering
\includegraphics[width=\textwidth]{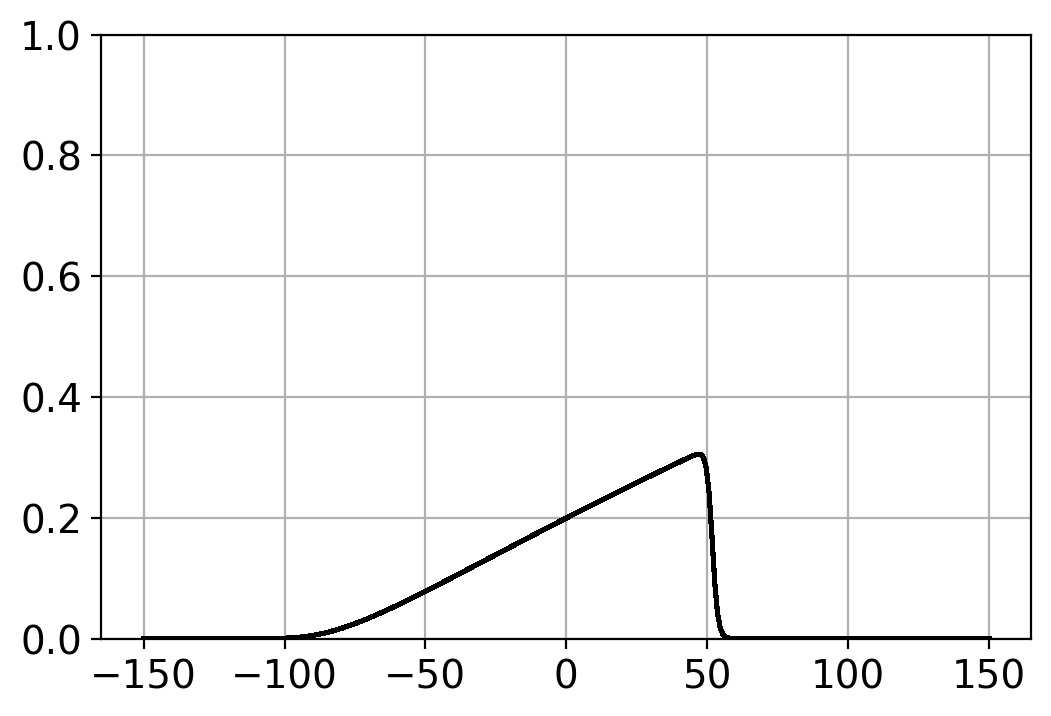}
\end{subfigure}
\begin{subfigure}[t]{0.19\textwidth}
\centering
\includegraphics[width=\textwidth]{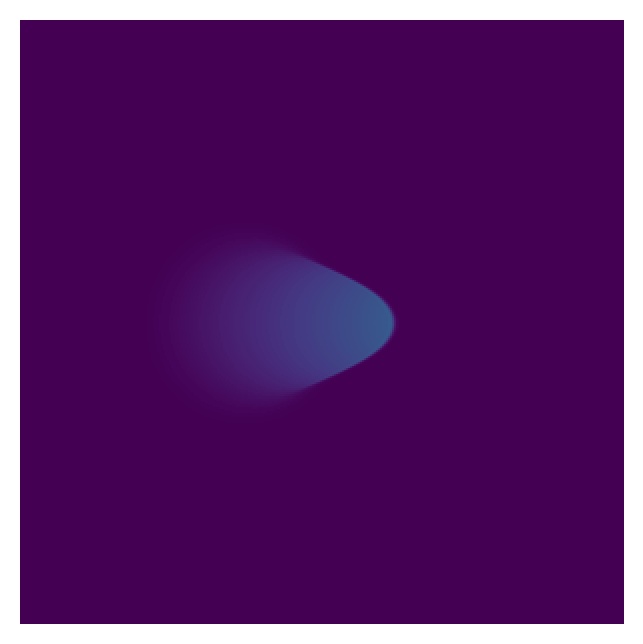}
\end{subfigure}
\begin{subfigure}[t]{0.27\textwidth}
\centering
\includegraphics[width=\textwidth]{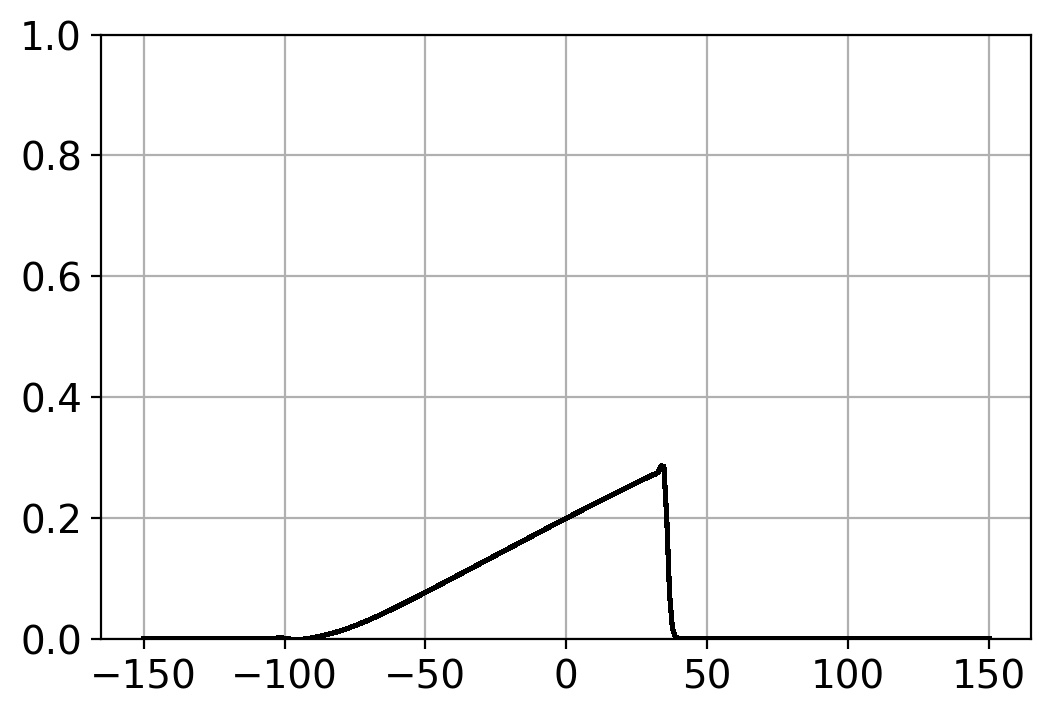}
\end{subfigure}

\begin{subfigure}[t]{0.19\textwidth}
\centering
\includegraphics[width=\textwidth]{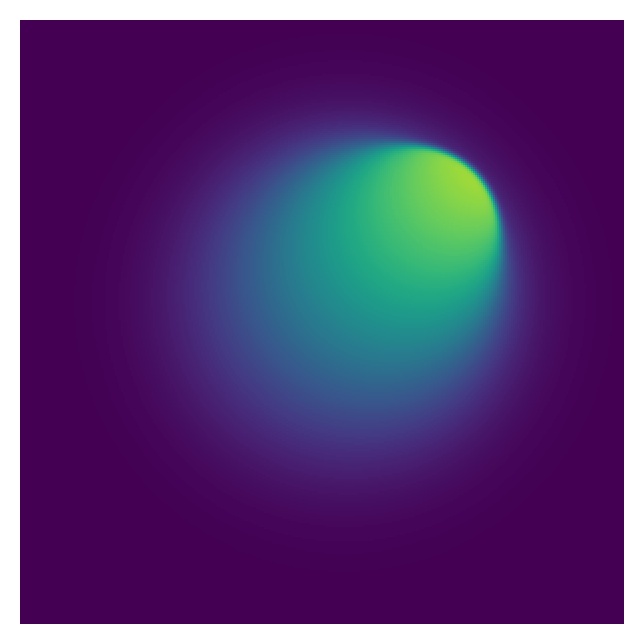}
\put(-110,30){\rotatebox{90}{\Large 10s}}
\end{subfigure}
\begin{subfigure}[t]{0.27\textwidth}
\centering
\includegraphics[width=\textwidth]{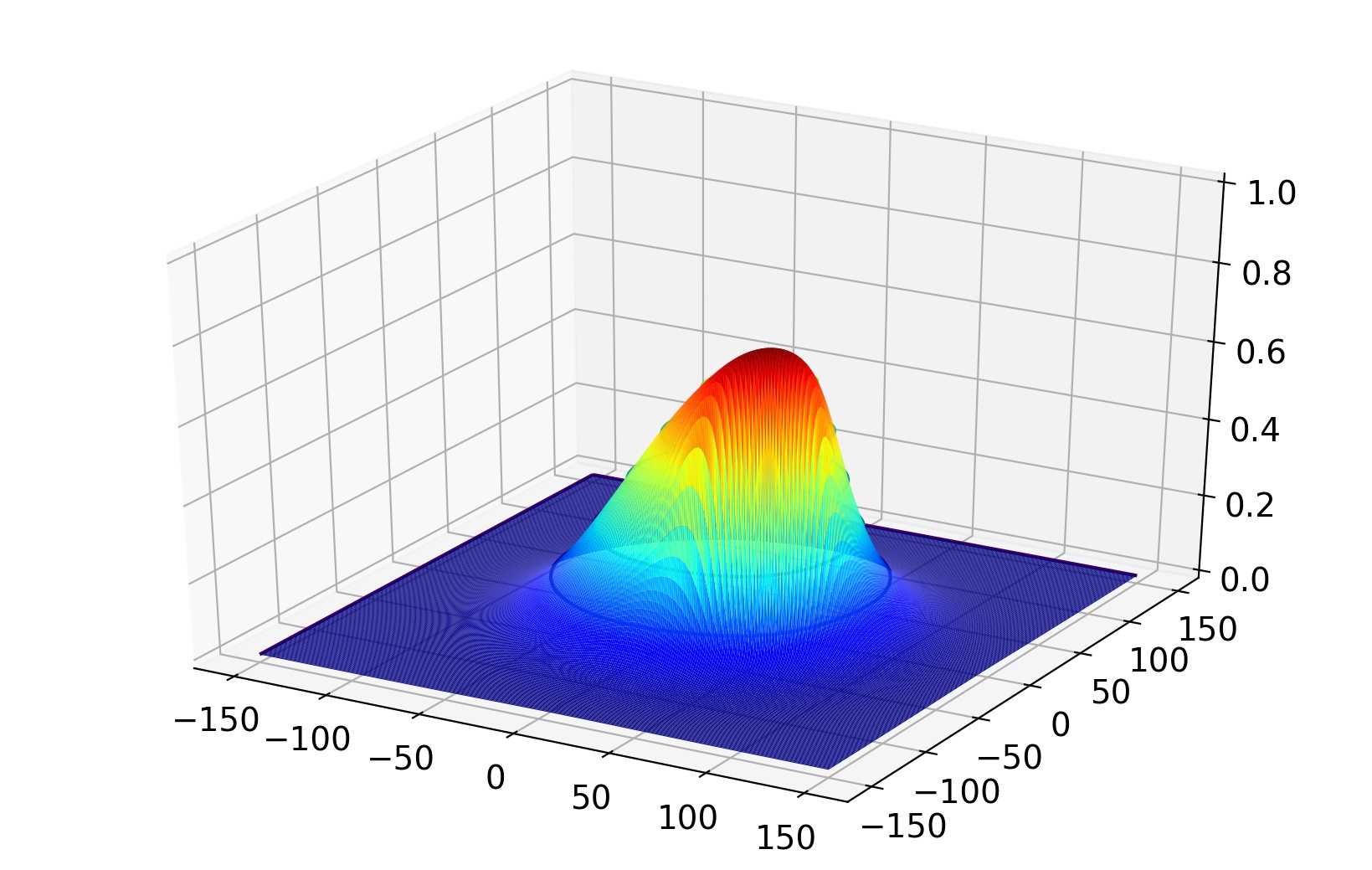}
\end{subfigure}
\begin{subfigure}[t]{0.19\textwidth}
\centering
\includegraphics[width=\textwidth]{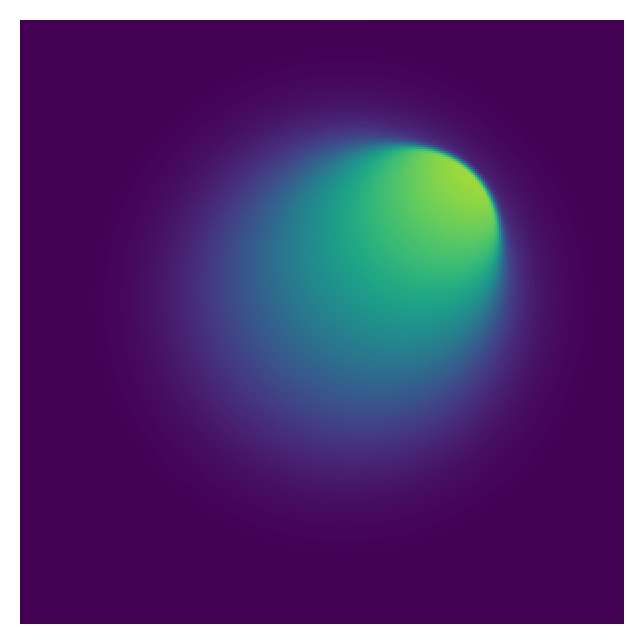}
\end{subfigure}
\begin{subfigure}[t]{0.27\textwidth}
\centering
\includegraphics[width=\textwidth]{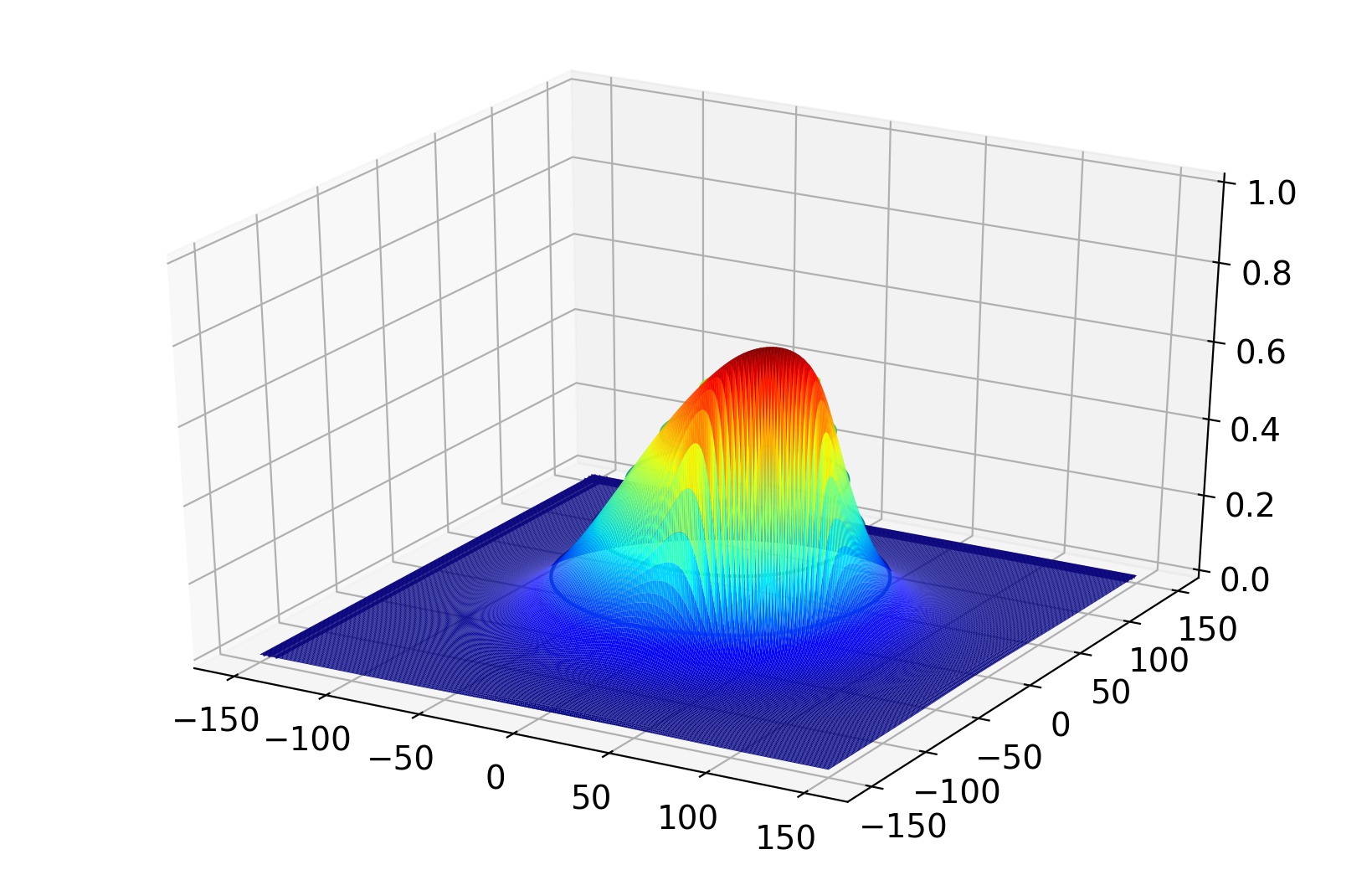}
\end{subfigure}

\begin{subfigure}[t]{0.19\textwidth}
\centering
\includegraphics[width=\textwidth]{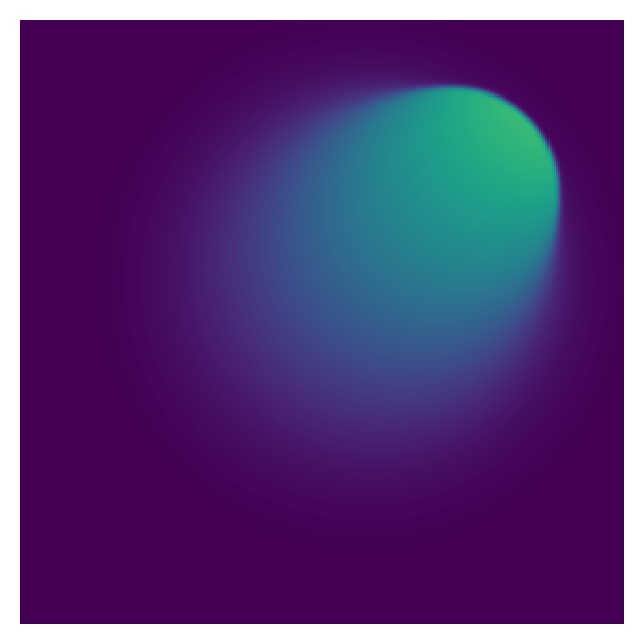}
\put(-110,30){\rotatebox{90}{\Large 30s}}
\end{subfigure}
\begin{subfigure}[t]{0.27\textwidth}
\centering
\includegraphics[width=\textwidth]{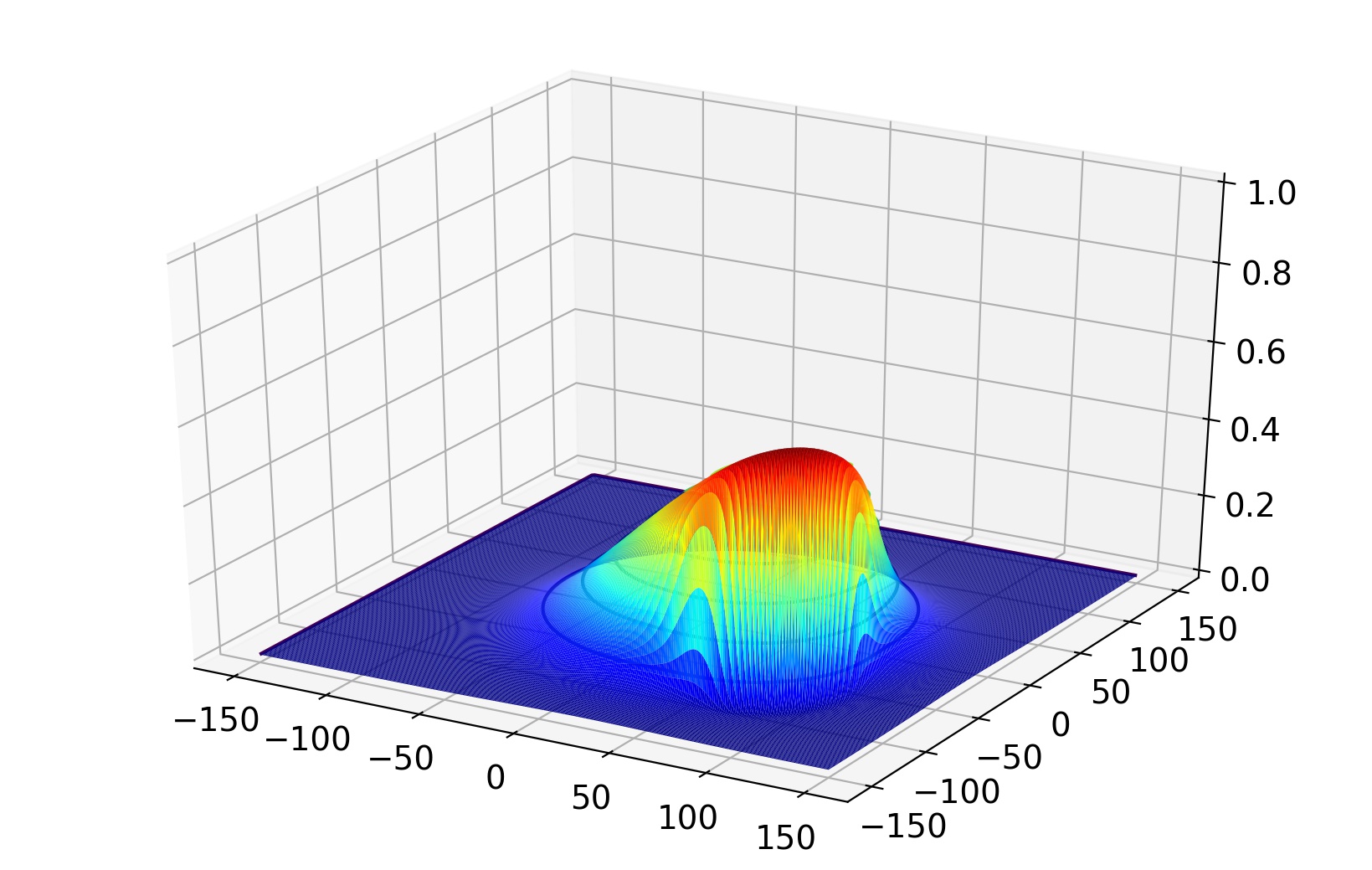}
\end{subfigure}
\begin{subfigure}[t]{0.19\textwidth}
\centering
\includegraphics[width=\textwidth]{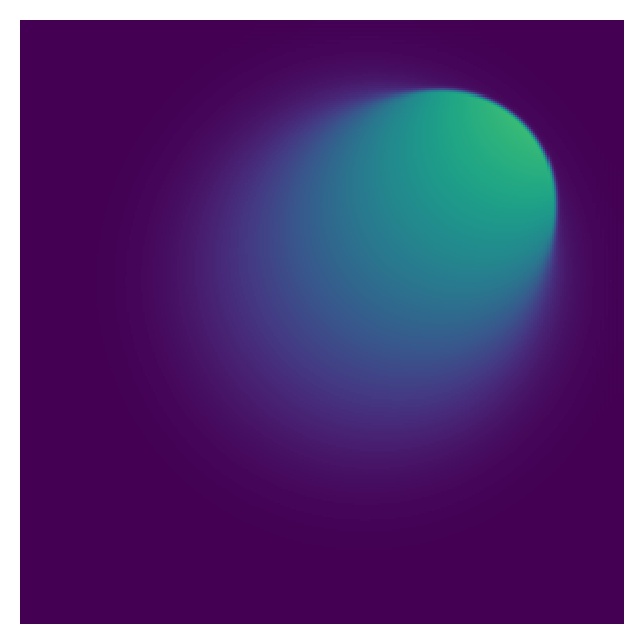}
\end{subfigure}
\begin{subfigure}[t]{0.27\textwidth}
\centering
\includegraphics[width=\textwidth]{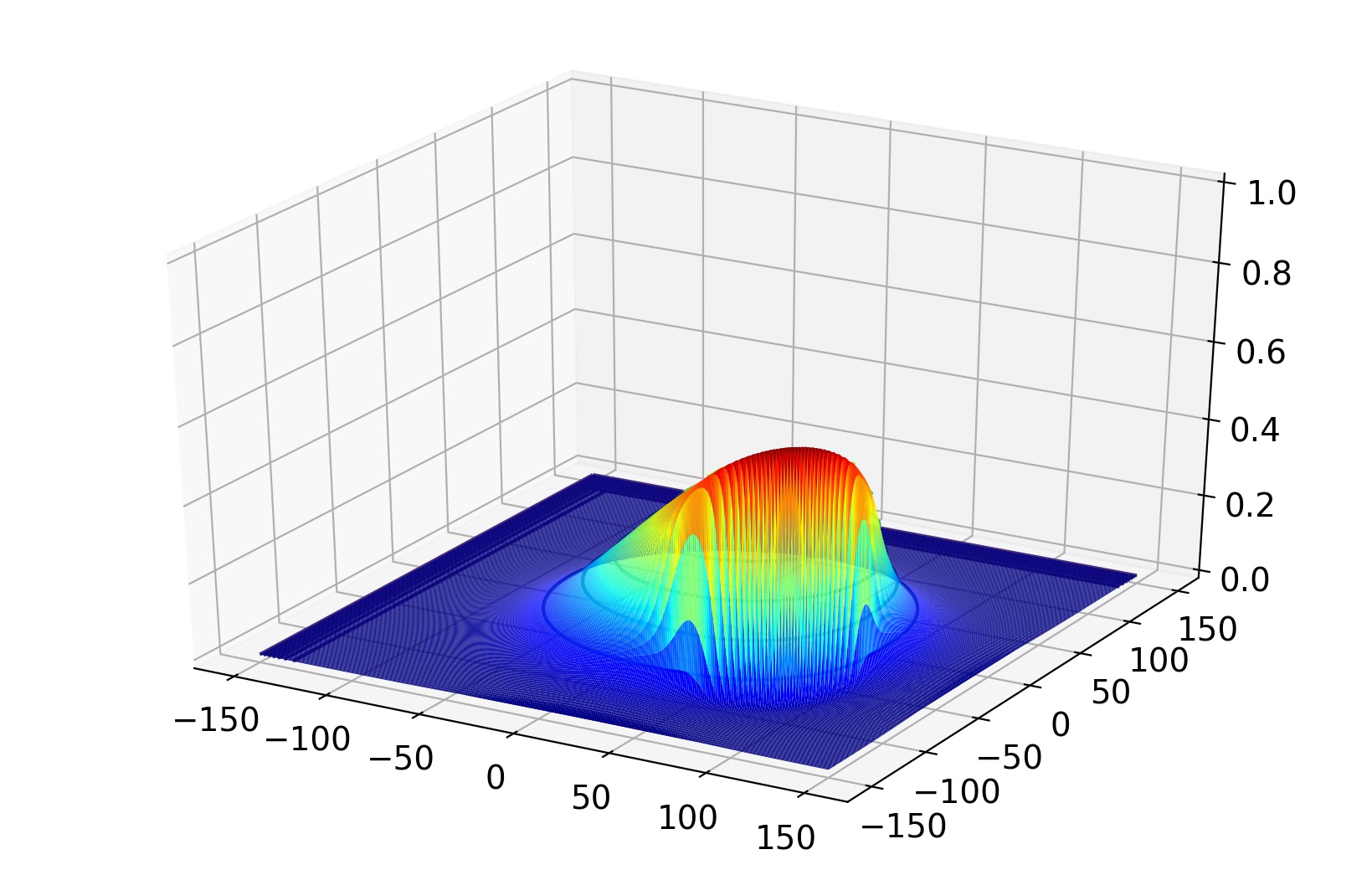}
\end{subfigure}

\begin{subfigure}[t]{0.19\textwidth}
\centering
\includegraphics[width=\textwidth]{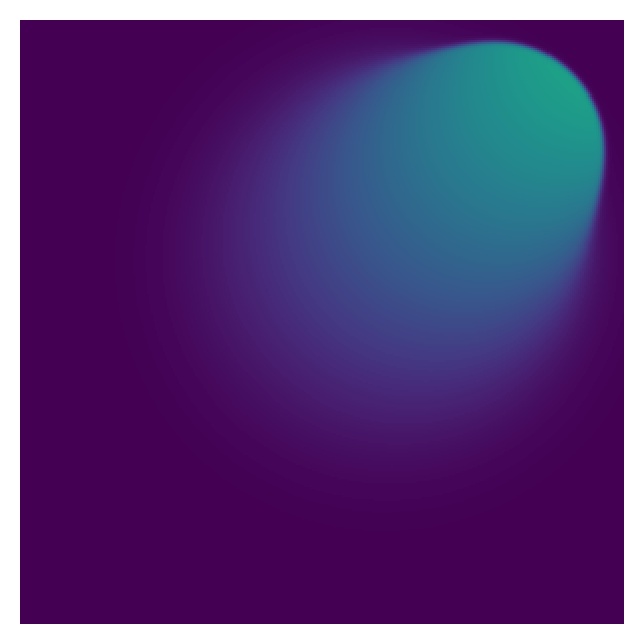}
\put(-110,30){\rotatebox{90}{\Large 60s}}
\caption*{\text{\large (a)}}
\end{subfigure}
\begin{subfigure}[t]{0.27\textwidth}
\centering
\includegraphics[width=\textwidth]{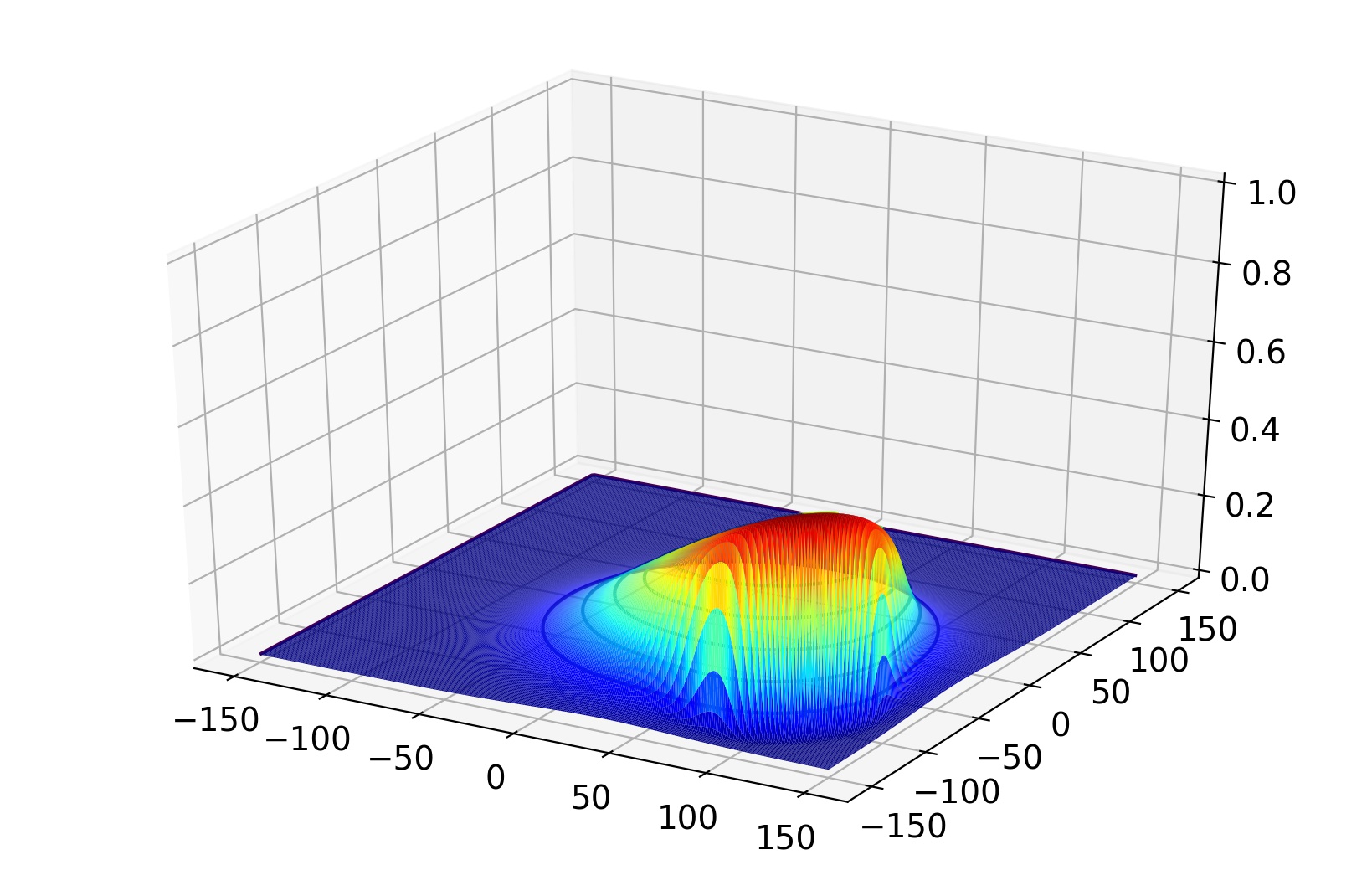}
\caption*{\text{\large (b)}}
\end{subfigure}
\begin{subfigure}[t]{0.19\textwidth}
\centering
\includegraphics[width=\textwidth]{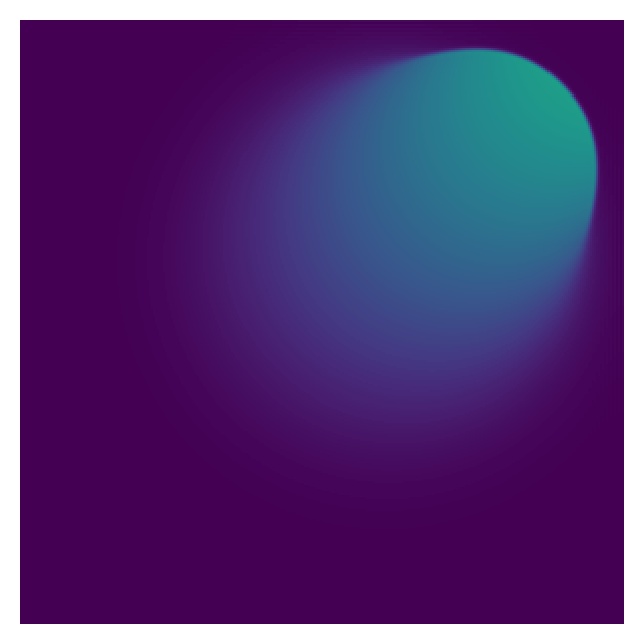}
\caption*{\text{\large (c)}}
\end{subfigure}
\begin{subfigure}[t]{0.27\textwidth}
\centering
\includegraphics[width=\textwidth]{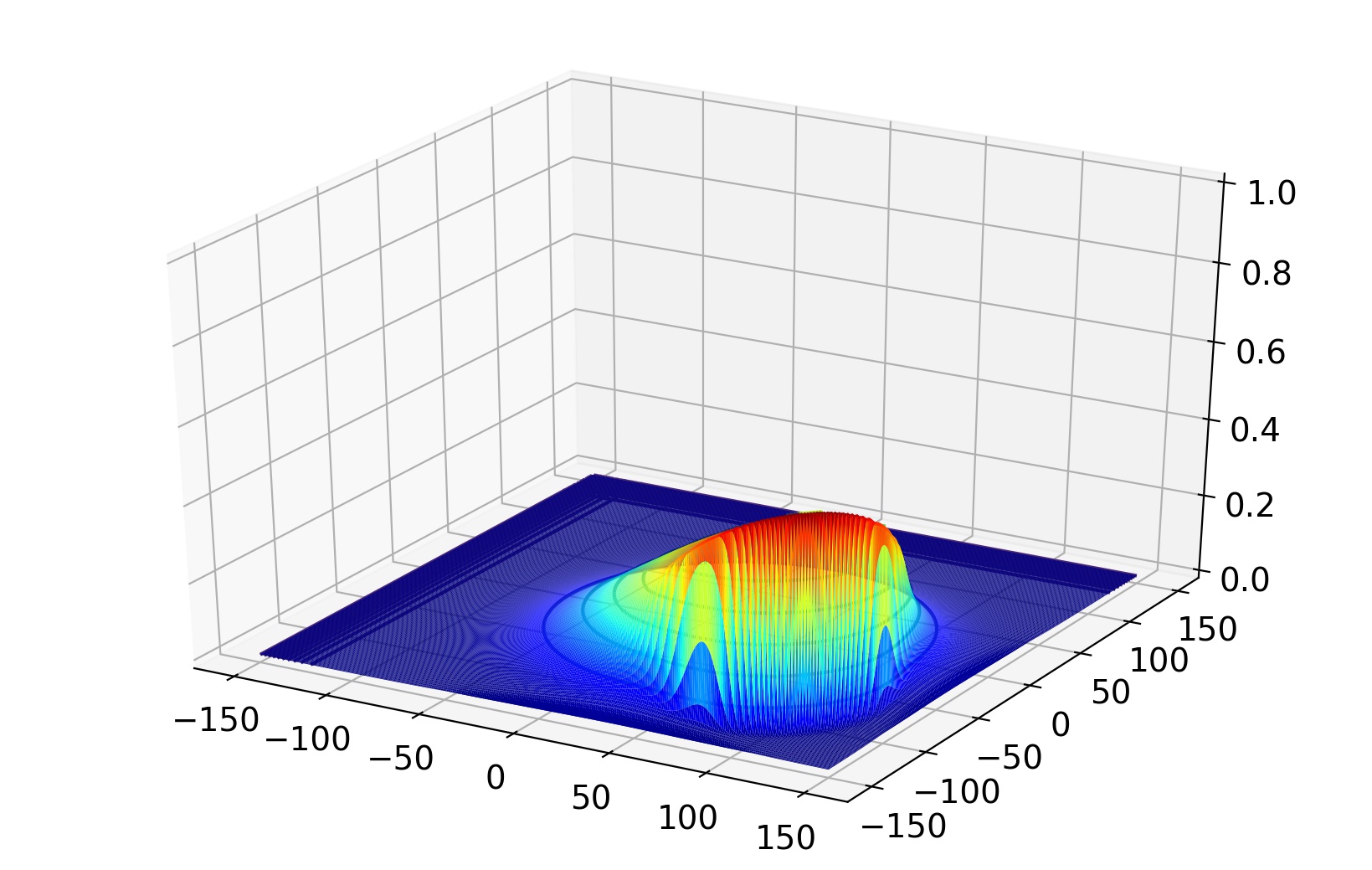}
\caption*{\text{\large (d)}}
\end{subfigure}
\caption{Numerical prediction of 2D nonlinear Burgers equation by a CNN. The top 12 plots relate to the flow field with only the $u$ component and a Gaussian-shaped wave initially centred at $(x_{0},y_{0})=(-100,0)$, at 10, 30 and 60. The bottom 12 plots relate to the flow field with the $u$ and $v$ components and a Gaussian-shaped wave initially centred at $(x_{0},y_{0})=(0,0)$, at 10, 30 and 60. Columns represent second-order time stepping scheme: (a) and (b)~upwind differencing for advection operator and central differencing for diffusion operator in space; (c) and (d)~upwind differencing for advection operator and second-order central differencing for diffusion operator in space.}
\label{fig:burgers_equation}
\end{figure}

For a domain measuring 300~by 300 with the origin at the centre of this, we validate the CNN which has been initialised in such a way that it can solve the Burgers equation and compare the results with a traditionally coded PDE solver. The results from the CNN and the PDE solver were found to be identical up to round-off errors. Subsequently, two tests were performed. The first with a $v$ velocity component of zero (which remains at zero throughout) and a $u$ velocity component that is initialised using the Gaussian distribution, (see the first term in Equation~\eqref{Gaussian_distribution_T}). The nodal spacing in the $x$ and $y$ directions is constant at $\Delta x=\Delta y=1$. The Gaussian distribution is centred at $(x_0,y_0)=(-100,0)$ with a width defined by $\gamma=15$. For the second example, the same domain is used with non-zero initial conditions for both the two velocity components. The Gaussian (associated with $u^0(x,y)$ and $v^0(x,y)$) is centred at $(x_0,y_0)=(0,0)$ with a width of $\gamma=\SI{40}{m}$ (see first term of Equation~\eqref{Gaussian_distribution_T}) and multiplied by $\frac{1}{\sqrt{2}}$. 
In these examples, the viscosity, absorption and source terms of the Burgers equation are set to zero. A predictor-corrector scheme with second-order accuracy is used in time, with central differencing in space for the diffusion term. In one simulation, first-order upwind differencing is used in space for the advection and in a second simulation, a central difference scheme is used. The Courant number is set to be $C=0.1$ using $\Delta t=0.1$. To solve the Burgers equation, the discretisation schemes are coded as proposed, by initialising the weights of the filter of a convolutional layer. The results can be seen in Figure~\ref{fig:burgers_equation}. We see that the method resolves the non-linearity of Burgers equation resulting in the expected shock formation. This happens very quickly during the simulation and all the plots shown in Figure~\ref{fig:burgers_equation} show a sharp front. No oscillations are observed, despite the abruptness of the sharp front, because of the use of first-order upwind differencing of the advection term.   

\subsection{Flow past a bluff body with Navier-Stokes equations}
Here we demonstrate the ability of a CNN to forecast 2D and 3D flow past a bluff body by solving the incompressible Navier-Stokes equations as described in Section~\ref{NS}. The domain has dimensions 512~by 512 with a 40~by 40 solid square block centred at $(x_{0},y_{0})=(128,256)$. The finite element nodes have a uniform spacing in the $x$ and the $y$ directions ($\Delta x$ and $\Delta y$) which is set to~1. The initial velocity and pressure fields are set to zero across the whole domain. A zero pressure boundary condition is imposed on the right-hand side of the domain and other three sides have a zero derivative boundary condition for the pressure field. Boundary conditions for the velocity field include the use of the slip boundary condition at the bottom and the top walls. Inflow and an outflow velocities at the left and the right boundaries are set to~1 in the $x$-direction and~0 in the $y$-direction. 

The Reynolds number used here is $200$ and is specified through the viscosity of the flow, assuming a density of one. For 2D, this Reynolds number yields unsteady flow, but for 3D the resulting flow is steady. We apply a predictor-corrector scheme, accurate to second order, to discretise the time derivative, see Section~\ref{AD}, with a fixed time step of $\Delta = 0.1$. To solve the Navier-Stokes equations, the discretisation scheme is coded by initialising the weights of the filter of a convolutional layer. We use a finite element discretisation based on bilinear rectangular (in 2D) and quadrilateral (in 3D) finite elements for the advection terms. We use the 27 point stencil for the diffusion and advection operators in 3D implemented with $3\times 3\times 3$ filters as described in the appendix. The two components of the velocity field $u,v$ for 2D and three velocity components $u,v,w$, in 3D, are treated as separate inputs of the neural network by each being assigned to a channel. The pressure field is obtained by applying a multigrid solver to the resulting Poisson equation, see Equation~\eqref{PC-DP}. 

\begin{figure}[htbp]
\centering
\begin{minipage}{1\textwidth}
\hspace{18em}
\includegraphics[width=0.4\textwidth]{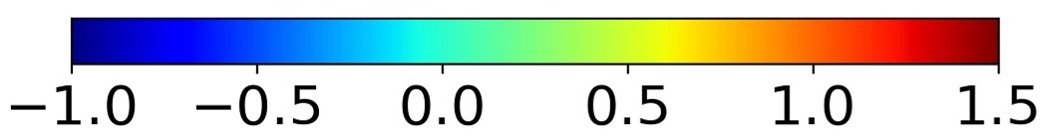}
\put(-300,10){{\large u-component (m/s)}}
\end{minipage}

\begin{subfigure}[t]{0.24\textwidth}
\centering
\includegraphics[width=\textwidth]{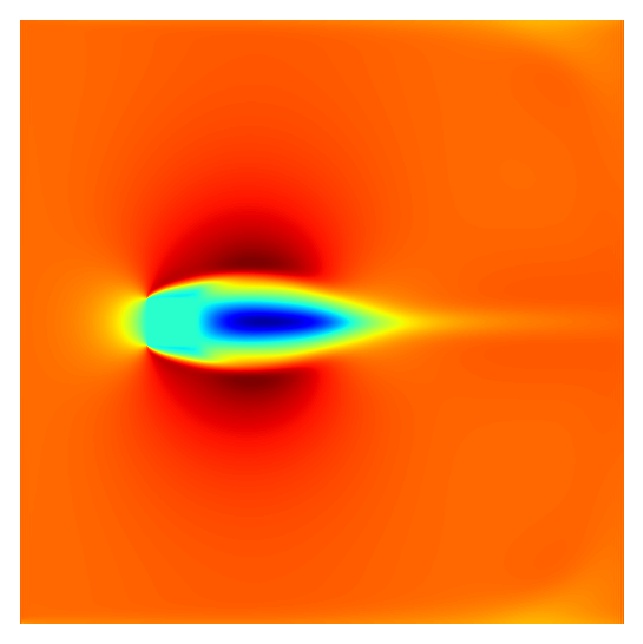}
\end{subfigure}
\begin{subfigure}[t]{0.24\textwidth}
\centering
\includegraphics[width=\textwidth]{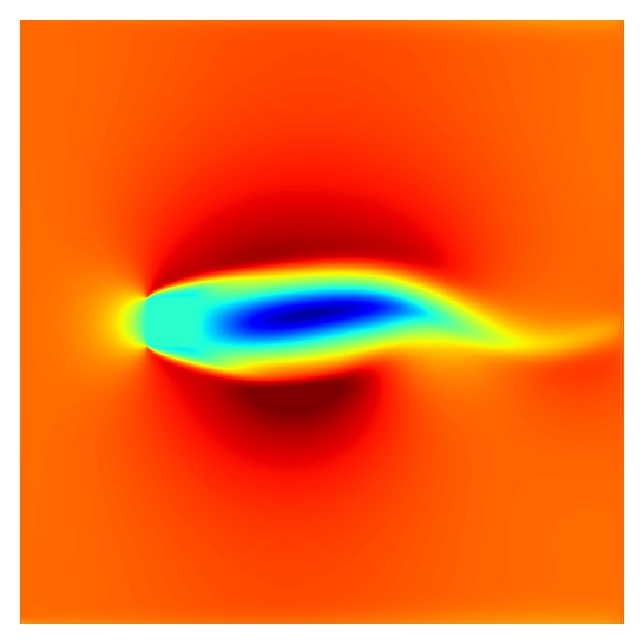}
\end{subfigure}
\begin{subfigure}[t]{0.24\textwidth}
\centering
\includegraphics[width=\textwidth]{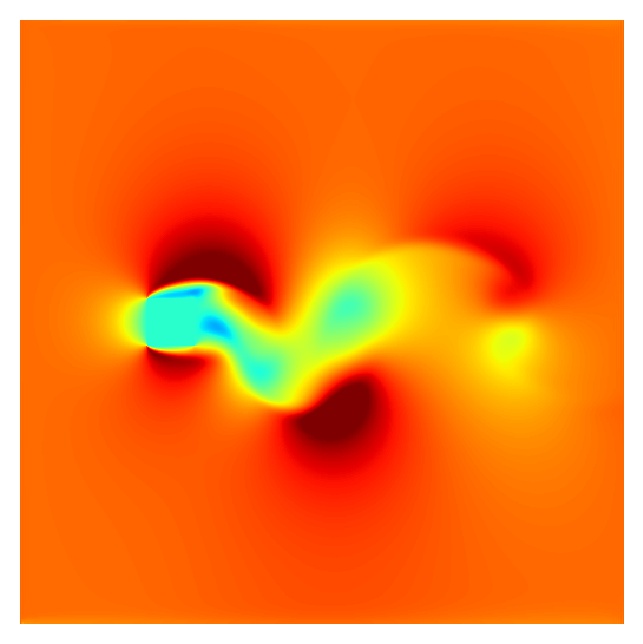}
\end{subfigure}
\begin{subfigure}[t]{0.24\textwidth}
\centering
\includegraphics[width=\textwidth]{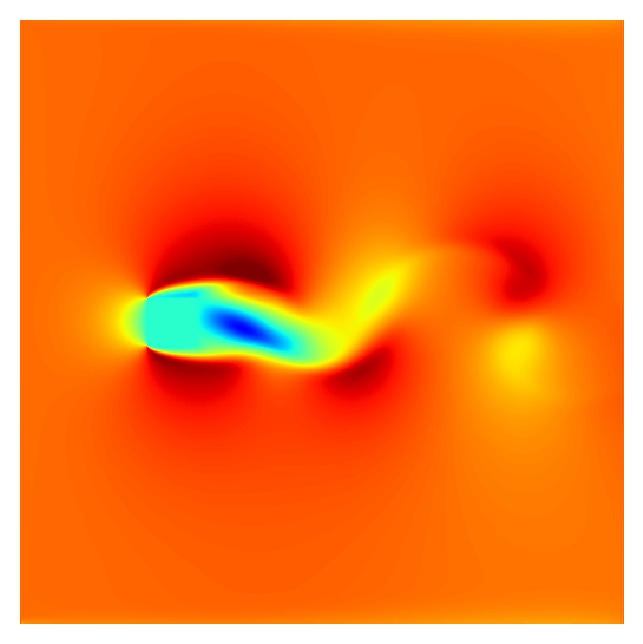}
\end{subfigure}

\begin{minipage}{1\textwidth}
\hspace{18em}
\includegraphics[width=0.4\textwidth]{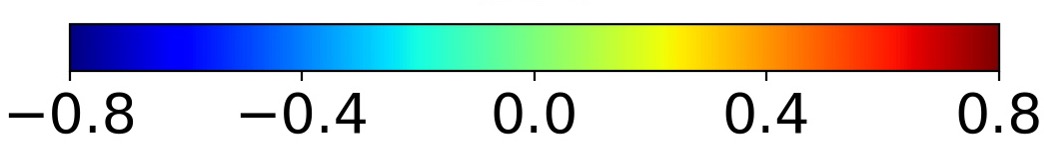}
\put(-300,10){{\large v-component (m/s)}}
\end{minipage}
\begin{subfigure}[t]{0.24\textwidth}
\centering
\includegraphics[width=\textwidth]{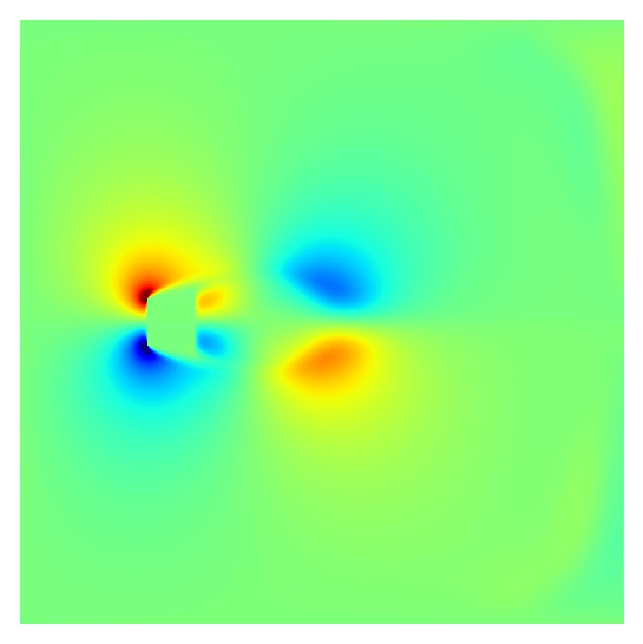}
\end{subfigure}
\begin{subfigure}[t]{0.24\textwidth}
\centering
\includegraphics[width=\textwidth]{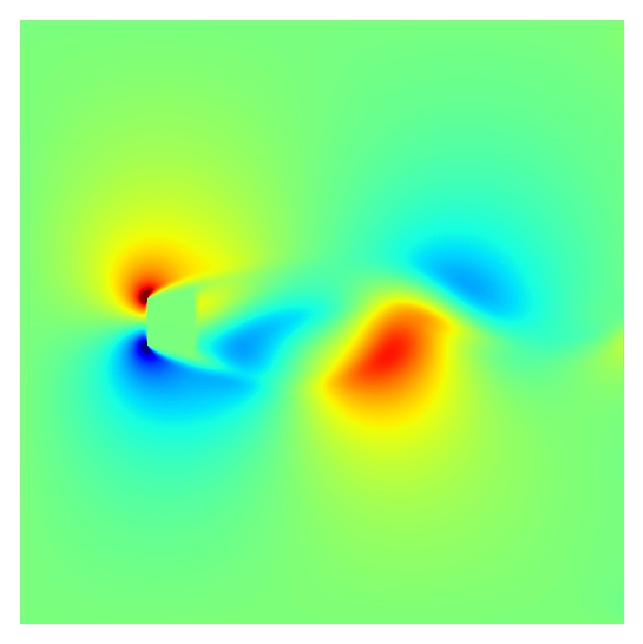}
\end{subfigure}
\begin{subfigure}[t]{0.24\textwidth}
\centering
\includegraphics[width=\textwidth]{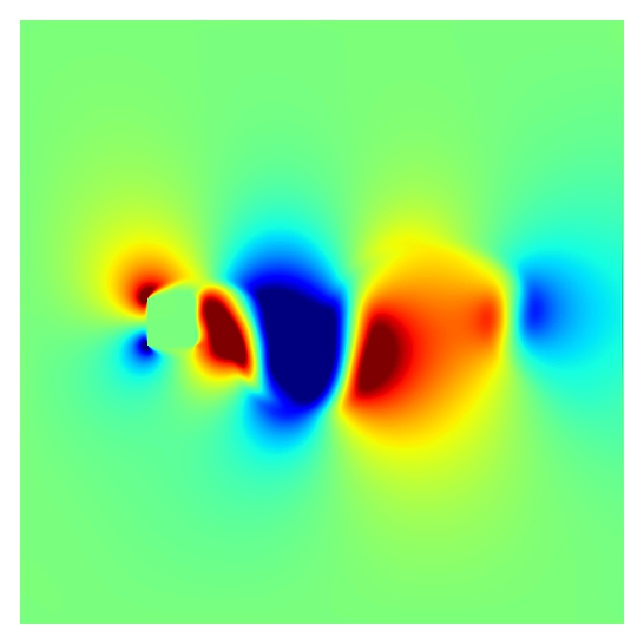}
\end{subfigure}
\begin{subfigure}[t]{0.24\textwidth}
\centering
\includegraphics[width=\textwidth]{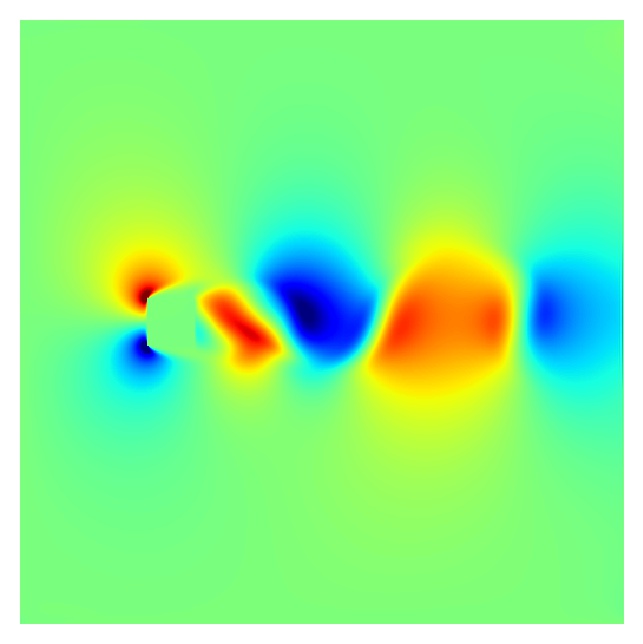}
\end{subfigure}
\begin{minipage}{1\textwidth}
\hspace{18em}
\includegraphics[width=0.4\textwidth]{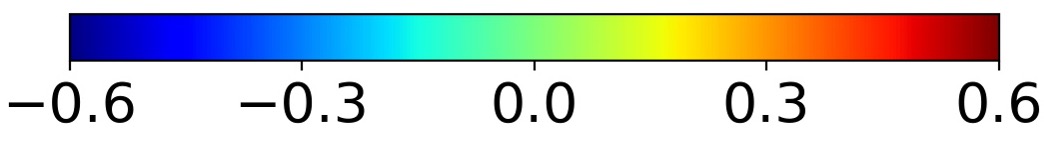}
\put(-300,10){{\large pressure (pa)}}
\end{minipage}
\begin{subfigure}[t]{0.24\textwidth}
\centering
\includegraphics[width=\textwidth]{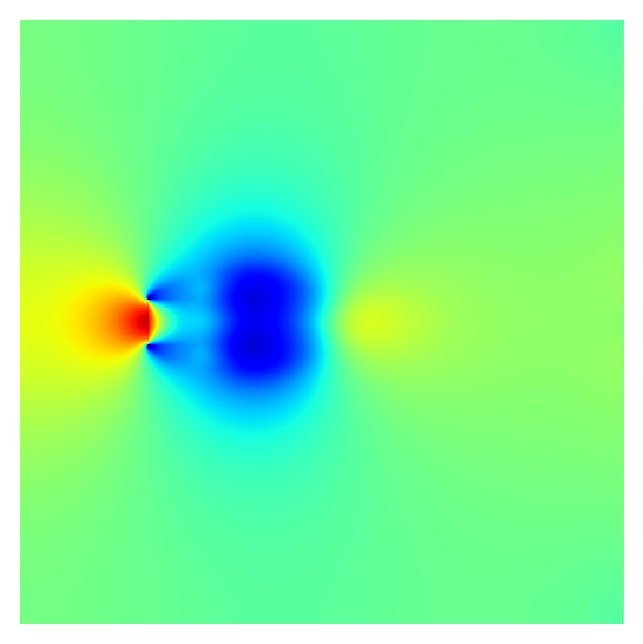}
\caption*{\text{\large 500s}}
\end{subfigure}
\begin{subfigure}[t]{0.24\textwidth}
\centering
\includegraphics[width=\textwidth]{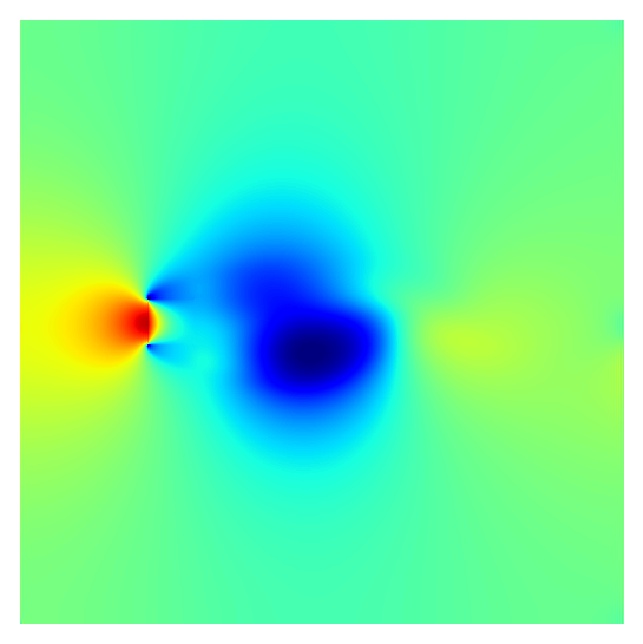}
\caption*{\text{\large 1000s}}
\end{subfigure}
\begin{subfigure}[t]{0.24\textwidth}
\centering
\includegraphics[width=\textwidth]{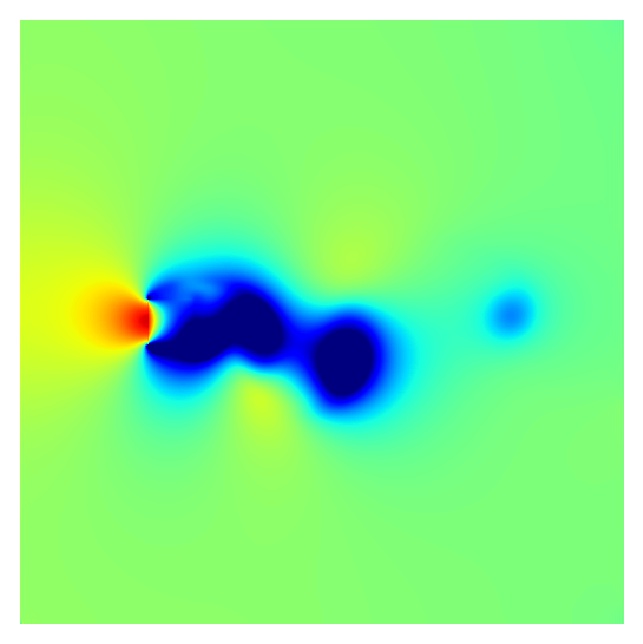}
\caption*{\text{\large 1500s}}
\end{subfigure}
\begin{subfigure}[t]{0.24\textwidth}
\centering
\includegraphics[width=\textwidth]{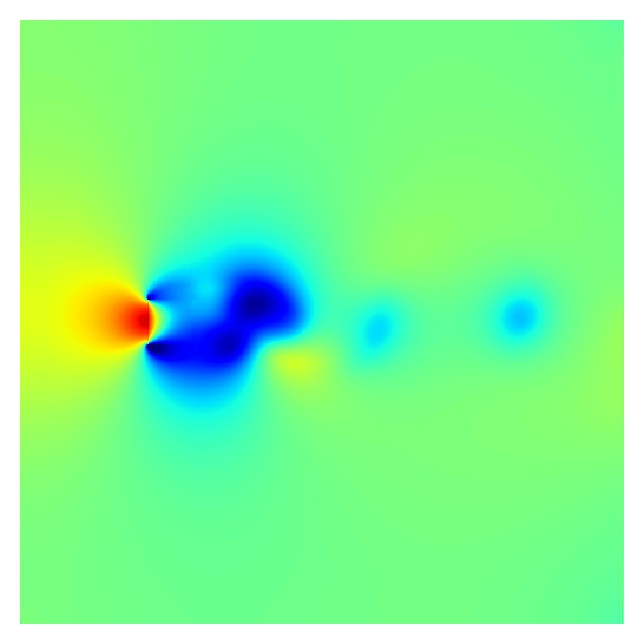}
\caption*{\text{\large 2000s}}
\end{subfigure}

\begin{subfigure}[t]{0.16\textwidth}
\centering
\includegraphics[width=\textwidth]{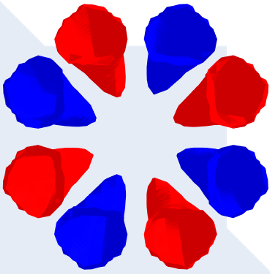}
\caption*{\text{\large (a)}}
\end{subfigure}
\begin{subfigure}[t]{0.42\textwidth}
\centering
\includegraphics[width=\textwidth]{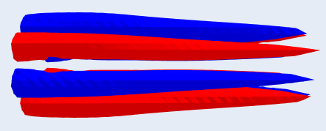}
\caption*{\text{\large (b)}}
\end{subfigure}
\begin{subfigure}[t]{0.354\textwidth}
\centering
\includegraphics[width=\textwidth]{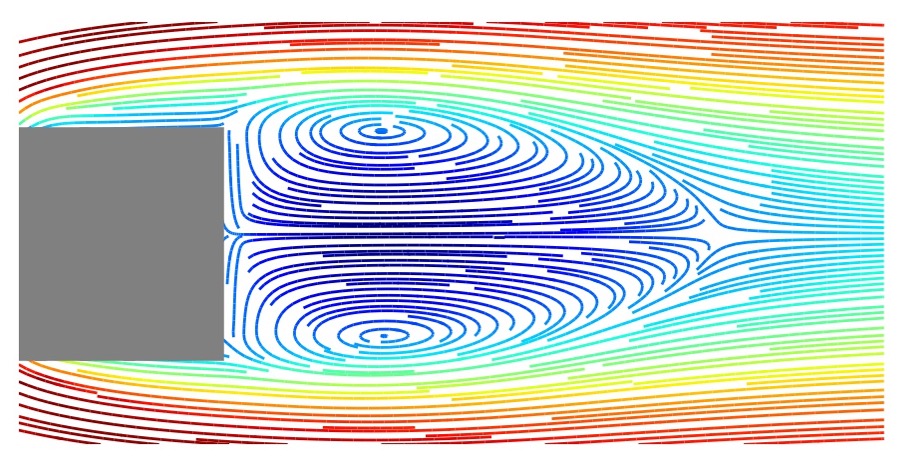}
\caption*{\text{\large (c)}}
\end{subfigure}
\caption{Numerical prediction of 2D and 3D flow past a bluff body by a CNN. The 12 plots relate to the two velocities and the pressure of the 2D flow field, at \SI{500}{s}, \SI{1000}{s}, \SI{1500}{s} and \SI{2000}{s}. The dimension of 2D computational domain is $\SI{512}{m}$ by $\SI{512}{m}$, where a solid square body is modelled as $\SI{40}{m}$ by $\SI{40}{m}$ and centred at $(x_{0},y_{0})=(\SI{128}{m},\SI{256}{m})$. The bottom 3 plots relate to the steady vortex structure in the 3D flow field at \SI{1000}{s}: (a) and (b) are the rear and side views of the isosurface representing $\omega_{x} = \pm 0.01$ with red and blue for positive and negative; (c) is 2D steady streamlines coloured by the u-component velocity. The dimension of 3D computational domain is $\SI{128}{m}$ by $\SI{128}{m}$ by $\SI{128}{m}$, where a solid square body is modelled as $\SI{10}{m}$ by $\SI{10}{m}$ by $\SI{10}{m}$ and centred at $(x_{0},y_{0},z_{0})=(\SI{32}{m},\SI{64}{m},\SI{64}{m})$. The Reynolds number of both cases is~200. Both sets of results use a second-order time discretisation scheme with a spatial discretisation based on bilinear rectangular and quadrilateral finite elements in 2D and 3D respectively.}
\label{fig:bluff}
\end{figure}

\begin{figure}[htbp]
\centering
    \centering   \includegraphics[width=0.7\textwidth]{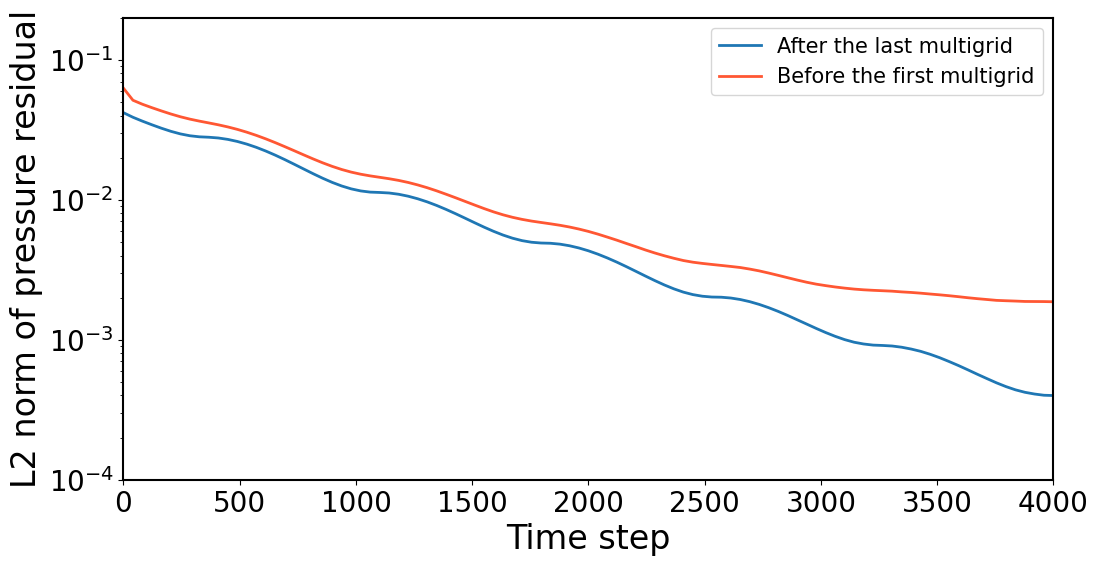}
    \caption{\label{fig:l2norm_over_time}A plot of the $L_2$ norm of the pressure residual in the multigrid solver over time for the solution of 3D flow past a bluff body.}
\end{figure}

The results for both 2D and 3D flows are shown in Figure~\ref{fig:bluff}. The top 12 plots correspond to the 2D results for velocity and pressure. We see that the method resolves the flow structures around bluff body, including the unsteady separation of flow and vortex shedding~\citep{Lekkala2022}. In the 2D simulation, the non-dimensional frequency of vortex shedding or Strouhal number is calculated to be $0.142$ from our results for $Re=200$, which is very close (relative error of 3.4\%) to the database value of 0.147 established by \citep{wiki:xxx}. Results for the 3D flows are shown in the three bottom plots of Figure~\ref{fig:bluff}. For this case, the dimension of the domain is 128~by 128 by 128 with a 10~by 10 by 10 solid square block centred at $(x_{0},y_{0},z_{0})=(32,64,64)$ and in which the solution domain is defined by $x\in[0,128],\; y\in[0,128], z\in[0,128]$. Figure~\ref{fig:bluff}(a)--(c) shows the steady vortex structure seen downstream of the bluff body. The length of the re-circulation bubble $l_{1}$ is $\SI{22}{m}$ and the distance between the centre of the circulation bubble where there is no velocity and the cube $l_{2}$ is $\SI{10}{m}$. The ratio of $l_{1}$ to $l_{2}$ can be computed as 2.20, showing a close agreement with the value of 2.22 calculated in~\citep{meng2021wake} (relative error of 0.9\%). For this test case, we use 3 multigrid cycles and show the temporal evolution of the $L_2$ norm of the pressure residual (Equation~\eqref{PC-DP}) in Figure~\ref{fig:l2norm_over_time}, both before the multigrid cycles and after. The reduction of this residual in time is partly explained by the fact that in 3D the flow is steady.  

The computational efficiency of the proposed solver is assessed on various computing architectures, with the Intel Xeon 2.3GHz CPU and NVIDIA Tesla T4 GPU (including 2560 CUDA cores) being chosen. The study involves testing three 3D cases comprising $128^{3}$, $256^{3}$ and $512^{3}$ finite element nodes, each running for five time steps with 20 multigrid iterations per time step. See Table~\ref{tab:timings} for the timings. 

\begin{table}[htbp]
\centering
\begin{tabular}{rcrr}
\toprule
& & \multicolumn{2}{c}{time taken}\\
\cline{3-4}
grid size & number of FE nodes & a single CPU & a single GPU\\
\toprule
$128^{3}$ & \num{2e6} & 165\phantom{100}        & 3 \phantom{100}\\
$256^{3}$ & \num{17e6} & \num{1275}\phantom{100} & 11 \phantom{100}\\
$512^{3}$ & \num{134e6} & \num{14376}\phantom{100} & 34 \phantom{100}\\
\bottomrule
\end{tabular}
\caption{\label{tab:timings}The time taken (in seconds) for the NN4PDEs approach to solve five time steps with 20 multigrid iterations per time step of flow past a bluff body.}
\end{table}

\section{Discussion}\label{sec:discussion}

We believe that the Neural Physics approach of programming PDE solutions using AI software is potentially very important for several reasons. (1)~The approach enables interoperability between GPUs, CPUs and AI processors, which, in turn, enables exploitation of AI community software. Furthermore, energy shortfalls and climate change mean that the use of energy efficient computers will become increasingly pressing. Running software on any new architecture can be difficult, however, as these machines are designed to run AI software, the proposed method, NN4PDEs, is already compatible with such computers. (2)~Another advantage of the approach is that the models are fully differentiable, due to the automatic differentiation functionality of the AI libraries. This makes optimisation tasks such as data assimilation and inverse problems much easier to formulate and brings the potential of running these problems on GPUs and AI processors. (3)~NN4PDEs is ideally placed to combine AI-based reduced-order models and physics-based discretisations of the differential equations, facilitating a number of approaches including AI-based Subgrid-Scale (SGS) models and physics-informed methods~\cite{Raissi2019,Zheng2020,Long2019}. It is now accepted that future SGS methods will be increasingly based on AI and that these often require implicit coupling to CFD: we believe that programming in environments compatible with AI is the only satisfactory approach. Also, model hierarchies, ranging from high fidelity CFD to both fine and coarse reduced-order models, can be embedded in a single neural network, making the methods much more computationally efficient. (4)~New features in modelling can be realised without having to write large quantities of code, for example, mixed arithmetic precision or the coupling of multi-physics. (5) Finally, using AI software could make model development more accessible than programming within existing CFD codes, as the AI libraries already abstract many of the complexities of low-level mathematical operations or optimisation details. The libraries offer long-term sustainability as code is based on long standing, community supported, software projects. 

The examples in this paper are all solved on regular structured grids. For unstructured meshes and parallel computing one can generalise the above approach. For unstructured meshes, one can generalise the above approach by using graph neural networks (GNNs)~\cite{Hanocka2019,Tencer2021}. When using multigrid methods, the graph can be coarsened, as often done in GNN approaches~\cite{Wu2020} or one could use space-filling curves to generate mapping to structured 1D grids~\cite{Heaney2020}. The latter could be highly efficacious as the 1D grid structure can map efficiently onto the memory of the computer, with no indirect addressing. To implement the parallelism, message passing (MPI) can be used to update the halos on the perimeters of the partitioned subdomains. In addition, this approach also allows blocks of different patterns to be used which enables semi-structured or block structured grids to be used and merges the use of parallelisation and unstructured meshes: more than one subdomain can be used on a given core using the MPI approach. We intend to explore these possibilities in the future.

\section{Conclusion}\label{sec:conclusion}
We have introduced the novel ``Neural Physics'' approach, which demonstrates how neural networks can be used to solve differential equations, and have presented some simple applications and theory to underpin the approach. Implementing discretisations of PDEs using convolutional layers in AI libraries, our Neural Physics approach has produced identical results, to within round-off error, to PDE solvers written in more conventional programming languages e.g.~modern Fortran. We have used finite difference and finite element discretisations in 2D and 3D, solved the resulting systems explicitly or with multigrid solvers written as a U-Net, and applied these methods to the advection-diffusion equation, non-linear Burgers equation and the incompressible Navier-Stokes equations. 

In this paper, we indicate how the power of both AI software and hardware can be brought into the field of numerical modelling by repurposing AI methods, such as Convolutional Neural Networks (CNNs) and U-Nets, for the standard operations required in the field of the numerical solution of PDEs. We show that the proposed methodology can solve PDEs using AI libraries in an efficient way, and present a new avenue to explore in the development of methods to solve PDEs and Computational Fluid Dynamics problems. 

In the future, we will explore distorting the structured grid so that the resolution is higher in areas of interest than other areas as well as parallelisation of the approach across multiple GPU nodes. The Neural Physics approach is closely tied to convolutional layers which themselves are dependent on structured grids. Being able to distort such a grid to focus resolution on areas of interest could have a significant impact on accuracy and efficiency. 
In preliminary tests, the NN4PDEs code runs in parallel on a single GPU node (with four \SI{40}{GB} NVIDIA RTX A100 GPUs) being able to model urban environments over a cross-sectional area of \SI{4}{km} by \SI{5}{km} with up to \SI{1}{m} resolution using 2 billion finite-element nodes. Although significant, running across multiple GPU nodes would yield the possibility of achieving higher resolution or simulating larger areas and we intend to investigate this further.

\section*{Acknowledgements}
The authors would like to acknowledge Dr Niki Loppi of NVIDIA AI Technology Center Finland for insightful discussions on aspects of GPU implementation. We also thank Dr Neil Ashton (NVIDIA), Dr Pablo Salinas (OpenGoSim) and Dr Arash Hamzehloo (Empa) for interesting conversations on the idea presented here. 

The particular form of the AI4PDEs/NN4PDEs code in \href{https://github.com/ImperialCollegeLondon/AI4PDEs}{this github repository} evolved during a Schmidt Sciences Hackathon in Oxford organised by Prof.~Ben Lambert; supported by RSEs Dr~Fergus Cooper, Dr~Jack Leland, Dr~Oliver King and Dr~Alasdair Wilson; with Hackathon team members Drs Jialun Chen, Claire Heaney, Milan Kl{\"o}wer, Pranav Mamidanna, Laura Mansfield and Jinzhao Sun, and observers Dr~Boyang Chen and Prof.~Christopher Pain.

Finally we would like to thank the reviewer for helping to improve the manuscript.

\section*{Funding Sources}
We would like to acknowledge the following UKRI grants: \textbf{ECO-AI}, ``Enabling \ch{CO2} capture and storage using AI'' (EPSRC EP/Y005732/1); \textbf{AI-Respire}, ``AI for personalised respiratory health and pollution (EPSRC EP/Y018680/1); \textbf{INHALE}, Health assessment across biological length scales (EPSRC EP/T003189/1); \textbf{WavE-Suite}, ``New Generation Modelling Suite for the Survivability of Wave Energy Convertors in Marine Environments'' (EPSRC EP/V040235/1); the \textbf{PREMIERE} programme grant, ``AI to enhance manufacturing, energy, and healthcare'' (EPSRC EP/T000414/1); \textbf{D-XPERT:} ``AI-Powered Total Building Management System'' (Innovate UK, TMF 10097909); \textbf{GP4Streets}, ``Pioneering Climate Adaptation in Urban Environments'' (EPSRC, NERC, AHRC, ESRC, MRC and DEFRA APP44894 / UKRI 1281).  Support from Imperial-X's Eric and Wendy Schmidt Centre for AI in Science (a Schmidt Futures program) is gratefully acknowledged.

\section*{CRediT authorship contribution statement}

\textbf{BC:} Investigation, Software, Writing (Review and Editing), Visualisation; 
\textbf{CEH:}; Methodology, Writing (Original Draft, Review and Editing), Funding Acquisition. 
\textbf{CCP:} Conceptualisation, Methodology, Writing (Original Draft, Review and Editing), Software, Supervision, Funding Acquisition.

\bibliography{references}

\appendix
\section{2D \& 3D Convolutional Filters for Diffusion and Advection Discretisations}
\label{A1}

The 3 by 3 filters (for 2D) and the 3 by 3 by 3 filters (for 3D) implemented in the proposed convolutional neural networks are shown below based on a constant grid spacing of $\Delta x=\Delta y=\Delta z$, time step $\Delta t$, constant diffusivity $\nu$ and constant advection velocities $u$, $v$ and $w$. 

2D 5-point stencils for the discretised diffusion and advection operators (with the advection operator split into directions $x$ and $y$) - that is, $A^\nu$, $A^u$ and $A^v$ respectively in Equations~\eqref{eq:burgers_point}, \eqref{AAA} and~\eqref{Tn+1}: 
\begin{equation}
    \frac{\nu \Delta t}{\Delta x^{2}}
    \left(\begin{array}{rrr} 0&-1&0\\-1&4&-1\\0&-1&0\\\end{array}\right), \ 
    \frac{u\, \Delta t}{2\Delta x}
    \left(\begin{array}{rrr} 0&-1&0\\0&0&0\\0&1&0\\\end{array}\right),\ 
    \frac{v\,\Delta t}{2\Delta x}
    \left(\begin{array}{rrr} 0&0&0\\-1&0&1\\0&0&0\\\end{array}\right).
\end{equation}

2D 9-point stencils for the discretised diffusion operator and advection operator (split into $x$ and $y$ directions):
\begin{equation}
    \frac{\nu \Delta t}{4\Delta x^{2}}
    \left(\begin{array}{rrr} -1&-2&-1\\-2&12&-2\\-1&-2&-1\\\end{array}\right),\ 
    \frac{u\,\Delta t}{12\Delta x} 
    \left(\begin{array}{rrr} -1&-4&-1\\0&0&0\\1&4&1\\\end{array}\right).
    \frac{v\,\Delta t}{12\Delta x}
    \left(\begin{array}{rrr} -1&0&1\\-4&0&4\\-1&0&1\\\end{array}\right),\ 
\end{equation}

3D 7-point stencil for the diffusion operator (1st, 2nd and 3rd slices):
\begin{equation}
    \frac{\nu \Delta t}{\Delta x^{2}}
    \left(\begin{array}{rrr} 0&0&0\\0&-1&0\\0&0&0\\\end{array}\right),\ 
    \frac{\nu \Delta t}{\Delta x^{2}}
    \left(\begin{array}{rrr} 0&-1&0\\-1&6&-1\\0&-1&0\\\end{array}\right),\ 
    \frac{\nu \Delta t}{\Delta x^{2}}
    \left(\begin{array}{rrr} 0&0&0\\0&-1&0\\0&0&0\\\end{array}\right).
\end{equation}

3D 7-point stencil for the advection operator in $x$ (1st, 2nd and 3rd slices):
\begin{equation}
    \frac{u\,\Delta t}{2\Delta x}
    \left(\begin{array}{rrr} 0&0&0\\0&0&0\\0&0&0\\\end{array}\right),\ 
    \frac{u\,\Delta t}{2\Delta x}
    \left(\begin{array}{rrr} 0&-1&0\\0&0&0\\0&1&0\\\end{array}\right),\ 
    \frac{u\,\Delta t}{2\Delta x}
    \left(\begin{array}{rrr} 0&0&0\\0&0&0\\0&0&0\\\end{array}\right),  
\end{equation}

in $y$ (1st, 2nd and 3rd slices):
\begin{equation}
    \frac{v\,\Delta t}{2\Delta x}
    \left(\begin{array}{rrr} 0&0&0\\0&0&0\\0&0&0\\\end{array}\right),\ 
    \frac{v\,\Delta t}{2\Delta x}
    \left(\begin{array}{rrr} 0&0&0\\-1&0&1\\0&0&0\\\end{array}\right),\ 
    \frac{v\,\Delta t}{2\Delta x}
    \left(\begin{array}{rrr} 0&0&0\\0&0&0\\0&0&0\\\end{array}\right), 
\end{equation}

and in $z$ (1st, 2nd and 3rd slices):
\begin{equation}
    \frac{w\,\Delta t}{2\Delta x}
    \left(\begin{array}{rrr} 0&0&0\\0&1&0\\0&0&0\\\end{array}\right),\ 
    \frac{w\,\Delta t}{2\Delta x}
    \left(\begin{array}{rrr} 0&0&0\\0&0&0\\0&0&0\\\end{array}\right),\ 
    \frac{w\,\Delta t}{2\Delta x}
    \left(\begin{array}{rrr} 0&0&0\\0&-1&0\\0&0&0\\\end{array}\right).
\end{equation}

3D finite element 27-point stencil for the diffusion operator (1st, 2nd and 3rd slices):
\begin{equation}
    \frac{\nu \Delta t}{26\Delta x^{2}}
    \left(\begin{array}{rrr} -2 & -3 & -2 \\-3 &-6&-3\\-2&-3&-2\\\end{array}\right),\ 
    \frac{\nu \Delta t}{26\Delta x^{2}}
    \left(\begin{array}{rrr} -3&-6&-3\\-6&88&-6\\-3&-6&-3\\\end{array}\right),\ 
    \frac{\nu \Delta t}{26\Delta x^{2}}
    \left(\begin{array}{rrr} -2 & -3 & -2 \\-3 &-6&-3\\-2&-3&-2\\\end{array}\right).
\end{equation}

3D finite element 27-point stencils for the advection operator in $x$ (1st, 2nd and 3rd slices):
\begin{equation} 
    \frac{u\,\Delta t}{72\Delta x}
    \left(\begin{array}{rrr} -1&-4&-1\\0&0&0\\1&4&1\\\end{array}\right),\ 
    \frac{u\,\Delta t}{72\Delta x}
    \left(\begin{array}{rrr} -4&-16&-4\\0&0&0\\4&16&4\\\end{array}\right),\ 
    \frac{u\,\Delta t}{72\Delta x}
    \left(\begin{array}{rrr} -1&-4&-1\\0&0&0\\1&4&1\\\end{array}\right),
\end{equation}

in $y$ (1st, 2nd and 3rd slices):
\begin{equation}
    \frac{v\,\Delta t}{72\Delta x}
    \left(\begin{array}{rrr} -1&0&1\\-4&0&4\\-1&0&1\\\end{array}\right),\ 
    \frac{v\,\Delta t}{72\Delta x}
    \left(\begin{array}{rrr} -4&0&4\\-16&0&16\\-4&0&4\\\end{array}\right),\ 
    \frac{v\,\Delta t}{72\Delta x}
    \left(\begin{array}{rrr} -1&0&1\\-4&0&4\\-1&0&1\\\end{array}\right),
\end{equation}

and in $z$ (1st, 2nd and 3rd slices):
\begin{equation}
    \frac{w\,\Delta t}{72\Delta x}
    \left(\begin{array}{rrr} 1&4&1\\4&16&4\\1&4&1\\\end{array}\right),\ 
    \frac{w\,\Delta t}{72\Delta x}
    \left(\begin{array}{rrr} 0&0&0\\0&0&0\\0&0&0\\\end{array}\right),\ 
    \frac{w\,\Delta t}{72\Delta x}
    \left(\begin{array}{rrr} -1&-4&-1\\-4&-16&-4\\-1&-4&-1\\\end{array}\right).
\end{equation}

\section{Multigrid methods based on CNNs} 
\label{appendi_MGMG}

\begin{figure}[htbp] 
\centering
\includegraphics[width=16cm,angle=0, trim=0mm 0mm 0mm 22mm, clip]{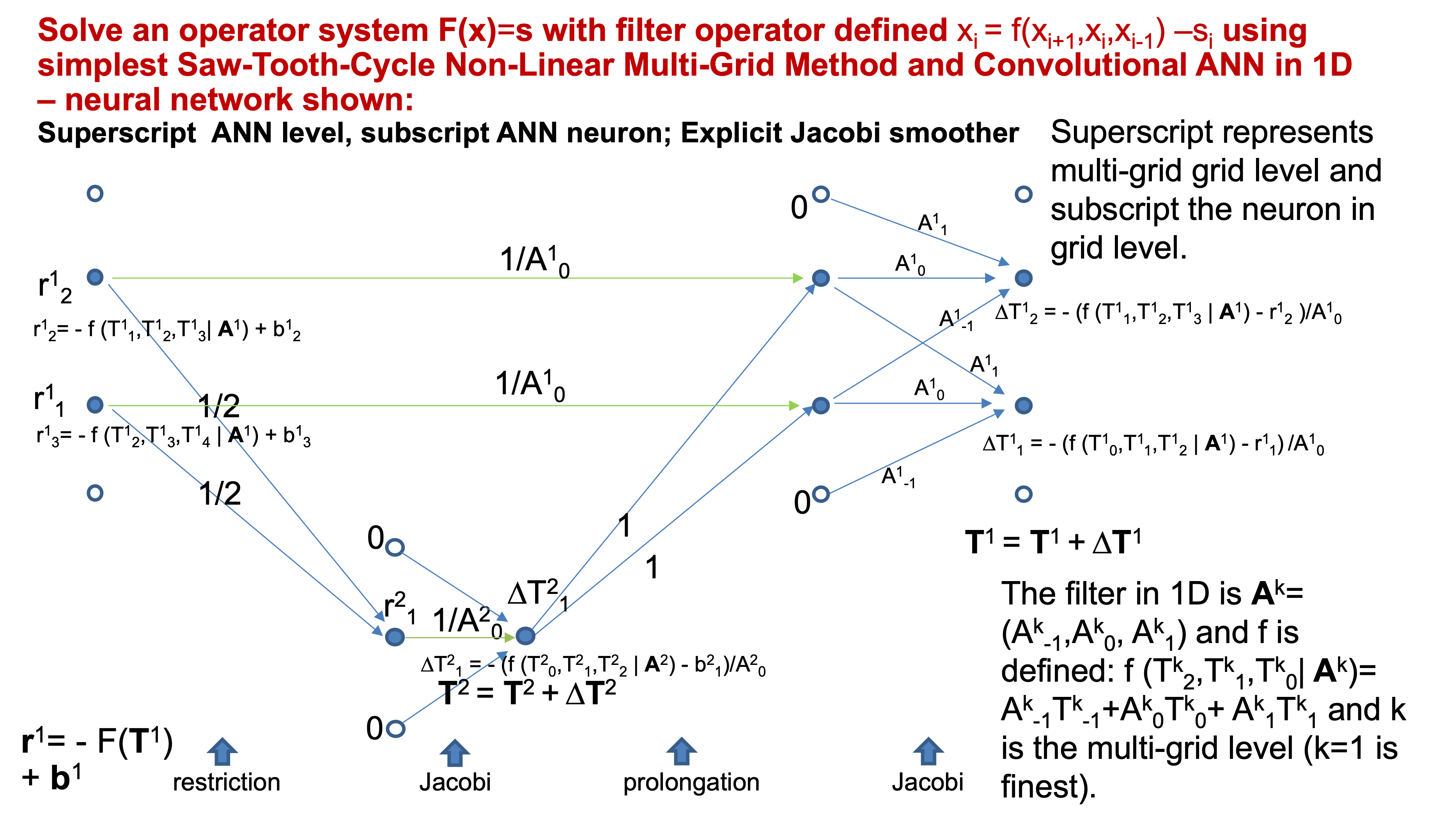}
\caption{Schematic diagram showing how a CNN can be repurposed to produce a multigrid method. The system $A*T=b$ is solved for tensor $T$ using a sawtooth multigrid method and convolutional neural network in 1D. The superscripts indicate the neural network grid level, subscripts the neural network grid neuron. The Jacobi smoother is also shown. Filled blue circles represent interior nodes, open circles represent halo nodes.}
\label{pages1-4-top} 
\end{figure}

\begin{figure}[htbp] 
\centering
\includegraphics[width=5cm,angle=0, trim=0mm 0mm 0mm 10mm, clip]{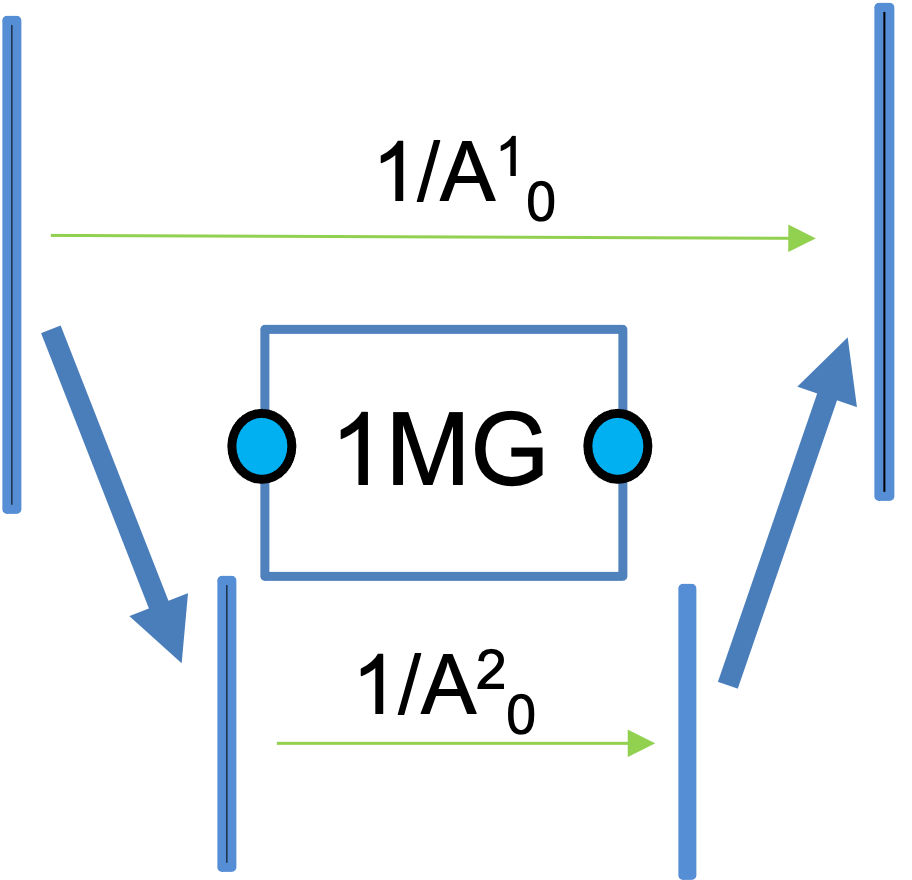}
\caption{Schematic diagram of U-Net or Saw-tooth multigrid cycle with two levels}
\label{pages1-4-top_new} 
\end{figure}

Figures~\ref{pages1-4-top} and \ref{pages1-4-top_new} illustrate how a multigrid method can be represented by a CNN. In Figure~\ref{pages1-4-top}, we have labelled the weights of a convolutional neural network which can mimic a multigrid solver.

for restriction, which can be effected by a single filter. 
The filters change between the layers, when we apply Jacobi relaxation, because the grid spacing $\Delta x$ changes between the layers. 

Suppose we wish to solve $A * \bm{T} = \bm{b}$ for $\bm{T}$ on grid level~$k=1$, then the residual is on grid level~$1$:
\begin{equation}
\bm{r}^1 = \bm{b} - A * \bm{T}
\end{equation}
and the restriction operator is: 
\begin{equation}
\bm{r}^{k+1} = I_R^k * \bm{r}^k, \quad \forall k\in\{1,2,...,M\}, 
\label{r-recur}
\end{equation}
in which $M$ is the number of multigrid levels and $I_R^k$ is the restriction operator. For 1D, this would contain two halves, in 2D it would contain four quarters and in 3D it would contain eight eighths, see Figures~\ref{c-and-r} and~\ref{pages1-4-top}.  Equation \eqref{r-recur} is applied to all grid levels, starting
from the finest grid level $k=1$ to the coarsest grid level $k=M$. 

For 1D, 2D and 3D, the Jacobi method results in:
\begin{equation}
 \bm{\widetilde{{\Delta T}}}^k = \bm{r}^k/A_{0}^k, \;\;\; \bm{\widetilde{{\Delta T}}}^k = \bm{r}^k/A_{0,0}^k, \;\;\; \bm{\widetilde{{\Delta T}}}^k = \bm{r}^k/A_{0,0,0}^k, 
 \label{e19}
\end{equation}
in which division `$/$' here implies that every element in the tensor is divided by the scalar $A_{0}$ or $A_{0,0}$ or  $A_{0,0,0}$. 
After prolongation, the solution correction becomes 
\begin{equation}
\bm{\widetilde{\widetilde{\Delta T}}}^k =\bm{\widetilde{{\Delta T}}}^k + I_P^{k+1} * \bm{T}^{k+1}, \quad \forall k\in\{1,2,...,M-1\}\,,  
 \label{e20}
\end{equation}
although for the final prolongation step, we have
\begin{equation}
\bm{\widetilde{\widetilde{\Delta T}}}^M =\bm{\widetilde{{\Delta T}}}^M, 
 \label{e20b}
\end{equation}
as the operator $I_P^{M}$ contains ones, see Figures~\ref{c-and-r} and~\ref{pages1-4-top}.  At each level, we correct the solution 
using 
\begin{equation}
\bm{T}^k \leftarrow\bm{T}^k + \bm{\widetilde{\widetilde{\Delta T}}}^k, 
 \label{e21}
\end{equation}
and follow this by a smoothing step or Jacobi iteration(s):
\begin{equation}
\bm{\widetilde{\widetilde{\widetilde{\Delta T}}}}^k = - A* \bm{T}^k + \bm{r}^k\,.
 \label{e22}
\end{equation} 
The final update to the solution is
\begin{equation}
\bm{T}^k \leftarrow\bm{T}^k + \bm{\widetilde{\widetilde{\widetilde{\Delta T}}}}^k. 
 \label{e23} 
\end{equation}
Equations~\eqref{e20}, \eqref{e20b}, \eqref{e21}, \eqref{e22} and~\eqref{e23} are applied at each grid level starting from the coarsest $k=M$ to the finest $k=1$.

\end{document}